\documentclass{fac}
\usepackage{longtable}
\usepackage{multirow}
\usepackage{booktabs}
\usepackage{color}
\usepackage{hyperref}
\usepackage{todonotes}

\def\FM{FM}
\ifprodtf \else \usepackage{latexsym}\fi

\newcommand\black{\ensuremath{\blacktriangleright}}
\newcommand\white{\ensuremath{\vartriangleright}}

\newif\ifamsfontsloaded
\ifprodtf
  \newcommand\whbl{\white\kern-.1em--\kern-.1em\black}
  \newcommand\blwh{\black\kern-.1em--\kern-.1em\white}
  \newcommand\blbl{\black\kern-.1em--\kern-.1em\black}
  \newcommand\whwh{\white\kern-.1em--\kern-.1em\white}
  \amsfontsloadedtrue
\else
  \checkfont{msam10}
  \iffontfound
    \IfFileExists{amssymb.sty}
      {\usepackage{amssymb}\amsfontsloadedtrue
       \newcommand\whbl{\white\kern-.125em--\kern-.125em\black}%
       \newcommand\blwh{\black\kern-.125em--\kern-.125em\white}%
       \newcommand\blbl{\black\kern-.125em--\kern-.125em\black}%
       \newcommand\whwh{\white\kern-.125em--\kern-.125em\white}}
      {}
  \fi
\fi

\title[A Survey of Practical Formal Methods for Security]
      {A Survey of Practical Formal Methods for Security}

\author[T. Kulik, B. Dongol, P.G. Larsen, H.D. Macedo, S. Schneider, P.W.V. Tran-J\o{}rgensen and J. Woodcock]
    {Tomas Kulik$^1$, Brijesh Dongol$^2$, Peter Gorm Larsen$^1$, Hugo Daniel Macedo$^1$, 
     \and Steve Schneider$^2$, Peter W\"urtz Vinther Tran-J\o{}rgensen$^3$ and Jim Woodcock$^{1,4}$\\
     $^1$Aarhus University\\
     $^2$University of Surrey\\
     $^3$Bankdata\\
     $^4$University of York}

\correspond{Tomas Kulik, Aarhus University, Department of Electrical and Computer Engineering,
            Finlandsgade 22, 8200, Aarhus, Denmark.
            e-mail: tomaskulik@ece.au.dk}

\pubyear{2020}
\pagerange{\pageref{firstpage}--\pageref{lastpage}}
\begin{document}
\label{firstpage}

\makecorrespond

\maketitle

\begin{abstract}
In today's world, critical infrastructure is often controlled by computing systems. This introduces new risks for cyber attacks, which can compromise the security and disrupt the functionality of these systems. It is therefore necessary to build such systems with strong guarantees of resiliency against cyber attacks. One way to achieve this level of assurance is using formal verification, which provides proofs of system compliance with desired cyber security properties. The use of Formal Methods (FM) in aspects of cyber security and safety-critical systems are reviewed in this article. We split FM into the three main classes: theorem proving, model checking and lightweight FM. To allow the different uses of FM to be compared, we define a common set of terms. We further develop categories based on the type of computing system FM are applied in. Solutions in each class and category are presented, discussed, compared and summarised. We describe historical highlights and developments and present a state-of-the-art review in the area of FM in cyber security. This review is presented from the point of view of FM practitioners and researchers, commenting on the trends in each of the classes and categories. This is achieved by considering all types of FM, several types of security and safety critical systems and by structuring the taxonomy accordingly. The article hence provides a comprehensive overview of FM and techniques available to system designers of security-critical systems, simplifying the process of choosing the right tool for the task. The article concludes by summarising the discussion of the review, focusing on best practices, challenges, general future trends and directions of research within this field.
% In today’s world, critical infrastructure is often controlled by computing systems, introducing risks for cyber attacks, capable of disrupting the functionality of these systems. One way to achieve high level of security assurance is using formal verification, which provides proofs of system compliance with desired cyber security properties. The use of Formal Methods (FM) in aspects of cyber security and safety-critical systems are reviewed in this article. We split FM into the three main classes: theorem proving, model checking and lightweight FM. We further develop categories based on the type of computing system FM are applied in. Solutions in each class and category are presented, discussed, compared and summarised. We describe historical highlights and developments and present a state-of-the-art review in the area of FM in cyber security. This review is presented from the point of view of FM practitioners and researchers, commenting on the trends in each of the classes and categories. The article hence provides a comprehensive overview of FM and techniques available to system designers of security-critical systems, simplifying the process of choosing the right tool for the task. The article concludes by summarising the discussion of the review, focusing on best practices, challenges, general future trends and directions of research within this field.
\end{abstract}


% !TEX root = ../main.tex
\section{Introduction}
Digital services are currently spreading to all aspects of society~\cite{Rana&15}. This in turn causes dependence of society on the cyber infrastructure needed to support these services. The heavy reliance on cyber infrastructure poses new challenges in the form of cyber attacks and potentially cyber terrorism~\cite{Kenney&15}, with threat actors encompassing the full range from interpersonal offenders, cyber criminals and ``hacktivists'' through to well-resourced state actors~\cite{cybok}.  Disturbances in financial, industrial or day-to-day consumer services could lead to significant financial and societal costs. As digitisation spreads further, the potential attack surfaces only grow larger, increasing the challenge of protecting digital services~\cite{Urbach&19,Williams&16}. As systems grow larger and more complex, significant resources have to be spent to secure these system against known cyber attacks. Often the protection mechanisms are incorporated to close vulnerabilities uncovered after a successful cyber attack, and hence are of a reactive nature. This approach relegates cyber security from a primary challenge to be solved within the system to an afterthought~\cite{Steward&12}.

Due to the wide spectrum of cyber attacks, it is difficult to directly quantify their impact on society~\cite{Gandhi&11}, however very often they involve significant financial costs as well as potential disruptions in quality of life. One example is a potential cyber attack against electricity infrastructure, including electricity marketplace, which could lead to destruction of generators and disclosure of confidential data~\cite{Pincetic&09}. Another example is attacks against manufacturing facilities causing delays or decrease in quality of production~\cite{Bracho&18},~\cite{Quarta&17}. These examples demonstrate that cyber threats should be considered as significant as physical threats against societal infrastructure.

The earlier the potential cyber security threats are discovered within new systems, the cheaper the mitigation for these threats will be~\cite{Wardell&16}. Formal Methods (FM) provide an opportunity for discovery and mitigation of cyber threats at all stages of the lifecycle of a system. Using FM brings mathematical rigour to the field of cyber security assurance. This is possible since FM are techniques that use model-based approaches, where the models are rigorously specified~\cite{Wing&90}. These models represent the software, hardware or a combination of the two for the system in question. The primary benefit of using FM stems from the mathematical proof of the internal consistency of the system design~\cite{Hall&05}. This proof provides strong assurances since it considers the entire system behaviour, and once proven true it remains true, whereas in traditional testing it is only possible to cover specific scenarios. FM can be seen as a tool well suited for providing assurances of cyber security for digital society~\cite{Wing&98}. Beyond the assurance of behavioural correctness of a system, the adoption of a fully fledged formal approach is known to reduce the number of implementation errors, which are the building blocks of exploits.

Within the area of FM there are distinct approaches. The main categories we consider are: 
\begin{itemize}
\item \textit{Theorem Proving}, analysing a formal description for important properties based on computer-based proofs.
\item \textit{Model Checking}, checking whether a finite-state model of a system meets a given specification in an exhaustive manner.
\item \textit{Lightweight FM}, using formal techniques to analyse a system either statically or dynamically (this concept was coined in \cite{Jackson&96c} but we have extracted the model checking from their characterisation into a category of its own).
\end{itemize}

In all cases, the methods are applied to determine if a system behaves in a correct way and many approaches have received significant tool support for automation of the verification and validation process~\cite{Alglave&11}. In this survey we consider all of the approaches and their application to specific areas of digital society. We further consider FM as applied to the specific level of abstraction of system behaviour ranging from the application level to the hardware level. By considering the state-of-the-art research in formal verification across these dimensions, we provide a non-exhaustive overview of application of FM in specific disciplines. The aim of this survey is to allow practitioners to identify a proven method applicable to a system in their domain, hence increasing the adoption of FM in the field of cyber security.

%Further reviews include works by Koblitz and Menezes \cite{KM19}, who describe some of the major pitfalls in provable security, including underspecification of requirements, incorrect (overly strong) assumptions, and errors in the proofs themselves.

% !TEX root = ../main.tex
\subsection{Methodology}

The amount of research publications within the area of applying FM towards cyber security challenges is significant. Therefore several constraints have been placed on the choice of research publications to be considered within this survey. The first important constraint is the recency of the research reported, considering the landscape of \emph{the last decade}, limiting the publication date to be no earlier than 2012. Furthermore all of the research work need to be \emph{published in scientific venues} such as journals, conferences or workshops. The next constraint is focus on computer-based tool supported formal methods, i.e.\ only formal methods with tools that can provide \emph{computer-based analysis} and often guide users on performing this analysis are considered. This consideration is in order to focus more on the FM that could be potentially applied outside of academia, bringing the benefits of the \emph{formal security analysis to industry}. This goes hand in hand with our focus on the applied FM, searching for research publications, where a tool supported FM is utilised to deal with \emph{a concrete cyber security problem}. Hence this survey does not focus on theoretical advances of FM in security or proposed processes that briefly mention use of formal methods, such as theoretical approaches to model checking algorithms, specification of hyperproperties and similar. Furthermore our survey does not cover the approach to security commonly referred to as {\em provable security}. This refers to a mathematical approach to analysing the security of cryptographic mechanisms or systems. The approach considers the system in the context of an attacker model, and expresses the security requirements within that model as a limitation on what the attacker should be able to achieve.  A proof consists of establishing that the attacker would need to break a known hard problem (such as the Quadratic Residuosity Problem \cite{gm84}) in order to break the security of the system.  Thus the security of the system is reduced to the difficulty of the underpinning hard problem.  This approach is typically used within the field of cryptography rather than secure systems, and so falls outside the scope of our survey. We point the reader to~\cite{Barbosa&19} providing the report within the area of FM in cryptography. Finally, we constrain our search to research that considers aspects of security explicitly, and not as a by-product of safety or correctness. The search for the research publications was carried out as a cross database search using Google Scholar, while focusing on research papers, excluding research abstracts or extended abstracts.

As this survey shall provide a reader with a quick overview of the research conducted, we have further decided to categorise different research publications by the industry (domain) on which they focus as well as the level of abstraction on which the formal method is utilised. In this way researchers and also potentially industrial users can quickly find the area of their interest within this survey. Furthermore, we classify the research based on the cyber security problem classification as elicited from the discovered research papers and inspired by existing literature~\cite{cybok}.

% !TEX root = ../main.tex
\subsection{History}

%\todo{Editor: JW}

This section presents a history of impactful research works within FM in security over the last 40 years. We choose four case studies where formal methods have been applied to secure systems:
\begin{enumerate}
\item The Needham-Schroeder Public-Key Protocol. Lowe used a refinement model checker to find a triangular attack on the protocol. This was a new attack on a protocol that had previously been proven correct by Burrows et al.~\cite{BurrowsAN1990}.
\item The Mondex smartcard. This was the first commercial product to be certified to ITSEC Level E6. There was considerable discussion at the time as to whether this was even possible.
\item The Tokeneer ID Station. There were similar questions about the feasibility of using FM to achieve the level of rigour required by the higher assurance levels of the Common Criteria. Tokeneer settled this matter.
\item The seL4 Microkernel. This system has the reputation of being the world's most assured microkernel. Significantly, it demonstrates that security and the use of formal methods do not lead to poor performance.
\end{enumerate}

\paragraph{The Needham-Schroeder Public-Key Protocol}

\begin{figure}
  \setlength{\unitlength}{0.8cm}%
  \begin{picture}(17,5)
    \linethickness{1pt}%
    \put(1,5){\vector(0,-5){4}}%
    \put(4,5){\vector(0,-5){4}}%
    \put(1,4){\vector(4,0){3}}%
    \put(4,3){\vector(-4,0){3}}%
    \put(1,2){\vector(4,0){3}}%
    \put(2.5,4.1){\makebox(0,0)[b]{$\{A,N_A\}_{pk_B}$}}%
    \put(2.5,3.1){\makebox(0,0)[b]{$\{N_A,N_B\}_{pk_A}$}}%
    \put(2.5,2.1){\makebox(0,0)[b]{$\{N_B\}_{pk_B}$}}%
    \put(1,5.2){\makebox(0,0)[b]{\large $A$}}%
    \put(4,5.2){\makebox(0,0)[b]{\large $B$}}%
    \put(2.5,0){\makebox(0,0)[b]{(a)}}%
    \put(10,5){\vector(0,-5){4}}%
    \put(13,5){\vector(0,-5){4}}%
    \put(16,5){\vector(0,-5){4}}%
    \put(10,4.1){\vector(4,0){3}}%
    \put(13,2.9){\vector(-4,0){3}}%
    \put(10,2.1){\vector(4,0){3}}%
    \put(11.5,4.2){\makebox(0,0)[b]{$\{A,N_A\}_{pk_I}$}}%
    \put(11.5,3.0){\makebox(0,0)[b]{$\{N_A,N_B\}_{pk_A}$}}%
    \put(11.5,2.2){\makebox(0,0)[b]{$\{N_B\}_{pk_I}$}}%
    \put(13,3.9){\vector(4,0){3}}%
    \put(16,3.1){\vector(-4,0){3}}%
    \put(13,1.9){\vector(4,0){3}}%
    \put(14.5,4.0){\makebox(0,0)[b]{$\{A,N_A\}_{pk_B}$}}%
    \put(14.5,3.2){\makebox(0,0)[b]{$\{N_A,N_B\}_{pk_A}$}}%
    \put(14.5,2.0){\makebox(0,0)[b]{$\{N_B\}_{pk_B}$}}%
    \put(10,5.2){\makebox(0,0)[b]{\large $A$}}%
    \put(13,5.2){\makebox(0,0)[b]{\large $I$}}%
    \put(16,5.2){\makebox(0,0)[b]{\large $B$}}%
    \put(13,0){\makebox(0,0)[b]{(b)}}%
  \end{picture}
  \caption{(a) Needham-Schroeder authentication protocol, and (b) attack} \label{fig:needhamschroeder}
\end{figure}
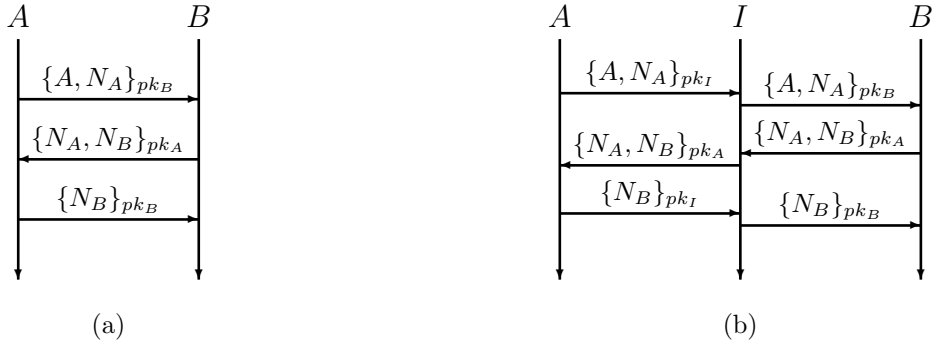

The Needham-Schroeder Public-Key Protocol is a transport-level protocol for communication between network devices~\cite{NeedhamS1978}, providing mutual authentication between two parties in a network. The protocol is visualised in Figure~\ref{fig:needhamschroeder}(a).  Simple and well known, it has become a popular benchmark for testing security protocol verification technology. We discuss it here because it is an important security protocol that nevertheless contained a significant error. This error was found using automated FM.

\cite{Lowe1995} showed that, contrary to its intention, the protocol fails to ensure authentication. In particular, he demonstrated that an intruder can impersonate an agent $A$ during a run of the protocol. The impersonator tricks another agent $B$ into thinking that they are talking to $A$.

The protocol uses public key cryptography. Each agent $A$ possesses a public key, which any other agent can get from a server. $A$ also possesses a secret key that is the inverse of its public key. Any agent can encrypt a message using $A$'s public key, but only $A$ can decrypt it, ensuring secrecy. %$A$ can sign a message by encrypting it with its secret key. Any other agent in possession of $A$'s public key can then decrypt this message. This ensures that the message did in fact originate from $A$.
The protocol also uses \emph{nonces}: random numbers coined for single runs of the protocol.

Lowe encoded the protocol in CSP \cite{Hoare85} and analysed it using CSP's model checker, FDR~\cite{FDR42019}. Lowe did not know in advance that an attack was possible, although he may have suspected it. He did not know where to look in the protocol for a vulnerability, but the exhaustive search carried out by the model checker found an attack in spite of this.

Suppose that $I$ (an intruder) is a network user who can take part in network sessions. $I$ can also intercept messages and inject new ones, but is not able to decrypt messages without the key. $I$ can produce a new message in two circumstances: if $I$ invents the nonce, or if $I$ already understands the message's contents. This intruder can also replay complete encrypted messages, even without understanding the contents \cite{MASP}. This approach is commonly known as the Dolev-Yao model \cite{DolevY81}.

The attack involves two simultaneous runs of the protocol, as shown in Figure~\ref{fig:needhamschroeder}(b). $A$ establishes a valid session with $I$. At the same time, $I$ impersonates $A$ to establish a fake session with $B$. The flawed run of the protocol could be explained as follows. $A$ sends a message with nonce $N_A$ to $I$, who decrypts the message with $I$'s secret key. $I$ relays the message to $B$, pretending to $B$ that $A$ is communicating. $B$ sends $N_B$ in response, encrypted for $A$, and so $I$ relays this encrypted nonce to $A$. $A$ decrypts $N_B$ and confirms it to $I$, who learns it. $I$ re-encrypts $N_B$ and returns it to $B$, which convinces $B$ that $A$ is the other party.  At the end of the attack, $B$ falsely believes that $A$ is the communication partner, and that only $A$ and $B$ know $N_A$ and $N_B$. This shows that the protocol is insecure. Protocol analysts call this a \emph{man-in-the-middle attack}. Here, it has been discovered automatically.

\paragraph{Mondex}

The Mondex application consists of smart cards with electronic purses (wallets) for electronic commerce~\cite{StepneyCW2000}. Customers use Mondex smart cards for low-value, cash-like transactions that need no third-party involvement. The Bank of England (the financial regulator in this instance) considered the requirements for Mondex to be security critical: Mondex must have no implementation or design bugs that could allow electronic counterfeiting. So the developers certified Mondex to the highest standard available at the time. This was ITSEC Level E6~\cite{ITSEC2006}, equal to Common Criteria Evaluation Assurance Level~7~\cite{CCRA2006}.%
\footnote{%
  The levels of the Common Criteria are:
  \begin{flushleft}
    \begin{tabular}{ll@{\quad}ll}
      EAL1: & Functional testing.                        & EAL5: & Semi-formal design and testing.            \\%
      EAL2: & Structural testing.                        & EAL6: & Semi-formally verified design and testing. \\%
      EAL3: & Methodical testing and checking.           & EAL7: & Formally verified design and testing.      \\%
      EAL4: & Methodical design, testing, and reviewing. &                                                    \\%
    \end{tabular}
  \end{flushleft}
} %
Mondex was the first commercial product to achieve ITSEC Level E6 (EAL7).

\cite{StepneyCW2000} further describe the development of the Mondex application, with its abstract and concrete models. The abstract model describes the world of electronic purses: atomic transactions exchange value and the abstract model expresses their required security properties. The concrete model is the purse design and the message protocol for value exchange.

\def\fuzz{{\large\it f\kern0.1em}{\normalsize\sc uzz}}%

The design team used the Z notation~\cite{Spivey1989,WoodcockD1996} to specify both models. They proved that the concrete model is formally a refinement of the abstract one.  This means that the concrete model respects all the abstract security requirements. The abstract model and its security properties is often easier to understand than the concrete model. Developers wrote manual proofs, believing that no efficient automated tools existed for such a large task. Instead, proof steps were type-checked using the \fuzz\footnote{See \url{spivey.oriel.ox.ac.uk/corner/Fuzz_typechecker_for_Z}.} and Formaliser tools~\cite{FlynnHB89}. Proofs were also checked by independent external evaluators.

There were four principal security properties:
\begin{itemize}
\item The system and its users may not create value.
\item The system must account for all value.
\item Purses must have enough value for their intended transaction.
\item All transfers must be between authentic purses.
\end{itemize}
The design team changed a secondary protocol after the proof revealed a bug. A detailed account of the project is given in~\cite{WoodcockSCCJ2008}. Mondex has proved to be a dependably secure system, guaranteed by its formal development.

\paragraph{Tokeneer}

The Tokeneer system was developed by the US National Security Agency (NSA)~\cite{BarnesCJWCE2006}.%
\footnote{%
  For comprehensive information on Tokeneer, see the AdaCore webpages \url{www.adacore.com/tokeneer}, where the entire project archive can be downloaded. AdaCore distribute the material generated by Altran under contract to the NSA under the terms of the Technology Transfer Agreement agreed between Praxis and the NSA. This material consists of all the core and support software for the Tokeneer ID Station, project documents, test cases derived from the system test specification, test tokens, and biometric data.
} It provides secure access to an enclave of workstations with controlled physical entry. Access control requires biometric checks and security tokens. These tokens describe a user's permitted actions within a particular visit to the enclave.

Developers needed to assure the security properties. They did this by conformance with the Common Criteria Evaluation Assurance Level 5~\cite{CCRA2006}. They also needed to show they could do this in a cost-effective way. NSA invited bids to use FM to develop a component of the Tokeneer system, and then monitored this experiment to measure the effort and skills needed to perform the development.

Praxis (a UK company) won the contract and wrote a formal specification in Z~\cite{Spivey1989,WoodcockD1996}, formally refining the specification to a SPARK program. SPARK is a subset of Ada with an accompanying tool-set~\cite{Barnes2012}. They proved key system properties and the absence of run-time errors, using traditional methods to develop extra software. These extra Ada programs provided interfaces with peripherals.

The project required 260 person-days, three people part-time, and nine months' elapsed time. It produced about 10k lines of SPARK code with about 16.5k contracts. About 200loc were written on average per day during the implementation phase, with about 40loc through the entire project. A further 3.5k lines of standard Ada code were produced, with about 200loc per day in the implementation phase or 90loc throughout the project. System testing took about 4\% of the project effort, much smaller than usual.

Two defects were found in Tokeneer. One was found using formal analysis, another was found by code inspection.\footnote{Diomidis Spinelli: \url{www.spinellis.gr/blog/20081018/}.} The testing team discovered two in-scope failures: missing items in the user manual.

The task set by NSA was to conform to Common Criteria EAL5. The Tokeneer development actually exceeded EAL5 requirements in several areas: configuration control, fault management, and testing. Although the main body of the core development work was carried out to EAL5, the development areas covering specification, design, implementation, correspondence were accomplished to EAL6 and EAL7.  Why? Because it was cheaper!

\paragraph{The seL4 Microkernel}

The third-generation microkernel seL4 provides abstractions for virtual address spaces, threads, and inter-process communication. It provides an explicit memory management model and capabilities for authorisation. There is a guarantee that the binary code of the ARM version of the seL4 microkernel is a correct implementation.%
\footnote{%
  On the ARM platform, there is a further proof that the binary code that executes on the hardware is a correct translation of the C code for sel4. This means that the compiler does not have to be trusted, and extends the functional correctness property to the binary. See \url{docs.sel4.systems/FrequentlyAskedQuestions.html}.%
} %
seL4 meets its abstract specification and \emph{does nothing else}. In particular, the seL4 ARM binary meets the classic security properties of integrity and confidentiality.

The seL4 micro-kernel has a formal proof of its C code against its abstract specification~\cite{KleinAEHCDEEKNSTW2010}. This proof is machine checked in Isabelle/HOL~\cite{NipkowK2014}. This assumes correctness of boot code, cache management, hardware, and hand-written assembly code.

The developers claim seL4 to be the only verified general-purpose operating system (OS) kernel. An operational model of the system forms as an abstract specification. A Haskell program prototypes the kernel. This prototype provides an automatic translation into Isabelle/HOL. The Isabelle code is then an executable, design-level specification of the kernel. This is hand coded in C to form a high-performance C implementation of seL4. Refinement proofs link the specifications and the C code. Developers proved that attackers cannot subvert the kernel. Not even if they use buggy encodings, spurious calls, or buffer overflow attacks.

%%% Local Variables:
%%% mode: latex
%%% TeX-master: "../main"
%%% End:

\subsection{Definitions/Background}

% \todo{BD: 1page}
% Here we will put the common terms definitions (e.g.: model, properties, process, channels \ldots).

Here we provide a set of common terms and definitions used throughout this article. This is particularly important since the fields of formal verification and security developed independently for many years, and hence, some terms are overloaded and have slightly different meanings depending on the context in which they are used. For example, in security (particularly cryptography), a \textbf{\em certificate} refers to a document that is used to bind an entity to a cryptographic key.  On the other hand, within FM, \textbf{\em certificate} is used as a proof of correctness of a system or protocol.

Throughout this article, we use \textbf{\em authentication} to refer to the process of \emph{identifying} and \emph{validating} whether a user (an entity or individual) accessing a system is who the user claims to be. This is in contrast with \textbf{\em authorisation}, which is the process of \emph{allowing} a user access to a system based on their identity.

The systems under consideration typically comprise a set of coordinated \textbf{\em processes}, which are program instances defining a set of instructions that are executed by one or more threads. We think of processes as being active entities in a system, as opposed to programs which are passive entities. Formal frameworks to describe the behaviour of processes include CSP, CCS, ACP, $\pi$-calculus, etc. Processes typically implement \textbf{\em protocols}, i.e., a set of rules for transmission of data, and may synchronise over \textbf{\em shared memory} or communicate over a \textbf{\em channel}, which is an abstraction of a physical communication network. Shared memory implementations are increasingly complex due to the use of intermediate processor caches and may implement many different consistency models~\cite{AdveG96}. Similarly, one may place many different assumptions on a channel, e.g., FIFO ordering of messages; whether the channel guarantees integrity, availability and confidentiality; whether the channel is error free; whether message types can be distinguished, etc. Security protocols are often designed to provide specific properties such as \textbf{\em isolation}, which is a design principle in which processes are separated and given privileged access to shared resources, e.g., shared memory (typically using techniques such as containerisation or virtualisation).

In general, verification is with respect to a \textbf{\em specification}, which is an abstract (formal or informal) description of the allowable behaviours of an entity, e.g., hardware, a system, computer program, data structure etc. Formal verification often proceeds with respect to a \textbf{\em model} of a system, which provides a precise formal description of an entity, capturing the key characteristics of the entity being modelled. One must ensure that every feature described by a model is an actual feature of the entity. Different models of the same entity may be developed depending on the properties that are of interest; a computer program for example may be modelled by relations between pre/post state; traces of states; functions between inputs and outputs, etc. A model may describe behavioural functionality, protocols etc. In FM, one typically develops models at several levels of abstraction, with precise descriptions of the relationship between these levels. FM for security also requires a model of an attacker, e.g., the Dolev-Yao model, which is used in the context of communicating systems. Within hardware verification, the term \textbf{\em co-verification} is used to prove that system software executes correctly on a \emph{representation} of the underlying hardware design. It enables integration of software with hardware, before any physical devices (e.g., chips or boards) are available.

The aim of verification is to ensure that it meets its \textbf{\em implementation} (of a specification), i.e., the physical manifestation of an entity. In some instances, one may refer to an implementation as a model that provides enough detail about an entity for the corresponding physical entity to be readily obtained.

%%% Local Variables:
%%% mode: latex
%%% TeX-master: "../main.tex"
%%% End:

%%% Local Variables:
%%% mode: pdflatex
%%% TeX-master: "../main.tex"
%%% End:

% !TEX root = ../main.tex
\section{Survey}\label{sec:survey}

\subsection{Categorisation and overview}
\label{subsec:overview}
% !TEX root = ../main.tex
Since the FM in security are applied across many domains, we structure the scope of the survey by presenting a categorisation based on a domain and a level of abstraction. This is done in order to provide a systematic overview of the wide field of FM in security. The labels for the four domains we have selected are: 

\begin{description}
\item[\textbf{Financial:}] Aggregates the works applying FM in the area of finance/money as payment systems, home banking, financial markets, crypto-currencies. Examples are mobile banking apps, ATM infrastructures, the FIX stock exchange protocol, smart-cards/hardware wallets. The financial domain is represented in section~\ref{subsec:fin}.
\item[\textbf{Industrial:}] This label agglomerates works dealing with computing systems applied in the production of goods or services, manufacturing and industrial control. Examples are a Water Treatment Management Panels, PLC control networks, Modbus/TCP, motor controllers. The industrial domain is represented in section~\ref{subsec:ind}.
\item[\textbf{Consumer:}] It categorises works focusing on the security of end-user/individuals personal computation devices and applications such as a command-line shell, a home operating system, a Voice over IP protocol, and an exercise smart appliance. The consumer domain is represented in section~\ref{subsec:con}.
\item[\textbf{Enterprise:}] This is the dual of the Consumer category, as it is used to group the works focusing on the security of corporate systems providing computing services satisfying the needs of organisations instead of individuals. Examples are email services, e-government systems, the sn2 protocol, data servers warehouses. The enterprise domain is represented in section~\ref{subsec:ent}.
\end{description} 

\begin{figure}
  \centering
  \includegraphics[width=1\textwidth,bb=0 0 530 530]{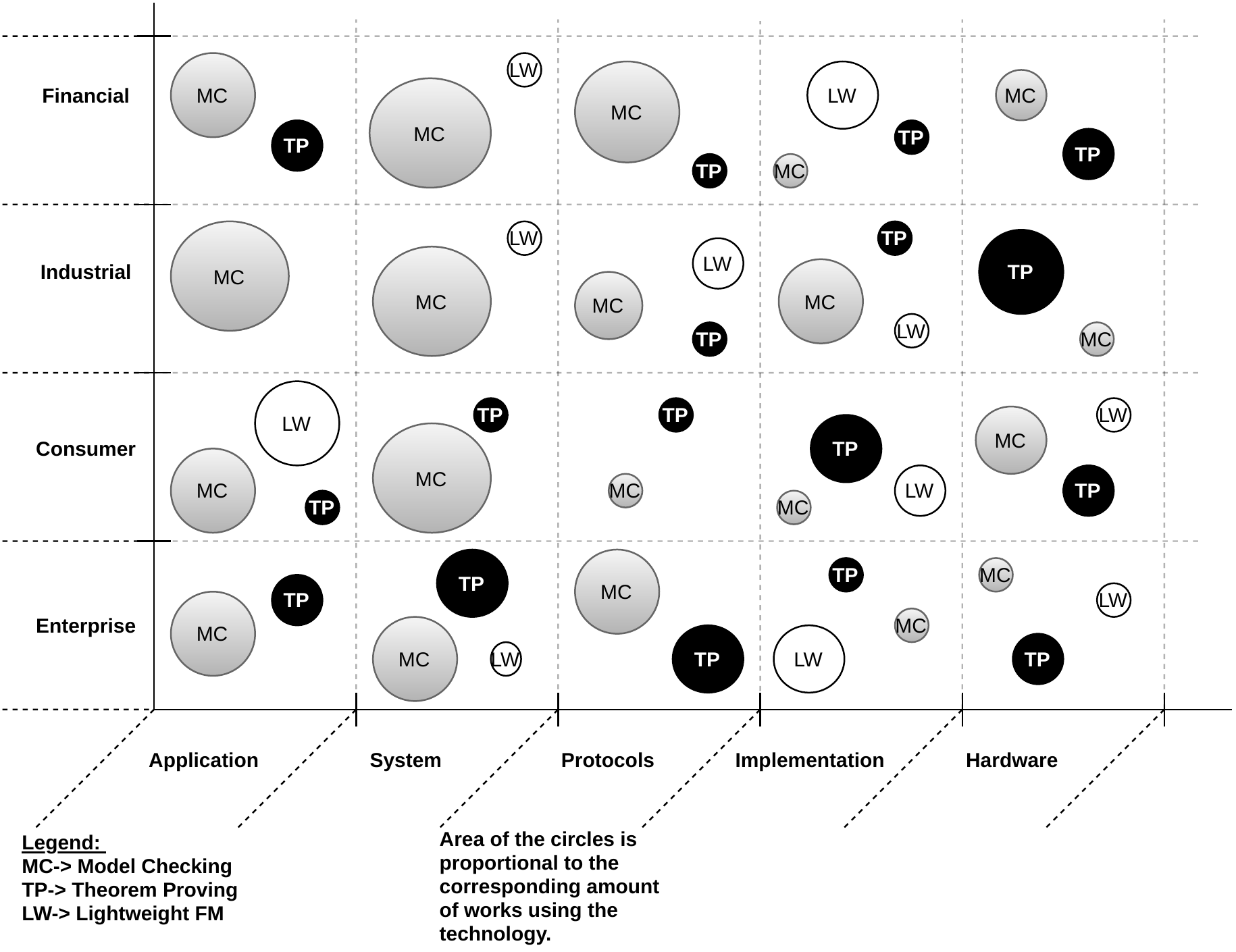}
  \caption{Classification of formal methods in security}
  \label{fig:taxonomy}
\end{figure}

As presenting the four domains would only separate the FM in security research by the field of application, we further present five levels of abstraction at which the formal verification is carried out. These levels of abstraction are:
\begin{description}
\item[\textbf{Application:}] Used for works that apply FM for security at the %on a purpose of 
application or purpose of computation %computation 
level.
\item[\textbf{System:}] Used for works that apply FM for security within the architectural level, often encompassing multiple subsystems. 
\item[\textbf{Protocols:}] It is used to apply FM to assert properties or analyse communication protocols between system components level. 
\item[\textbf{Implementation:}] This is a cross-cutting category encompassing all the works that focus on application/usage of FM directly on the resulting system (e.g.: run-time monitoring) instead of emphasis on designs and specifications.
\item[\textbf{Hardware:}] Used to classify works applying FM in the process of hardware development.
\end{description}
This categorisation allows us to systematically review the state-of-the-art-research and provide an overview based on this. In order to provide a clear overview of FM in security we further apply a third dimension, defining the type of the FM used, i.e.\ model checking, theorem proving and lightweight FM. This provides a quantitative overview of different research works within Figure~\ref{fig:taxonomy}.

% PGL the next paragraph is already in the methodology subsection In order to limit the scope of the survey, focus is on applied FM in security. This means that in the research works considered, FM is used to verify security properties of either existing or reference applications, systems, protocols, implementations or hardware. We do not consider purely theoretical research aiming at theoretical advancement of FM themselves. It is also important to note that specific research works often do not fit completely under one domain, hence a distinction is made based on how close each given work is to a given domain.

The sections within this survey follow a logical organisation, where the research works are grouped together by the type of the application, system, protocol, implementation or hardware that they are applied to. Within this grouping, the research works are further organised into paragraphs following logical categorisation. As an example the first paragraph could consider works related to manufacturing while the next paragraph considers works related to industrial control. In both cases the research is aimed at industrial domain but with a different scope. Each section within the survey represents a single domain, where we present a systematic summary of research works belonging to different levels of abstraction.

% As an example, first paragraph within the section~\ref{sec:indSystem} discusses formal methods in security of control aspects of industrial control systems, i.e. the internal control network, while the second paragraph considers control system including aspect of connectivity towards the Internet and cloud systems. Each section represents a single domain and under each domain the survey presents one page summary of research works belonging to single level of abstraction.

%%% Local Variables:
%%% mode: latex
%%% TeX-master: "../main.tex"
%%% End:

% \newpage

% \newpage
% \subsection{Survey quick view}
% \label{subsec:table}
% \input{sections/table/tableOverview}

% \newpage{}
% \clearpage
\subsection{Financial}
\label{subsec:fin}

Financial computing including banking systems, independent budgeting applications and mobile payment applications is a rapidly developing field. This section provides an overview of how FM have been used to analyse the security of banking mobile applications, alternative currencies, such as cryptocurrencies, smart contracts, banking backend systems, electronic trading systems, payment protocols, cryptocurrency hardware and wallets. On a different levels of abstraction it could be seen within this section that the \textit{system} level consists of most research works. This section also mentions the legal challenges when applying FM to financial systems. These legal challenges arise from limited access to these systems and a certain level of avoidance of publication of potential vulnerabilities by vendors of these systems, i.e.\ often making it difficult for researchers to get deep insight into these systems.

\begin{figure}
  \centering
  \includegraphics[width=.8\textwidth]{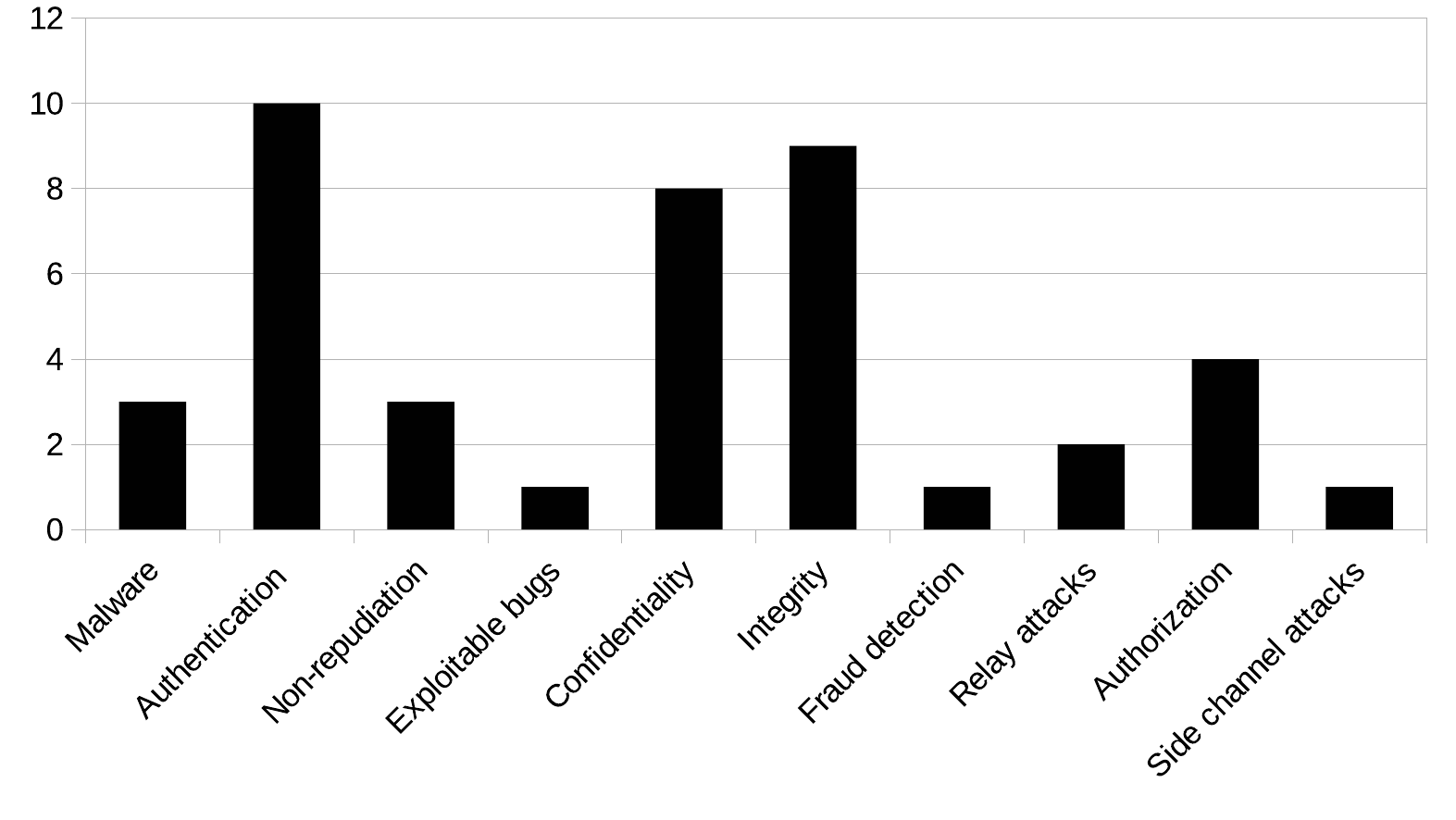}
  \caption{Cyber security topics in the financial section}
  \label{fig:finTopics}
\end{figure}

The method most used to analyse security within the financial domain is model checking, and authors apply different model checkers to specific problems. This could be attributed to the fact that the different entities whose security is being analysed lend themselves well to be modelled in a state transition representation and also that their state space is sufficiently limited to be analysed without issues such as the state space explosion. The cyber security topics present within the financial section are shown in Figure~\ref{fig:finTopics}.

% The field of financial applications is expanding rapidly. Not only do banks provide mobile applications, but whole alternative currencies are being developed. This rapid growth provides many opportunities for application of FM to security on an application level.

\subsubsection{Application}
% !TEX root = ../../main.tex
The field of financial applications is expanding rapidly. Not only do banks provide mobile applications, but whole alternative currencies are being developed. This rapid growth provides many opportunities for application of FM to security on an application level.

% The application category purports the most momentum in the financial area.  Whereas hardware, systems, and protocols move at the slow pace of institutional changes and regulations, the new gains are expected to be in the recent possibilities to interacting with clients via apps relying on TLS/web security, developing alternative currencies, and the new Fintech that thrives in the yet to be regulated, shady but dynamic domain.

The introduction of apps and reliance on web-based security paves the way to the application of FM to them. A recent example of that is found within~\cite{Chothia&17}. The authors discovered several vulnerabilities while analysing apps from 15 leading UK banks. The usage of ProVerif to formally verify their proposed correction to one of the identified flaws markets the usage of the techniques in the optimal approach. Applying FM to banking apps provides potential for further research, but legal restrictions apply.

Another area of FM applicability is banking malware. The major cybersecurity companies report an increase in financial malware, and expect to see more attacks and growth in this area.  One of the main categories is banking Trojans, which hide in subsystems such as the Android Accessibility Service API and may steal users' credentials. The authors of~\cite{Kumar&19} have analysed Android applications by first using the Krakatau bytecode tool\footnote{\url{github.com/Storyyeller/Krakatau}} to generate the Java byte-code of the application, then generating the Calculus of Communicating Systems (CCS)~\cite{Milner&89} model from the byte-code and finally dispatching it to the Concurrency Workbench of New Century (CWB-NC) model checker~\cite{Cleaveland&96}, searching for malware properties. The authors note that their approach provides over 98\% malware detection rate. Similarly the authors of~\cite{Iadarola&19} applied model checking to the CCS models generated from Java byte-code of mobile banking Android applications. The authors have used the $\mu$-calculus~\cite{Kozen&83} to further specify malware behavioural properties such as stealing money in background operations, intercepting Short Message Service (SMS) messages and password resetting. The model was then verified against the malware properties using the CWB-NC model checker. The authors carried out a case study determining if an application exhibits an overlay malware behaviour, where the application overlays a screen indistinguishable from an honest application screen with intent of stealing banking information, demonstrating the viability of their approach.

In the alternative currency area, participants access the currency exchange network via applications, serving as virtual wallets storing the funds and authentication secrets (private keys) that prove fund ownership.  The authors of~\cite{Turuan&16} analyse the properties of the Electrum Bitcoin wallet. Specifically the authors focus on the two-factor authentication used by this wallet. The authors have created a model in ASLan++~\cite{vonOheimb&12} and verified the behaviour of the two-factor authentication scheme using the Cl-Atse protocol analyser~\cite{Turuani&06} for a bounded number of sessions. The analysis has uncovered a potential vulnerability in the user registration process.

New applications include \emph{smart contracts}, which are contracts between two parties written in a programming language.  They define the flow of money based on rules and conditions that, when met, trigger the underlying financial infrastructure to perform an exchange of funds. The automatic contract enforcement feature opens the way to great losses if the program or the execution platform contains exploitable bugs, thus one finds an expanding number of FM applications in this field, specifically use of certified contract languages~\cite{Annenkov&18} and the certification of the virtual machine byte-code~\cite{Park&18}.
For a more detailed account, and a specific survey on the area and the current opportunities for FM we refer the readers to the work~\cite{Miller&18}.

% Maybe move to protocols \citeN{Chaudhary&15} present a formal model of the Bitcoin protocol in UPPAAL with emphasis on evaluating the possibility of double spending in the case of collusion of malicious peers.

%%% Local Variables:
%%% mode: latex
%%% TeX-master: "../../main.tex"
%%% End:

% \newpage

% !TEX root = ../../main.tex
\subsubsection{System}
\label{subsubsec:system}

The financial systems area evolves at two speeds. The legacy core banking systems delivering services such as in-house internal bookkeeping, the SWIFT inter-bank network system, and the Europay Master Visa (EMV) standard that underpin the world economy progress slowly. On the other hand, fierce competition by outsiders or new regulatory demands~\cite{Wich&17} push the sector to move faster. A sign of the two speeds appears in the discrepancy between the percentages of COBOL running the systems.  In 2017, Reuters reported that while only 43\% of the general US banking systems are built on top of COBOL, it is a staggering 95\% in legacy based subsystems such as ATM swipes or in-person transactions.
%\footnote{\url{fingfx.thomsonreuters.com/gfx/rngs/USA-BANKS-COBOL/010040KH18J/index.html}}.

The two speeds are also apparent in the application of FM.  Most of the FM references mentioning legacy systems are decades old % , great
% achievements from the past,
as the case of the CICS Z specification or the applications of BAN logic. Few newer works exist in this area, as for example, the case of the reuse of BAN logic to verify a mobile payment system~\cite{Ahamad2012} and the use of the SPIN tool to model check networks of ATM systems, as the case of 1-link in Pakistan~\cite{Obaid2017}. Furthermore, the authors of~\cite{Zhang&12} also apply SPIN to verify a model of internet payment systems.

Finally, we highlight the work in~\cite{Santone&13}, as it shows the legal barriers faced by the proposer of FM in the domain. While the work has potential for security improvemets in the financial field by finding vulnerabilities within the banking processes, wide acceptance is yet to be observed. Another exception to the standstill in legacy systems is the domain of traditional Electronic Trading Systems (ETSs).
% \todo[inline]{Remove} \textcolor{red}{The authors of~\cite{Passmore&17} have proposed a new stack based perspective on the financial systems. In this view, the stock exchanges and market venues provide an Instruction Set Architecture (ISA) for the upper-stack elements (derivatives, smart-contracts). This is akin to microprocessor design, which provides an ISA that higher-level compilers, operating systems and user-mode programs rely on. The stack based perspective is coupled with a formal verification system, \emph{Imandra}, that is written in OCaml and relies heavily on SMT solving}

% .\todo[inline]{Remove}\textcolor{red}{The authors of~\cite{Cervesato&19} propose a specification of an automated trading system and provide proofs for two high-level properties regarding transactions liveness and regulation-compliant pricing. Nevertheless, both works are far from practice and do not make it clear whether they imprint any displacement on production systems.  It is unclear whether the enforcing of regulations and security of the traditional electronic trading systems goes beyond standard software testing and quality assurance.}

Within the ETS domain, we also observe a boom in the application of FM, specifically in non-traditional areas of
%non-traditional field of
blockchain-based cryptocurrencies and unregulated/decentralised systems. There exists a large number of works in the area of blockchain-based financial services. The scope of such works range from the verification of an algorithm in a cryptocurrency platform~\cite{Yoo&19} and the verification of the Etherium virtual machine system underlying the smart contract concept~ \cite{Hildenbrandt&18}, to the formal verification of a whole blockchain system \cite{Duan&18}.
% \todo[inline]{Remove}\textcolor{red}{The authors of~\cite{Sarswat&19} have developed a complete Coq formalisation for analysing trades in financial markets.}

We must further mention outstanding works that illustrate the usage of FM beyond plain system analysis.  For instance, the authors of~\cite{Aarts&13} provide an example on how to build the usually labour intensive models of systems by applying automata learning techniques, in this case applied to the EMV standard. Another pattern is the adoption of formal models as components in fraud detection such as within the work of~\cite{Rieke&13} by use of a lightweight formal specification.
% \todo[inline]{Remove}\textcolor{red}{The authors of~\cite{Zhao&19} explore the Petri nets formalism to build a fraud detection system in online shopping processes. The authors note that the lightweight process could be widely used to provide security assurances for online shop customers.} In another example, the authors of~\cite{Rieke&13} perform fraud detection using a lightweight formal specification.

%%% Local Variables:
%%% mode: latex
%%% TeX-master: "../../main.tex"
%%% End:

% \newpage
\subsubsection{Protocols}
\label{sec:fin:proto}
% !TEX root = ../../main.tex

The world of financial transactions forms a pillar of modern society. To protect these transactions FM could be used to provide strong security assurances.

Near Field Communication (NFC) is a short range radio technology often used within smartphones, bank cards and payment terminals, to facilitate contactless payments. These transactions are secured by use of the EMV protocol~\cite{Khu-smith&02}, which has been found to contain significant security vulnerabilities where the attacker can obtain user's payment details, present within the NFC payments~\cite{Mehrnezhad&16}. To address this, the authors of~\cite{Madhoun&16} have proposed a new protocol, specifically aimed at mitigating the vulnerabilities of the NFC part of the EMV protocol. The proposed protocol provides mutual authentication with non-repudiation between the bank card and the terminal, the integrity of the banking data and ensures that the bank card is valid. In order to ensure that the protocol provides the expected security the authors specify the protocol in the Security Protocol Description Language~\cite{Cremers&12} and express confidentiality and authentication claims within the Scyther tool~\cite{Cremers&08}. The verification has demonstrated that the specification satisfies both claims. Moreover, the proposed solution is applicable for online and offline payments. Similarly, the authors of~\cite{Abughazalah&14} have proposed a protocol for securing of NFC enabled mobile payments. This is important as vulnerabilities such as a possibility of relay attacks~\cite{Roland&13} have been discovered in the popular NFC payment system Google Wallet. The proposed protocol uses mutual authentication between entities involved in transactions and leverages single-use passwords. In order to verify security of the protocol, the authors create a high-level description of the protocol, translate it to CSP and express several security requirements as formal properties. The verification is then carried out using the FDR model checker. The protocol was shown to be resistant to replay attacks and no feasible attack was found against mutual authentication and tag anonymity. The authors note that they plan to implement the protocol and again use formal verification against the implementation. Similarly~\cite{Alsheri&13} provided the first formal analysis of the NFC mobile coupon application proposed by~\cite{Dominikus&07}. The analysis used the Casper tool to translate the protocols directly into CSP for direct model-checking with FDR, to consider security with respect to the security requirements of forgery protection and user authentication.  The formal analysis identified attacks against the authentication properties, giving rise to a proposed solution for which the analysis in~\cite{Alsheri&13} found no further attacks.

One of the most widely used e-commerce transaction protocols is the secure electronic protocol. The authors of~\cite{Xiao&14} have modelled a simplified version of the protocol~\cite{Lu&99} in Promela and verified mutual authentication properties using the SPIN model checker. The authors have expressed authentication properties as LTL formulas and discovered several vulnerabilities. The authors further proposed an improvement to the protocol addressing the vulnerabilities and note the effectiveness of model checking in security analysis of protocols. Similarly, the authors of~\cite{Hartung&12} have analysed a biometric transaction authentication protocol~\cite{Hartung&10} providing authentication and non-repudiation of the origin in insecure environments. The protocol uses a biometric transaction device, providing the biometric data of the user, where the authors have created an idealised model of this device. Furthermore, the authors have modelled the protocol in $\pi$-calculus~\cite{Abadi&99} and verified the security properties of authentication and non-repudiation using ProVerif~\cite{Blanchet&18}. The analysis uncovered that in case a malware on the client blocks messages from the server, the user would not know if the transaction was successful. The authors proposed an extension to the protocol addressing this issue.

SMS is also often used to facilitate mobile payments. To secure these payments~\cite{Bojjagani&15} have proposed a secure SMS based protocol. The protocol utilises elliptic curve cryptography providing signatures and encryption. In order to demonstrate that the protocol is secure the authors have verified it using AVISPA~\cite{Armando&05}, specifically model checking AVISPA endpoints and BAN logic. The analysis has shown the protocol complies with properties of confidentiality, message freshness and third party trust while utilising the current SMS infrastructure.

%%% Local Variables:
%%% mode: latex
%%% TeX-master: "../../main.tex"
%%% End:

% \newpage
\subsubsection{Implementation}
% !TEX root = ../../main.tex

Vulnerabilities in financial software could lead to financial losses or disclosure of sensitive information. Preventing these vulnerabilities at the implementation level is an area where FM could be of substantial benefit.

The spread of smartphones has allowed for rapid introduction of digital financial services in the developing countries. The authors of~\cite{Ibrar&17} have considered static code analysis as a tool for assessing the vulnerabilities of the Android applications used to access the financial services. The authors have selected seven applications from developing countries and three applications from developed countries. The applications have been analysed by three static analysis tools, MobSF\footnote{github.com/MobSF/Mobile-Security-Framework-MobSF}, Quark~\cite{Kotipalli&16} and AndroBugs~\cite{Lin&15}. Each of the tools could analyse the applications against a predetermined set of vulnerabilities. The analyses have considered thirteen vulnerabilities and has shown that the applications from developing countries suffer from more vulnerabilities than those from developed countries. The authors recommend development of static analysis tools aimed specifically at financial applications. Similarly, the authors of~\cite{Taylor&17} have used static analysis on 10400 Android applications in order to determine how security of financial applications compares to other applications. The authors have collected quarterly snapshots of the Google Play Store for two years in order to analyse the evolution of security of financial applications by use of the AndroBugs and MobSF tools. The selected vulnerabilities were based on the top ten OWASP list\footnote{owasp.org/www-project-mobile-top-10/}. The results show that the financial applications have been gaining more vulnerabilities over time. The authors note this is a worrying trend and suggest that the developers further employ static and dynamic analysis tools.

{\em EMV 2}, the next generation of EMV, integrates new security schemes, such as biometric security. The author of~\cite{Freitas&18} has modelled 80\% of the EMV 2 specification using VDM in order to determine possible security issues and provide a formal model for implementation. The size of the model (50 KLOC) has brought difficulties to the VDM IDE Overture~\cite{Larsen&10}. Furthermore in order to model EMV 2 with a high level of fidelity the author had to solve several corner cases brought by the VDM SL semantics. The author has attempted automatic Java code generation, finding that the code generator needs improvements in order to handle such models, opting instead for manual implementation. The author has created a wish-list of nine improvements that can improve the modelling and validation experience of VDM tools.

The financial industry is moving towards the use of open APIs, where third party companies can access banking data. To this extent, the authors of~\cite{Fett&19} have created a highly detailed, close to implementation formal model of the popular OpenID Financial-grade API. The model is based on the Web Infrastructure Model proposed in~\cite{Fett&14}, where the behaviours are captured as a set of theorems. Within the model the authors have considered security properties covering the authorisation, authentication and session integrity. Manual theorem proving has uncovered several vulnerabilities, specifically the ability of an attacker to access protected resources owned by legitimate actors. The authors have proposed several fixes based on the analysis.

Bitcoin is a peer to peer cryptographic currency system allowing for mutually distrusting parties to perform financial transactions. This property of Bitcoin contracts could lead to their wider use. One thing that makes security assurance hard in this setting is the difficulty in analysing security of these contracts. The authors of~\cite{Andrychowicz&14} have modelled and analysed two types of these contracts in UPPAAL against several security properties. One of the results was the determination of a time limit after which the contract protocol shall abort in order to remain secure. The authors note that UPPAAL is a suitable tool for analysing the different types of Bitcoin contracts.

%%% Local Variables:
%%% mode: latex
%%% TeX-master: "../../main.tex"
%%% End:

% \newpage
\subsubsection{Hardware}
% !TEX root = ../../main.tex

The rise of cryptocurrencies and financial technology have led to creation of a specific financial hardware. While this field is in its infancy, FM have been utilised to improve security of these devices.

One of the important aspects of technology in finance is the ability to securely approve transactions. This is invaluable in the field of cryptocurrencies where \$1B has been stolen in 2018 alone\footnote{\url{ciphertrace.com/wp-content/uploads/2018/10/crypto_aml_report_2018q3.pdf}} mainly due to the theft of private keys from users' computers. The authors of~\cite{Athalye&19} have proposed the \textit{NOTARY} hardware device used for approval of security critical operations. The device provides multi-agent capabilities, allowing for different approval agents. The authors demonstrate this, by implementing a Bitcoin wallet and a general-purpose approval agent. In order to switch between the agents without potential leak of sensitive information, the device is based on principles of reset based switching, ensuring that the device always enters a deterministic start state and provides resiliency against side-channel attacks~\cite{Kocher&19}. The authors designed the device by continuously verifying the property of a deterministic start by modelling the RTL description of the device then dispatching the model to a SMT solver. In the case where the SMT solver found a counterexample the authors tweaked the design to ensure no counterexample is found. The authors note that while the multi-agent device provides higher security than single-purpose hardware wallets, it is comparable in retail cost. The hardware wallets are often considered more secure than software wallets. This is due to the assumption that the hardware device contains low complexity firmware making it less prone to bugs and vulnerabilities.The authors of~\cite{Marcedone&19} asked a question: what if the hardware wallet manufacturer cannot be trusted? It could be that the manufacturer installs a back-door to the pseudorandom generator. The authors suggest use of a two-out-of-two signature scheme~\cite{Desmedt&89} using signatures partially generated by the hardware wallet and partially known by the user (password). The authors focus on the unforgeability property, where an adversary corrupting the hardware wallet cannot forge the signature, while considering a malicious client, a malicious hardware token and selective access to legitimate tokens. The authors verify the property against their scheme using theorem proving, while suggesting that the scheme allows for use of untrusted hardware wallets. Also within the area of hardware wallets the authors of~\cite{Arapinis&19} have attempted to formally prove security properties of several wallets. The authors note that the wallets are the only means to interact with the cryptocurrency assets and as such, FM shall be used to verify their security. To do this the authors create a formal model of a hardware wallet in the universal composable framework~\cite{Canetti&01} and identify ``all potential attack vectors" and conditions under which the wallet is secure. Furthermore the authors analyse implementations of three popular hardware wallets by mapping them to their model and prove theorems about the essential security properties. This has shown that these hardware wallets are only secure under specific assumptions with an important note that perfect cryptography alone does not guarantee security.

Falling costs of hardware such as NFC-enabled mobile phones and off-the-shelf NFC USB readers create security challenges for contactless card payments. One of the potential attacks is a relay attack where an attacker relays communication between the card and a terminal to another terminal using commodity hardware. To mitigate this attack, the authors of~\cite{Chothia&15} have introduced a time bounding scheme into the payment process. In order for the attack to be successful the attacker has to relay messages at the speed of light from any large distance. This is prevented by the proposed scheme, based on a distance bounding protocol~\cite{Boureanu&14}. The authors model their scheme in applied $\pi$-calculus and verify it using ProVerif. The authors note that their scheme offers good protection against attacks using commodity hardware.

%%% Local Variables:
%%% mode: latex
%%% TeX-master: "../../main.tex"
%%% End:

% \newpage
\subsection{Industrial}
\label{subsec:ind}

Industrial processes are a backbone of a modern society as they provide control not only for production of necessary goods, but also utilities such as electricity and water treatment. This section provides an overview of how FM have been utilised in security analysis of automotive control applications, robotic applications, PLC software, industrial communication protocols such as Modbus and OPC UA, SCADA systems and hardware devices underpinning industrial computing. An interesting note is that the research works are distributed uniformly across the different levels of abstraction, demonstrating that all aspects of industrial computing have been scrutinised using FM in order to provide either security analysis or security assurances. In the industrial application of FM the problem is often considered domain specific, i.e. cyber security properties are based on whether the considered industrial system is for example an automotive controller or a water treatment plan.

\begin{figure}
  \centering
  \includegraphics[width=.8\textwidth]{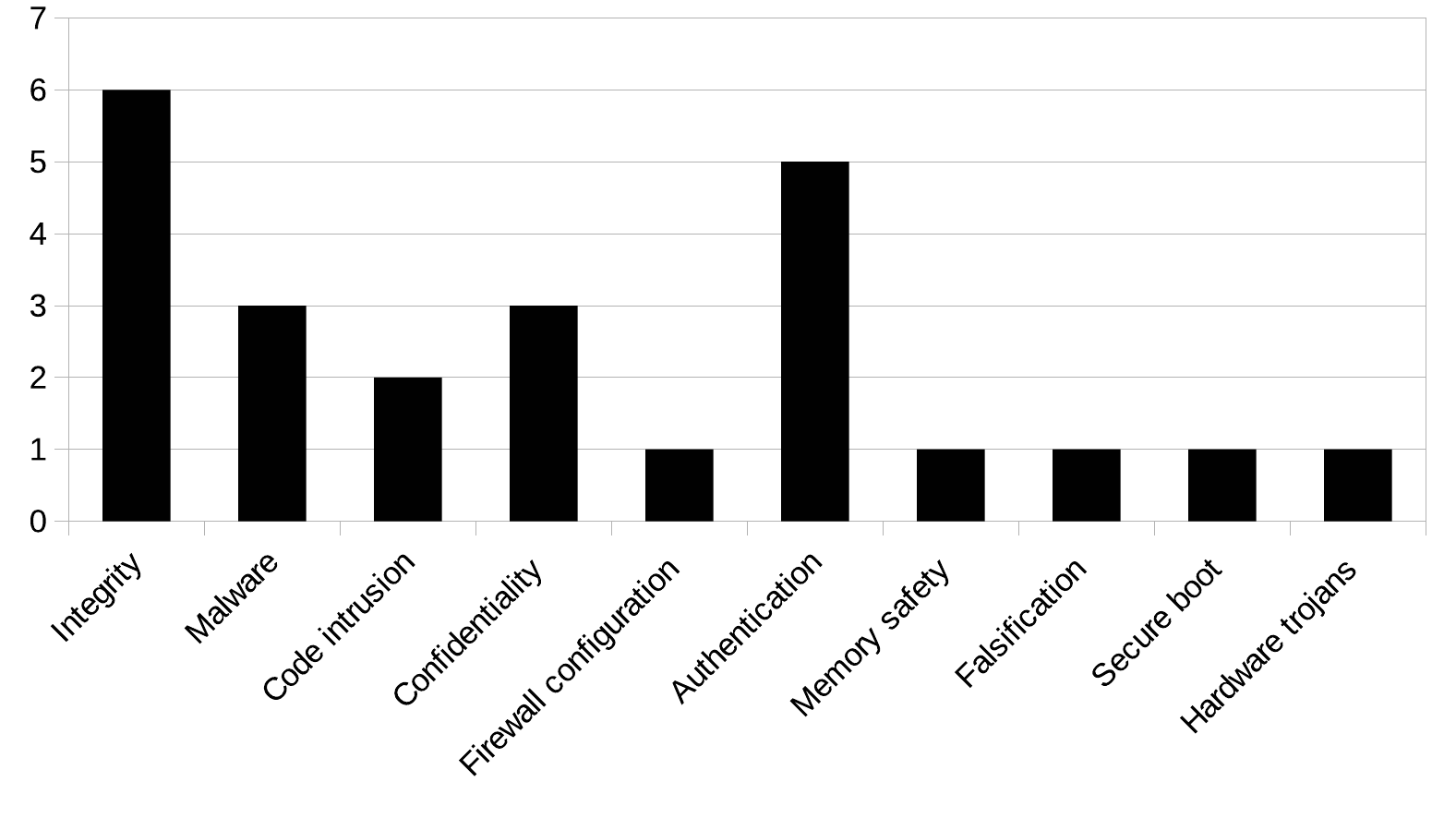}
  \caption{Cyber security topics in the industrial section}
  \label{fig:indTopics}
\end{figure}

As within the \textit{financial} section, the most used FM to analyse the security properties is model checking. This could once again be attributed to the nature of the problem where for example PLC programs and industrial processes lend themselves to be easily modelled using state transition systems. As some of the industrial computing is complex, the problems are often modelled more abstractly in order to avoid the state space explosion problem. Within the hardware level of abstraction however in industrial computing, theorem proving is often the FM of choice, as it allows description of the hardware in more detail. The cyber security topics within the industrial section are shown in Figure~\ref{fig:indTopics}.

\subsubsection{Application}
\label{sec:indApp}
% !TEX root = ../../main.tex

Industrial applications often control a critical industrial process, where security assurance provided by FM could bring massive benefits.

In the automotive area, the authors of~\cite{Heneghan&19} have proposed a model checking approach to security assurance of the Electronic Control Units (ECUs) in order to provide secure advanced driver assistance systems. The authors have proposed a method for automated translation of the ECU applications to CSP~\cite{Hoare&78}. The CSP model is then dispatched to the FDR model checker, verified against several cyber-security properties and potential counter examples are fed back to the ECU application developers. The automated translation approach was developed for the proprietary programming language CAPL. The approach has indicated large cost reduction in security assurance, however the authors state the need to extend their work with the notion of timing.

Industry 4.0 is a concept bringing interconnectivity and data sharing to production facilities. It is in this area that the authors of~\cite{Nigam&19} have investigated formalisation of the behaviour of industry 4.0 applications to formally analyse their security. The application that was formalised was controlling a cap attaching robot. The authors have created a model using Maude~\cite{Meseguer00} and provided two models of an attacker. The first model is a symbolic intruder capable of injecting messages, while the second model is of a symbolic intruder capable of tampering with messages. The analysis found four distinct messages an attacker can inject that would put the system into an undesired state. This has been used to update the security policies, preventing these attacks. The use of FM for the security of robotic applications was also investigated by~\cite{Wang&19} in order to enable model based design for robotics. The authors consider applications based on the Robotic Operating System (ROS)~\cite{ROS}. The first step in the method is the creation of timed automata models based on the specification of the application and security, safety and liveness properties expressed in CTL. The UPPAAL model checker~\cite{Larsen&97} is then used to verify the model against the defined properties. The authors further create a code generator to generate the implementation of the application using the ROS C++ code.

SCADA systems consist of several applications, including the SCADA application itself or applications distributed on PLCs and micro controllers. The authors of~\cite{Mercaldo&19} have proposed a method based on model checking for security analysis of water treatment system SCADA logs. The method consists of modelling the system behaviour using timed automata and expressing the properties based on the logs in timed temporal logic~\cite{Alur&90}. The authors express the properties representing the overflow and underflow attacks and validated their method against predetermined scenarios with 100\% attack detection rate based on the log data. On the PLC component level, attacks are often created as a self-propagating malware. The authors of~\cite{Zonouz&14} have proposed a method for malware detection by providing a symbolic execution of a PLC scan cycle and then combining multiple scan cycles into a temporal execution graph. The Z3 SMT solver~\cite{deMoura&08} is then used as a model checker, verifying security and safety properties related to application genuinity. The counter-examples generated by the model checker are used for application debugging purposes. The authors of~\cite{Kottler&17} consider an application created in the graphical Ladder Logic (LL) PLC programming language. Due to the graphical nature of LL it is difficult to find intruder code, which could find its way to the PLC due to the connected nature of modern PLCs. To find the intruder code, the authors model the application and use the NuSMV model checker~\cite{NuSMV} while expressing six possible intrusions as CTL properties. The verification shows that applications created in LL could be checked automatically for intrusion code within the PLC application.

%%% Local Variables:
%%% mode: latex
%%% TeX-master: "../../main.tex"
%%% End:

% \newpage
\subsubsection{System}
\label{sec:indSystem}
% !TEX root = ../../main.tex

Industrial systems are often used to control critical infrastructures requiring strong guarantees of security and safety. Cyber attacks against these systems could have severe consequences~\cite{Weinberger&11}. It is possible to decrease the attack surface of the industrial systems. The authors of~\cite{Rocchetto&17} have used formal design and analysis to verify security properties of a real world industrial control system. They have created a model in ASLan++ covering the system database, network, PLC and SCADA logic. Further the attacker was modelled as a Cyber Physical Dolev-Yao attacker~\cite{Rochetto&16}. The verification was carried out using the CL-Atse protocol analyser. As the target was an existing water treatment plant, the author has compared the formal analysis with practical assessment results, finding that the formal analysis discovered seven out of eight possible attacks, while the last one was not discovered due to the chosen level of abstraction. Due to the specialised nature of the PLCs, running anti-virus solutions is not possible. In~\cite{Denzel&17} the authors proposed a framework for malware-tolerant, self-healing control systems. The main idea is to distribute trust to ensure that a single compromised component does not break the security policy. The authors have added specialised reset circuits with two out of three voting to determine if a PLC is producing correct results. If the result is incorrect the system reloads the image to the PLC from network storage. To asses their framework, the authors have modelled the reset and voting protocol using ProVerif and utilized a Dolev-Yao attacker with access to different devices and carried out formal analysis to determine if the system remains compliant with the security policy. The analysis has proven the security aspects of the system. The authors of~\cite{Hailesellasie&18} have considered verification of the PLC network within an industrial control system by creating graphs of the potentially compromised PLC program and a trusted version of this program. They do this by first creating a formal model of both programs in UPPAAL, then translating this model into attribut graphs. The matching comparison of the graphs provides guarantees that the system has not been compromised. The approach has been demonstrated on a case study of industrial water level control system. Similarly the authors of~\cite{Wang&17} have chosen formal methods as an effective technology to avoid errors in security design of cyber-physical systems. The authors utilised timed automata to model a water treatment system consisting of multiple layers. The considered layers were the supervision layer, monitoring the controllers, the real-time control layer, executing the control and a physical layer consisting of a water tank, pump and necessary sensors. This was carried out as different layers are prone to different attacks. The authors further modelled recovery mechanisms for different attacks and the attacker. The system was then analysed against several cyber security properties using the PAT model checker~\cite{Sun&08}, showing that the recovery mechanisms can bring the system under control if the attack frequency is lower than a detection frequency of the recovery mechanism. The authors further plan to apply their method to more complex systems.

Since modern industrial control systems connect to multiple networks such as the PLC control network, enterprise network and even the Internet, firewall configuration is an important area of system security assurance. The authors of~\cite{Rysavy&13} have demonstrated formalisation of firewall rules to rule tuples representing the actions firewall should take in communication. Further, these rules and policies have been translated into logical formulas that were verified using the Z3 SMT solver. The authors note that the formal analysis has high potential to improve the error-prone process of firewall configuration. As control systems become more interconnected, connectivity towards cloud systems is often considered. In~\cite{Kulik&19}, the authors have verified several proposed mitigation strategies against attacks on cloud connected industrial control systems such as attacks where a remote client tries to push a malicious firmware update or execute brute force attack against the login system. A system architecture consisting of the control system, remote clients and a cloud intermediary was modelled in PlusCal~\cite{PlusCal} and TLA+~\cite{TLA}, while the properties have been expressed using LTL. The verification has shown the effectiveness of the mitigation strategies. Since the cloud connected architectures grow in complexity, using model checking is often difficult as verification at high level of detail leads to state space explosion. The authors of~\cite{Tran-Joergensen&19} analysed several cyber security properties of a cloud connected industrial control system corresponding to validity of system access tokens and firmware update files using combinatorial testing in conjunction with formal modelling. The model was created using VDM-SL~\cite{ISOVDM96,Fitzgerald&09} and properties expressed as a set of invariants, pre and post-conditions. Using the combinatorial testing feature of VDMJ~\cite{Larsen&10c}, the authors have generated 145 million tests, with a practical execution time of less than 20 hours.

%%% Local Variables:
%%% mode: latex
%%% TeX-master: "../../main.tex"
%%% End:

% \newpage
\subsubsection{Protocols}
% !TEX root = ../../main.tex

Industrial communication protocols often carry critical data and commands, hence require a high level of security. One of these protocols is Modbus/TCP. The authors of~\cite{Siddavatam&17} have formally investigated the security aspects of the Modbus/TCP using Coloured Petri Nets (CPN)~\cite{Jensen&09} in conjunction with Formal Component Analysis (FCA)~\cite{Priss&06}. The authors have created a formal model of the client, server and an attacker, where the attacker has the capability to tamper with the messages using an iterative approach, where they first create a CPN block diagram of the protocol. Based on the block diagram the authors create an initial model used to validate the protocol functionality without any threats present. Finally, the authors add the attacker model and carry out several protocol runs with normal and malicious behaviours. Data collected from these runs were used in an FCA with the ConExp tool~\cite{Rancz&08}. The authors discovered a possible attack and implemented it against a test bed to demonstrate the viability of the approach. Similarly, in~\cite{Nardone&16} the authors have focused on the security of the Modbus protocol by modeling several aspects of the protocol as a Dynamic State Machine~\cite{Nardone&15}, which was then automatically translated~\cite{Nardone&16b} to Promela for verification using the SPIN model checker. The authors have expressed the impossibility of a man-in-the-middle attack and data tampering as an LTL property. The verification showed that this property is broken and the authors investigated the counter-examples provided by the model checker. The authors note that compaction of counter-examples would simplify their analysis.

OPC UA is another widespread industrial protocol. The authors of~\cite{Puys&16} have formally analysed the sub-protocols of the OPC UA responsible for authentication and secrecy. The authors have modelled both sub-protocols in ProVerif and analysed them against the Dolev-Yao intruder model. The authentication sub-protocol has been shown to contain a vulnerability since even in the sign and encrypt mode, the message does not contain explicit identity of the receiver. The authors have proposed a fix by adding the public key of the receiver to the message and rerun their analysis without any security violations. The secrecy sub-protocol has only been proven secure if credential encryption is used, however authors note that this is not the case for all implementations of OPC UA. The authors of~\cite{Dreier&17} provide a formal definition of \textit{flow integrity}, ensuring security by preserving a flow order of messages. The authors express the OPC-UA and Modbus protocols as a set of theorems and define nine authenticity and integrity properties expressed as trace formulas within the TAMARIN theorem prover~\cite{Sharygina&13}. The authors demonstrate that the flow integrity can only be satisfied if the channels are assumed secure, i.e. the attacker cannot simply delete all messages and also note a weakness within the secure version of the Modbus protocol due to the insufficient use of cryptography. The DNP3 protocol is used to facilitate communication within the electrical grid. In~\cite{Amoah&16} the authors have formally analysed the authentication properties of the \textit{secure authentication} for the DNP3 protocol, the DNP3-SA. The authors have used CPN and security properties expressed in CTL within the CPN state space analysis tool~\cite{Jensen&07} to verify that the system against several cyber attacks. The authors have uncovered a flaw in one of the protocol modes, allowing for successful replay attack and proposed two potential fixes. The fixes were then verified using CPN and have demonstrated the capability of withstanding the attack. The authors note that the benefits of verification outweigh the time spent on modelling.

CAN is a broadcast, real-time protocol used in modern vehicles. The authors of~\cite{Bruni&14} have analysed MaCAN, an authenticated CAN protocol, using ProVerif, discovering a flaw in session initiation key exchange and a possibility of replay attack with partially modified data due to limited use of cryptography. The authors have proposed improvements limiting the impact of discovered flaws, noting that their security solution is applicable within the bandwidth limitations of the CAN network.

%%% Local Variables:
%%% mode: latex
%%% TeX-master: "../../main.tex"
%%% End:

% \newpage
\subsubsection{Implementation}
% !TEX root = ../../main.tex
Due to the critical nature of some of the industrial control systems, formal verification has been used to verify the security properties of their underlying implementations.  One of the major obstacles to formal verification of an implementation is often the size of the source code.

% \todo[inline]{Remove}\textcolor{red}{The authors of~\cite{Heitmeyer2008} have considered the issue of the large code base of a separation kernel by splitting the code into three categories, where only one category requires rigorous formal verification, while the proofs needed for the other two categories are trivial. First, the code was annotated with pre and postconditions using Hoare logic. The authors then created a top level specification of the relevant behaviour and expressed the desired context separation security properties. The top level specification and properties were translated to Timed Automata Modelling Environment (TAME) and verified by the PVS theorem prover~\cite{Owre&92}. Finally, the authors mapped the top level specification back to the code. While the authors have successfully demonstrated their approach, they present new research challenges for full automation of the process.}

The authors of~\cite{Apvrille&16} propose a method using SysML-Sec, a modelling system based on UML, for working with large code bases. The authors suggest creating models in SysML and iteratively refining behaviour of the model by adding more behavioural and security properties using SysML-Sec. The model can be automatically translated to $\pi$-calculus by use of the TTool~\cite{Apvrille&14} and formally analysed by ProVerif. In case that the refinement leads to a behaviour with a level of detail that cannot be practically formally verified the authors suggest usage of model to code transformations to perform security and safety code analysis. The authors note that their approach can lead to integration of formal verification into a software design and development process. 

In the aeronautics industry, a successful cyber attack could lead to loss of life. The authors of~\cite{Cofer&18} have created an implementation for unmanned air vehicle verified by FM. The focus was set on four assumptions, 1. the architecture is correct, 2. the components are correct, 3. the system execution semantics matches the model and 4. the system implementation corresponds to the model. To satisfy 1, the authors have developed a model using the Architecture Analysis and Design Language (AADL) to capture the important behavioural contracts and verified them using the jKind model checker~\cite{Gacek&17}. To satisfy 2, the authors have created a domain specific language Ivory following the principles of memory safety and avoiding undefined behaviour. To satisfy 3, the authors have used the seL4 microkernel, which was verified using the Isabelle/HOL theorem prover~\cite{NipkowK2014}. Finally to satisfy 4, the authors have created a tool that generates system images from the AADL models. Once the implementation has concluded, the attempts to attack the vehicle by a separate team were unsuccessful.

Software present on PLCs directly impacts the control process. In~\cite{Shrestha&18}, the authors propose a method for formal security verification of PLC control programs. The authors create state transition diagrams based on the control program and define the desired control loop security properties. The state diagrams are then manually translated to formal specification for the NuSMV model checker and the security properties are expressed in LTL. The authors have demonstrated their approach on a simple temperature control module design and express the need for automated translation of the PLC program specification to a formal specification in order to bring model checking to technicians for simple module design. The authors of~\cite{Tsukada&16} have considered the nature of PLC programs and their affinity to Petri Nets. The authors consider falsified software being loaded onto the PLC and present a method using Petri Net modelling followed by Kalman Decomposition (KD) to determine the properties of falsification. The authors then demonstrate automatic detection of falsification using the SPIN model checker, where the Petri Net was translated to Promela specification and falsification properties were expressed using LTL.

%%% Local Variables:
%%% mode: latex
%%% TeX-master: "../../main.tex"
%%% End:

% \newpage
\subsubsection{Hardware}
% \todo{BD}
% !TEX root = ../../main.tex

FM for hardware verification is well established, with numerous tools and domain-specific description languages, e.g., Verilog and VHDL regularly used in industry. Hardware specifications and the safety properties they must satisfy translate naturally to satisfiability problems, which is the type of problem that is amenable to automation. Is this also true for security properties?  This section considers this question in the context of industrial hardware.

% \bd{Summarise main outputs?}{}
% Domain specific: Hardware description languages ()

% Terms to consider 
% \begin{itemize}
% \item co-verification: Verification that embedded system software executes correctly on a representation of the hardware design. It performs early integration of software with hardware, before any chips or boards are available.
% \end{itemize}

The authors of~\cite{Huang&18} study co-verification of Intellectual Property Blocks (IPs) within System on Chip (SoC) architectures. Each IP comprises firmware accessing hardware through a memory-mapped input/output interface. % Thus, verification requires
% modelling of IP/IP and FW/HW interfaces.
The authors assume that an attacker can spoof commands% when no access control
% is available
, but that the attacker has no physical access to internal registers etc, i.e., the attacker can only attack the system through the communication interface. Multiple security considerations were taken in to account including \emph{secure boot} (all components of the firmware image must be authenticated), \emph{concurrency} (in particular time-of-check-to-time-of-use considerations \cite{BratusDSS08}), and integrity and confidentiality of on-device firmware.
%(e.g., that no secure memory can be read by untrusted IPs).

The main challenges encountered are the scalability of reasoning via suitable abstractions, the inherent concurrency of the systems, and extension of standard techniques to cover bit-precise reasoning (e.g., to handle shifting and masking, which is used to access hardware states stored in aligned hardware registers). Moreover, standard model checking tools (e.g., CBMC) could not be used since such tools do not naturally capture the behaviour of hardware.  Their solution is a semi-automatic co-verification methodology that proceeds via instruction level abstraction, which is combined with a toolchain comprising of Boogie~\cite{BarnettCDJL05} (an intermediate verification language) to Corral software verifier~\cite{LalQL12} together with SMACK~\cite{RakamaricE14} to support bit-precise checking.

The authors of~\cite{Lerner&14} consider hardware security verification for Industrial Control Systems (ICS), e.g., as present in factories. They cover a threat model that assumes all software layers can be compromised and are capable of attacking the system while keeping the attacks hidden from human operators. The aim is to develop an application-specific hardware monitor, TECEP (Trust Enhancement of Critical Embedded Processes), that acts as the final authority. TECEP includes two trusted components: an FPGA-based \emph{hardware monitor} (synthesised from formally analysed C code) and a \emph{junction box} that is validated in a hardware description language. The monitor's job is to test whether the outputs (of the physical plant, plant model and prediction unit) are compatible (with respect to predetermined tolerances). If not, a backup controller is used to override the production controller. % The junction box
% independently manages system components.
Their framework models the hardware monitor in
Frama-C~\cite{CuoqKKPSY12} with the Jessie plugin to
enable automatic deductive verification via
Why~\cite{FilliatreM07}.
% \begin{itemize}
% \item Verification: Junction box -  verified using model checking. 
% \item 
% \item Motor controller used as an example.
% \end{itemize}

There are several works on the formal verification of security of integrated circuits.  The authors of~\cite{Love&11} consider the trustworthiness of hardware using the \emph{proof carrying code} (PCC)~\cite{Necula11}.  The authors extend the semantic model of permissible Verilog at the, so-called, register-transfer level, then derive Coq definitions and theorems for the hardware descriptions extended with PCC annotations. More recently, the authors of~\cite{Guo&17} develop a notion called \emph{proof carrying hardware}, which is used to verify security properties of IP cores that may be supplied by untrusted vendors. Their techniques extend VHDL and is built on top of Gallina (the functional programming language of Coq). Their tool supports translation of informal security specifications into Coq and their case study examines a proof of correctness of AES encryption core.  In~\cite{Abbasi&17}, the authors consider the introduction of hardware trojans within Integrated Circuits (ICs) during manufacturing. To this end they use the nuXmv model checker with hardware properties specified using LTL reduced to behavioural traces indicating attack paths potentially exposing side channels in the analysis.

%%% Local Variables:
%%% mode: latex
%%% TeX-master: "../../main"
%%% End:

% \newpage
\subsection{Consumer}
\label{subsec:con}

Consumer computing such as use of personal computer, smartphones and underlying connected services is an integral part of modern life. Consumer computing has been often characterised as of less critical nature than for example industrial systems, however this view is changing as the society introduces more digital technologies to everyday life. This section provides an overview of use of FM in analysis of cyber security of consumer computing, ranging from consumer electronics for fitness equipment, mobile operating systems, web browsers, consumer Internet of Things (IoT) devices to commodity hardware for devices such as personal electricity meters. An interesting fact within the consumer domain is significant use of so called lightweight FM, utilised often not only on the application level of abstraction, but also considering implementation and hardware. One challenge in formal analysis of consumer systems is a rapid nature of evolution of these systems, where the competition in consumer markets often forces fast adoption of new technologies.

\begin{figure}
  \centering
  \includegraphics[width=.8\textwidth]{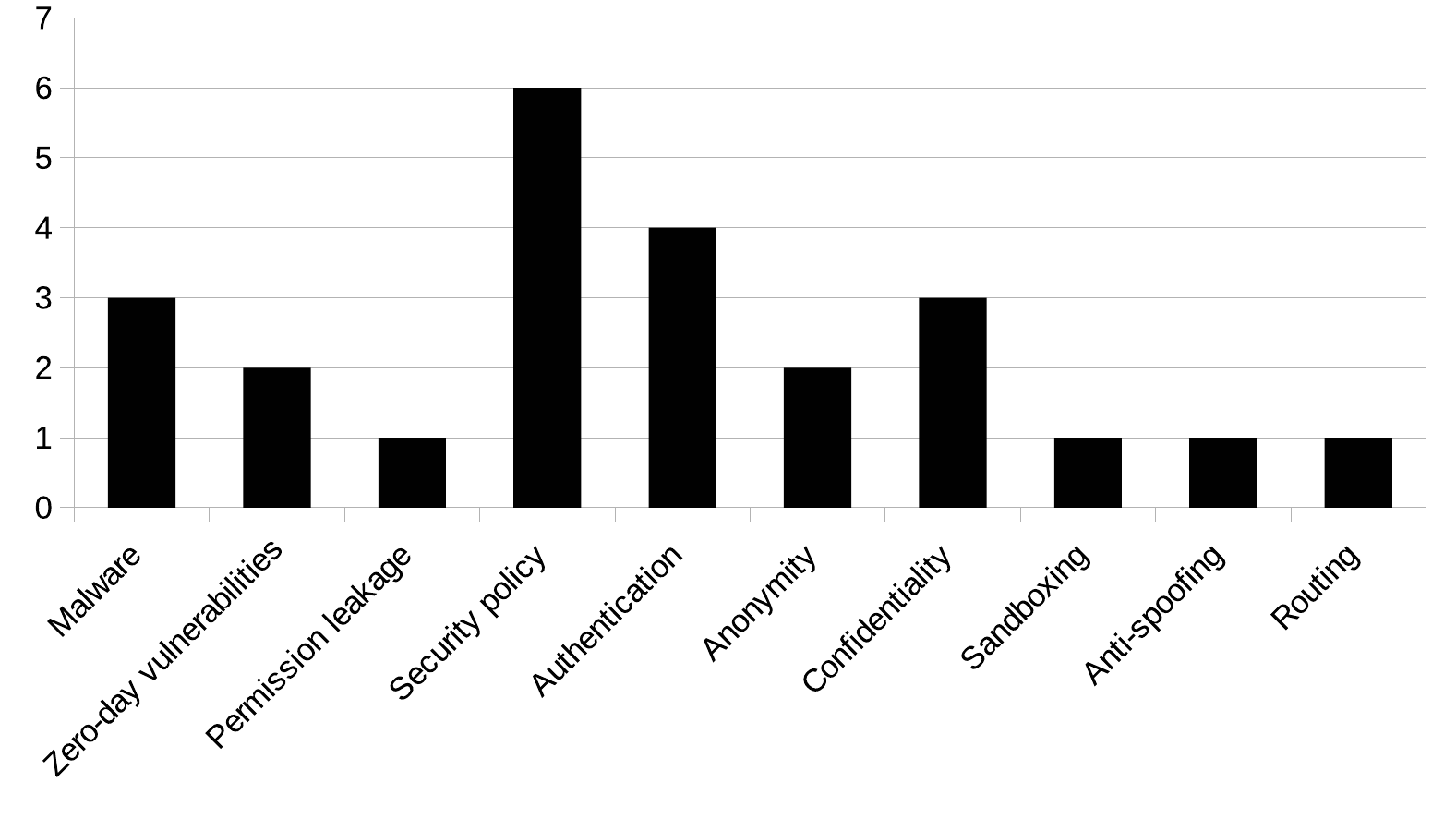}
  \caption{Cyber security topics in the consumer section}
  \label{fig:conTopics}
\end{figure}

Once again the most utilised FM is model checking. It could be argued that this is due to the significant model checking experience gained in other domains. Many consumer computing entities are however complex, interconnected chains of services, which could explain wider utilisation of lightweight FM, especially as some of these are used directly to build chains used to construct the consumer computing entities. It should also however be stated that theorem proving has also been utilised on all the different levels of abstraction within the consumer computing domain. The cyber security topics present within the consumer section are shown in Figure~\ref{fig:conTopics}.

%Introductory paragraph:
%\begin{itemize}
%\item Mobile: Android, \ldots
%\item 
%\end{itemize}

\subsubsection{Application}\label{sec:application}

% !TEX root = ../../main.tex
% Consumers applications are compromised mainly through malware

% Removed by JCPW

% The nature of the software in the consumer application area provides a thriving ecosystem for cyber-attacks \footnote{Software consisting of a multitude of loosely coupled software components with different maturity levels, used by a vast population of heterogeneous end-users, and successful attacks lead to significant reward.}. The majority involves the usage of malware (malicious software) to either collect user information or to compromise and control user's devices.  Although too expensive to be applied during the development of this kind of software, FM techniques have proven to be very efficient in the area of malware research, in particular, the usage of model-checking. \todo[noinline]{Currently FM are being adopted in consumer software developed by giants\ldots Need to adapt this text.}

Sound infrastructure for security applications is an enabling technology for secure consumer electronics. For example, Peloton is an American exercise equipment and media company, launched with Kickstarter funding in 2013. They produce consumer electronics for personal fitness, particularly exercise bikes. They couple the physical equipment with live exercise classes held in New York and they have a large library of past exercise classes. All this brings significant challenges, such as handling latency problems operating at such a large scale. They are particularly concerned about upgrade rollouts affecting different versions of their equipment. Their business model requires that this whole operation is secure. To achieve this, their infrastructure is based on Amazon Web Services (AWS), with its guarantees of security and scalability\footnote{\url{aws.amazon.com/solutions/case-studies/Peloton/}}.

% Argue on the MC approach and why it succeeds 
\FM\ thrives in checking consumer applications for malware. A practical definition of malware (one that can be used to classify executable files) intersects the perspective shift \FM\ advocate: focus on ``what'' is computed instead of the ``how''.
Quoting \cite{Kramer&10}: 
``any (formal) definition of the concept of malware depends on the definition of the concept of software system correctness''.
%
%
% Mention detection and variants of known attacks
Also, a majority of malware is the product of tools generating variants of known vulnerabilities/attacks or known malware. The authors of~\cite{Cimitile&18} show variants are easy to hide syntactically, but not semantically.

Model checking based approaches provide malicious behaviour semantic signatures by providing counterexamples.
% \todo[inline]{Remove: potentially move to history}\textcolor{red}{The early approaches combined temporal logic and control-flow graphs extracted from executable files \cite{Singh&03,Kinder&05}.}
Recent approaches as in~\cite{Song&12b,Song&14} extract push-down automata as models. A promising area is the application of the techniques to the realm of the Android operating system \cite{Song&14b}. The successful recipe is to choose a mathematical model of the executables and a logical formalism to describe the malicious behaviours in terms of the semantics of the program.

% Mention to theorem proving approaches and their limitations..

There are a few works where theorem proving is applied in malware~\cite{Smith&16}, but the number of publications is small, and it is difficult to ascertain if there is an effective gain from it.  Perhaps model checking is more appropriate to the domain due to its non-interactive nature since malware is inherently a game between attackers and an algorithm.

% Mention limits in detection (evolving nature of the game) Learning behaviours

Although successful, the \FM\ techniques provide no panacea to consumer malware protection. The malware game advances with discovery of zero-day (latent) vulnerabilities. FM have been argued to avoid these vulnerabilities in the first place
% \todo[inline]{Remove}\textcolor{red}{\cite{Chen&02}}
~\cite{Mayo&15}, but practically, new malicious behaviours are expected to appear, thus the problem becomes to learn malicious behaviours. There have been several successful proofs of concept where FM leverage the signature learning either in terms of 
% \todo[inline]{Remove or move to history}\textcolor{red}{data-flow extraction \cite{Christodorescu&07,Fredrikson&10}, tree automata learning \cite{Babic&11},} 
or using push-down automata reachability in the process \cite{Macedo&13,Dam&17,Dam&18}.

% Beyond malware 
% Addition of the marketplaces

Beyond detection, alternative approaches work to prevent attacks and malware to reach consumer systems in the first place. For instance, the adoption of the software marketplace paradigm reduces trivial attacks based on luring end-users to inadvertently install malware or on gross software implementation malpractice.  \FM\ backed approaches are also being applied in the field of marketplace vetting and attack mitigation \cite{Mercaldo&16}.
In addition, automated checks to user settings enforcing security policies and reducing the attack surface are already in place. This is a field with a lot of potential for \FM\ approaches~\cite{Bagheri&15}.

% Add new  fuzzing as a trend\ldots?

%%% Local Variables:
%%% mode: latex
%%% TeX-master: "../../main.tex"
%%% End:

% \newpage % TODO Will remove latter\ldots Currently is excellent to frame the one page text into ``one page''
\subsubsection{System}
% !TEX root = ../../main.tex

As consumer digital systems become more prevalent, formal verification has potential to provide the necessary security assurances.  Android is one of the most popular OSs in the world. Android enforces permissions at the application level, allowing applications to combine their privileges, which could potentially lead to a privilege escalation~\cite{Davi&11}. The authors of~\cite{Bagheri&15} propose a tool-based approach, called COVERT, for compositional analysis of Android inter-app permission leakage. COVERT assesses the security of a system as a whole by inferring the security properties of the individual applications. The tool uses a model extractor, generating Alloy specifications~\cite{Jackson12}, capturing the security properties of the applications. The Alloy analyser is then used to verify these properties against a manually created model of a specific Android framework. The authors have analysed more than 500 real world applications and have confirmed the findings previously found within~\cite{Felt&11,Au&12}, showing that many Android applications are over-privileged. The authors of~\cite{Bagheri&18} have moved towards analysing the permission protocol itself, leading to identification of design flaws. The authors have specified the permission protocol in Alloy and, among others, the analysis has discovered a previously theorised design flaw where two applications can apply the same custom permission, leading to the application which was installed first being able to access the resources of the application installed second. The authors have also discovered that many popular Android applications contain this vulnerability.

Operating systems usually contain an underlying security model. In~\cite{Devyanin&14} the authors have verified an proposed access and integrity control for a Linux-like OS. The authors have formalised the security model in Alloy and Event-B~\cite{Abrial&10}, where the Alloy analyser was used to provide constraint based checking of operation contracts within the security model. While the authors have experienced scalability issues with Alloy, the analysis have uncovered bugs that could become more serious if discovered in the implementation phase. The authors of~\cite{Mai&13} present the ExpressOS, a secure OS alternative to Android. The aim of this OS is to provide formally verified mechanisms used to enforce the security policies. This is achieved by expressing invariants representing security properties of the OS and annotating the source code of the OS with formal specification, mainly using code contracts. These abstractions are then discharged for verification using automated theorem provers. The authors note that the approach is feasible for verification of security invariants with only 2.8\% source code annotation overhead.

Consumer expectations on the digital technology led to the rise of a smart home, an ecosystem of small connected devices, often remotely controllable, bringing security vulnerabilities. The authors of~\cite{Kumar&17} have proposed Anonymous Secure Framework (ASF), a framework for ensuring anonymity, authentication and integrity in smart home environments. The authors have verified their framework for its security strength and anonymity using the model checking tools within AVISPA and by BAN logic. The ASF demonstrated resilience against several well-known attacks against smart home environments and demonstrated suitability of their framework for next-generation secure smart-home environments. In~\cite{Moshin&17} the authors have developed the \textit{IoTRiskAnalyzer} tool used to help the IoT engineers apply the most fitting security policies for their IoT environment. This has been achieved using a Markov Decision Process~\cite{Puterman&94}, formalisation of risk properties as probabilistic CTL formulas and verification using the PRISM model checker~\cite{Prism}. Car manufacturers are also taking advantage of connected devices, especially smartphones. Finally, the authors of~\cite{Busold&13} have developed a smartphone based immobiliser with formally verified protocol and hardware fulfilling specific security assumptions. The protocol has been verified using ProVerif against a Dolev-Yao attacker model to ensure strong guarantees of security requirements.

%%% Local Variables:
%%% mode: latex
%%% TeX-master: "../../main.tex"
%%% End:

% \newpage
\subsubsection{Protocols}
% !TEX root = ../main.tex
The area of consumer communication protocols covers text and multimedia communication lending itself to formal verification of security.

In~\cite{Cohn-Gordon&17} the authors have created a formal model of the Signal protocol in terms of predicates and theorems and have applied theorem proving, resulting in improvements in the use of the protocol's random generator.

Security of the consumer communication systems often depends on the mechanisms introduced in the Needham-Schroeder~\cite{NeedhamS1978} authentication protocol and the Denning Sacco protocol~\cite{Denning&81} for secret key distribution. The authors of~\cite{Chen&16} have created a simplified model of the Needham-Schroeder NPSK protocol and the Denning Sacco protocol and expressed security properties using LTL. The authors have provided an efficient model for model checking of the security properties using the Spin model checker.

%%% Local Variables:
%%% mode: latex
%%% TeX-master: "../main.tex"
%%% End:

% \newpage
\subsubsection{Implementation}
% !TEX root = ../../main.tex

The recent adoption of FM tools by large technology companies has shaken up the field. If in the past FM were tied to niche safety critical domains (e.g., aerospace, railway, medical), ``big government regulation'', the current panorama shows that the future brings the usage of FM tools in the daily practice of software engineering. No matter the intention behind the usage of FM tools, the outcome has demonstrated a contribution to increasingly secure implementations.

According to recent reports, when a developer commits a code modification to one of the large technology companies' codebases, a static analysis tool is invoked and a code review is provided. The author of~\cite{OHearn&18} describes the process as continuous reasoning, and any change to a Facebook product is analysed by the Infer static analysis tool, which checks ``small theorems'' on large codebases. This approach has been shown to improve the security of the company's own codebase and library implementations (e.g., OpenSSL). The same is reported about the software engineering practice inside Google \cite{Sadowski&18}, although it is not clear whether FM is used by Project Zero, its elite security team. However, the authors of~\cite{Babic&19} report on how numerous security vulnerabilities were fixed by applying FUDGE, a static analysis tool based on fuzzing developed in house.

Particularly targeted to the security domain, the authors of~\cite{Distefano&19} report a static analysis tool, Zoncolan, in collaboration with the Facebook App Security team. Zoncolan uses abstract interpretation to analyse and issue security alerts for the implementations of the applications in the company's codebase: Messenger, WhatsApp, Instagram, or Facebook. This level of application of FM shows the implementations of software used by millions of consumers has been swept by a FM tool. %Can the same be stated about the software of flying commercial airliners?

FM is also being applied to secure implementations of web browsers, which are designed with security in mind because they mediate a vast amount of personal information (e.g., credentials, banking details).  Nevertheless, due to a large attack surface of a web browser, attacks are possible, and implementation flaws are not uncommon.  In a bolder move, the authors of~\cite{Jang&12} propose a new browser, QUARK, that follows the ``kernel architecture''\footnote{Termed multi-process architecture in Google Chrome with sandboxing of untrusted code which accesses resources through a trusted broker.}  of modern browsers, but QUARK's kernel is formally verified. The formal verification yields to the Coq theorems to assert properties as tab non-interference, or cookie confidentiality and integrity.  According to work in~\cite{Fisher&17}, the price to pay for such a prime example of functional correctness verification (above airline runtime error-free level) is 25\% increase in overhead, affecting performance.

Increasing browsers' security risks, users demand and make use of various browser extensions. Browser extensions are developed using web technologies, but, when compared with a traditional web page, have access to more APIs and features therefore extensions can spy and exploit users as demonstrated in ~\cite{Guha&11}.
% \todo[inline]{Remove}\textcolor{red}{in the work in~\cite{Guha&11}. In the same work, the authors propose a verification tool, IBEX, to serve as an automated curator of extensions.}
In the same vein, but beyond verifying extensions, the authors of~\cite{Morrisett&12} show that x86 native code executed by arbitrary clients conforms with a predefined sandbox policy when using Google Chrome's Native Client service.

To complete the coverage of research on browsers developed using FM, we must mention the case of the Illinois browser operating system. In~\cite{Sasse&12} the authors report a verified design of an experimental browser and operating system reducing the size of the trusted code base to 42K lines. The work provides proven routing guarantees and anti-spoofing of the address bar URL relying on the Maude tool and rewriting logic to achieve this.

%%% Local Variables:
%%% mode: latex
%%% TeX-master: "../../main.tex"
%%% End:

% \newpage
\subsubsection{Hardware}
% !TEX root = ../../main.tex

In contrast to critical-system hardware (e.g., fly-by-wire hardware) attackers cannot be prevented from physical access to the consumer hardware, which provides a large attack surface.

% It is possible to divide this category into two classes: ad-hoc untested platforms (powering IoT solutions: lighting fixtures, cameras, \ldots) and general purpose trusted plaforms (main vendor solutions implementing standards as TPM)

Modern consumer hardware provides hardware-level protections for critical software components. An example of this is ARM TrustZone~\cite{Ngabonziza&16} providing separation between trusted and rich software providing potentially untrusted interfaces. The authors of~\cite{Ferraiuolo&17} propose verification of hardware security properties by use of information flow control at the level of the Hardware Description Language (HDL) such as SecVeriLog~\cite{Zhang&15}. The authors create \textit{SecVeriLogBL}, an extension to SecVeriLog, by adding new types for security labels defined in SecVeriLog. This allows for static analysis at the design time, providing a lightweight verification with small effects on the hardware performance. To demonstrate this approach, the authors have designed an implementation of TrustZone, including 10+ security bugs. The analysis has detected most of these bugs with exception of bugs that required type lowering, i.e., downgrading, to be expressed. The authors note that vulnerabilities that are not affected by downgrading are always detected. Similarly the authors of~\cite{Letan&16} present a formally defined hardware security enforcement for x86 architecture. In this setting, the software relies on underlying hardware for security enforcement, for example memory paging features of an x86 CPU. The authors note that incorrect implementations of hardware enforcement policies often lead to vulnerabilities~\cite{Kallenberg&14}. In order to avoid this, a formal framework \textit{SpecCert} is introduced to model the hardware architecture and then specify the software security requirements to be satisfied by the trusted software components that implement the hardware security enforcement mechanisms. The authors then use the framework to prove that the hardware provides the security assurances provided by the security policy. The authors use Coq to model the architecture and the Coq theorem prover to prove the soundness of the security policy. As an example they verify security policies regarding code execution isolation against an abstract minimal model of an x86 architecture. The authors note that in the future they plan to extend the proofs to physical hardware platform.

In the area of commodity hardware the authors of~\cite{Tabrizi&16} have used model checking to determine possible attacks on smart meters, which are considered critical devices~\cite{Khurana&10}. The authors have created a model of the smart meter using rewriting logic, formal definition of the attacker's actions and used the Maude tool to check that the attacker's actions are not able to break the security invariants. % . In case that an
% attacker's action breaks a security invariant it is considered an attack.
The discovered attacks were then mapped to an implementation of a smart meter, the SEGMeter, to investigate the practicality of these attacks. The authors determined that many attacks discovered by the model checker are indeed practical, despite the model being abstract and not specifically refined towards the SEGMeter implementation.

Today, hardware is often packaged as an SoC. In~\cite{Guo&16} the authors proposed a method for combining integrated theorem proving and model checking in order to verify security properties of complex SoCs. Due to the hierarchical nature of SoCs the authors propose that the design expressed in HDL is decomposed into sub-modules and security specifications into sub-specifications. The sub-specifications are then verified using the Cadence IFV model checker~\cite{Cadence}. These verified sub-specifications are then used as proven lemmas in the Coq theorem prover~\cite{Coq19}, removing the need to prove these lemmas by hand. This simplifies the model checking as well by providing only a small specification to the model checker, avoiding the state space explosion. The authors present their approach on a case study of a 32-bit CPU and note that their method provides significant reduction in the amount of effort compared to manual theorem proving. The authors extend their method by automated code conversion from HDL to verifiable specification~\cite{Guo&2017hw}. SoC complexity increases in Multi Processor SoCs (MPSoC), where multiple processors exchange data via Network on a Chip (NoC) routers. The authors of~\cite{Sepulveda&18} have used unbounded model checking to verify security properties of an NoC, which was practical due to the highly sequential behaviour of NoCs. The authors formalise the security and functionality correctness properties using LTL and use the CIP unbounded model checker~\cite{Kupferschmid&11} to verify them. As a proof of concept, the authors have analysed six different router implementations, determining the feasibility of their approach for NoC security analysis in early design stages.

% \todo[inline]{Remove}\textcolor{red}{The design process for hardware, especially consumer hardware is an important aspect of security assurance. The authors of~\cite{Khattri&12} present HSDL a Security Development Lifecycle for Hardware technologies. This approach calls for several designs for security steps, importantly a formal verification or assurance sub-step under the implementation review step. This method is presented by Intel in order to inspire development of security tools aimed at hardware such as CPUs, chipsets and SoCs.}

%%% Local Variables:
%%% mode: latex
%%% TeX-master: "../../main.tex"
%%% End:

% \newpage
\subsection{Enterprise}
\label{subsec:ent}

Enterprise and large corporate computing is the backbone of large international business. In recent years, there is a trend in enterprise computing to utilise cloud solutions, while still often operating on premises data centers. These data centers and cloud clusters are utilised for a plethora of enterprise tasks such as virtualisation of collaboration platforms, company management and hosting of corporate web portals. This section provides an overview of utilisation of FM to address security challenges of enterprise computing, ranging from secure data storage through virtualisation and software-defined networking security to strong authentication using hardware tokens. As enterprises are larger entities changes are often slower and need to be well managed. To this end, the FM have been utilised as a booster in cloud adoption by enterprises as several FM-based solutions have been proposed to enable enterprises secure switch from on premises data centers to federated cloud solutions.

\begin{figure}
  \centering
  \includegraphics[width=.8\textwidth]{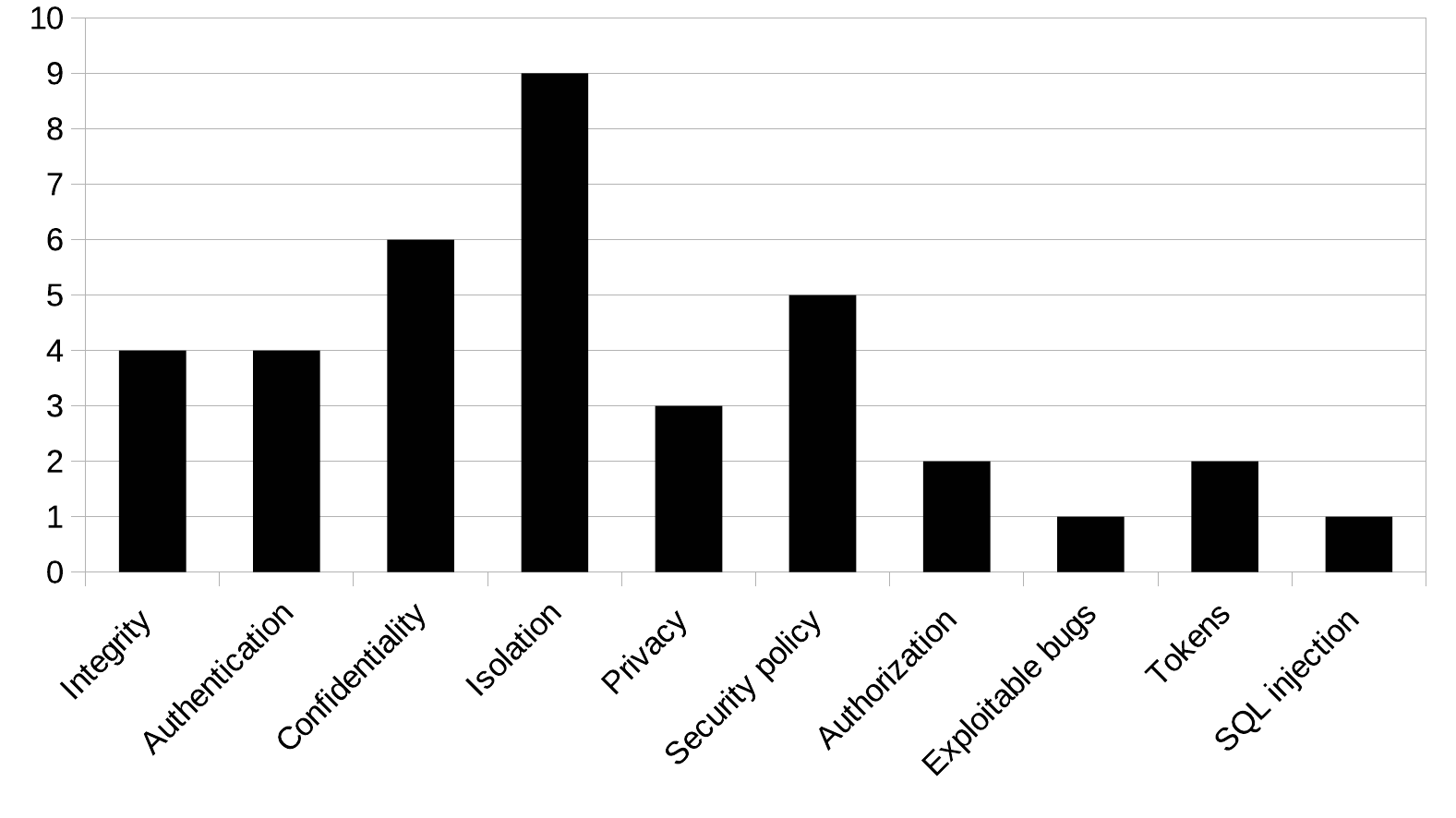}
  \caption{Cyber security topics in the enterprise section}
  \label{fig:entTopics}
\end{figure}

Similarly to previous sections, model checking is the most used tool in formal analysis of security in enterprise computing. Theorem proving is however not far behind especially within analysis of hardware such as Trusted Platform Module chips within enterprise servers. Lightweight FM have also been significantly utilised on the implementation level of abstraction, since they are often provided as plugins to software development environments, making them easily accessible. The cyber security topics present within the enterprise section are shown in Figure~\ref{fig:entTopics}.

% Mention broad adoption of FM by the cloud providers. Cite Byron paper where the case is surveyed

\subsubsection{Application}
% !TEX root = ../../main.tex

Enterprise applications often process and store data critical for an organisation. The data between these applications is often carried by networks defined in software, known as Software-Defined Networking (SDN). The authors of~\cite{Skowyra&14} have created a verification platform for SDN-enabled applications, i.e., applications capable of adjusting their network performance via APIs. The proposed verification platform \textit{Verificare} provides a modelling language VML, for composition of the application with the SDN. The tool translates the VML models to a LTS that is understood by the PRISM, Spin and Alloy model checkers. The counterexamples generated by the model checkers are translated back to statements about the VML model. The tool can use security requirements from a variety of formal libraries. The authors note that \textit{Verificare} can handle complexities above those that a single tool could. Like SDN, Service-Oriented Architectures (SOAs) are becoming a standard in deployment of enterprise applications. In SOA, applications are seen as interconnected services, increasing the deployment complexity. In~\cite{Armando&12}, the authors have created the \textit{AVANTSSAR} platform for automated validation of security in SOAs, which provides a formal specification language ASLan and is proposed as a successor of the AVISPA tool. The platform dispatches the models to CL-Atse, OFMC and SATMC tools and model checkers while providing the properties as LTL formulas. The authors have applied the platform to a large number of industrial case studies and discovered several security issues and vulnerabilities, most notably an issue with SAML SSO integration with Google Apps.

Relational databases are often used as a data storage for enterprise applications. In recent years, it has become popular to deploy these databases to untrusted clouds. The authors of~\cite{Chen&14} have proposed a construction based on cryptography for secure outsourcing of databases to untrusted servers. While the notion of Verifiable Databases was coined in~\cite{Benabbas&11}, the authors expand on it by providing an efficiency improvement based on a notion of incremental updates. Furthermore, the authors introduce an accountability property, expressing the ability of the server to tamper with the database. In order to demonstrate that the proposed construction is secure, the authors use theorem proving, providing proofs for several security and correctness properties. The authors note that their construction does not support all of the update operations.

In order to rationalise the use of resources, enterprise applications are often deployed using virtualisation. The virtualisation is controlled by hypervisors, applications providing virtual hardware and orchestration.
% \todo[inline]{Remove}\textcolor{red}{The authors of~\cite{Freitas2011} describe how \textit{Xenon}, a hypervisor based on a popular open source \textit{Xen}, has been developed with usage of FM in the area of security. The authors have expressed the behaviour and security policies in Z, CSP and Circus~\cite{Woodcock&01}. The primary focus of the authors was to prove the isolation properties of the hypervisor, i.e., two virtual machines not being able to tamper with each others resources, and the security of the hypervisor's external interfaces. The Z/Eves theorem prover~\cite{Saaltink&97} and CZT Circus tools~\cite{Malik&05} were used to carry out automated verification of security. The authors note that only by application of FM it was feasible to gain deep understanding of security for a large system.}
Similarly, the authors of~\cite{She&2013} propose a formal analysis scheme for security of a \textit{Xen} hypervisor called \textit{UVHM}. The authors introduce a known bug, previously reported to the Xen team and using their scheme (re)discover this vulnerability. The scheme consists of four steps, combining static analysis and model checking.

The work within~\cite{Cook18}, provides an interesting overview of how a leading cloud provider uses several FM tools for security assurance, many of which are used on the application level. As the adoption of cloud computing increases, the author notes the benefits, such as privacy protection, that the formal security analysis brings to the customer.

%%% Local Variables:
%%% mode: latex
%%% TeX-master: "../../main.tex"
%%% End:

% \newpage
\subsubsection{System}
% !TEX root = ../../main.tex

As enterprise computing systems are of a significant importance to modern life, formal verification could enable secure enterprise computing minimising financial and safety risks.

Rather than developing their own infrastructure, companies often choose to host their systems using third-party cloud platforms. Sometimes however, there are instances where companies prefer to keep part of the data in their private datacenters, creating federated cloud systems~\cite{Moreno&12}. The authors of~\cite{Zeng&16} have proposed an approach for analysing the dynamic behaviour of federated clouds. This was done by creation of formal models describing the security of information flow within federated clouds. The authors have used CPN to model the behaviour and carried out the formal analysis using CPN Tools~\cite{Jensen&07}. The authors have created different models for federated cloud security verification and stress the ability to use existing tools. Some of the benefits of using a public cloud for enterprise computing are high performance and scalability. The scaling is often carried out via live migration of virtual machines between different hosts to balance the load. This operation however also needs to guarantee consistency of security policy post migration. Similarly the authors of~\cite{Wang&12} have proposed a formal security assurance framework combining formal verification with security functional testing. The framework provides assurance by formal analysis of a security model and then utilises functional testing against this model once the system is implemented. In the first phase the security requirements are formalised in Z and the Z/EVES theorem prover is utilised to carry out the formal analysis. In the second phase, the authors automatically generate functional tests based on their formal security model. To assess their framework the authors apply it to a case study of an enterprise data exchange system, analysing properties of confidentiality and integrity. The authors further compare their test generation scheme to domain theory based~\cite{Blackburn&01} test generation and random test generation. The authors conclude that the case study has demonstrated feasibility and efficiency of their framework, when compared to other functional test generation methods and plan on utilising their framework to assess security of more advanced systems such as Z specified role based access control. In~\cite{Jarraya&12} the authors have created \textit{Cloud Calculus}, a formal approach for expression of firewall security rules and cloud topology. The approach is build on top of the Mobile Ambients~\cite{Cardelli&00} and the non-interfering Boxed Ambients calculus~\cite{Bugliesi&05}. The authors demonstrate security policy verification by a test of equivalence relations. Furthermore, the authors apply their approach to a case study inspired by Amazon Cloud. The authors express a desire to extend their system to take into account intrusion detection and secure tunnelling. Another benefit of a public cloud is the cost of operation enabled by resource sharing among multiple tenants. This could lead to Virtual Network (VN) isolation failures where non authorised tenants could obtain secret information. Furthermore, the authors of~\cite{Madi&18} created an offline framework capable of auditing the cloud infrastructure management system, detecting VN isolation failures. The authors have expressed security properties based on cyber security standards and created a model of multi layered VNs within a cloud system. This was done by expressing the model in first order logic, formally expressing 11 VN isolation properties and with use of data from an actual system verifying the model against the properties. The verification was carried out using the constraint satisfaction solver Sugar~\cite{Tamura&08}. The authors have further integrated their auditing system to the OpenStack cloud infrastructure management system and note that their solution is the only one that works among multiple cloud layers. The large amount of subcomponents that the clouds consist of leads to significant complexity. In order to manage this complexity, the cloud often uses agent systems~\cite{Gutierrez-Garcia&10}, where software agents are used from implementation of intelligence within the cloud system. To this end the authors of~\cite{Masmoudi&14} propose a security framework for agent based cloud systems consisting of three steps. First the authors model the NIST~\cite{Bohn&11} cloud reference architecture using the agent paradigm, then the authors enrich this model with security concepts for cloud computing in Z and finally the authors analyse the architecture against the security concepts using the Z/EVES theorem prover. The authors have so far considered three agents, the consumer, the cloud broker and the provider and included considerations for communication between these agents. Within their work the authors focus on analysing the property of isolation, i.e. ensuring that the data is not communicated to unauthorised agents. The authors conclude that their framework can feasibly improve security of agent based cloud systems and plan to extend it to consider more security properties. Similarly the authors of~\cite{Souaf&18} propose a broker solution, where the broker finds a cloud provider satisfying security requirements defined by customers. The security requirements are defined using first order relational logic~\cite{Jackson&00} and the broker then analyses the requirements while matching it against the offers from different cloud providers. This is based on customers specifying the functional requirements, such as OS, memory, etc. and security requirements in terms of relations between virtual machines and cloud clusters. The analyses of these relations against the cloud offers is carried out using KODKOD finite model finder~\cite{Torlak&07}. The authors currently focus on security challenges related to virtualisation and state that currently the approach is viable for smaller deployments.

Virtualisation is a key technology in cloud computing allowing the enterprise users to save costs on deployment. One of the downsides of virtualisation is the large attack surface it presents. The authors of~\cite{Hao&13} have proposed \textit{vTRUST}, a formal framework for verification of trust and security of virtualised systems. The framework is used to compose a hardware model together with model of an adversary, which then combined with the software model is verified against these properties that stem from service requirements. The models are expressed in CSP\#~\cite{Sun&09}, an extension of CSP and verified using the PAT model checker. The authors have used their framework to analyse a real world cloud system, where they discovered a subtle but critical bug and proposed a fix. Similarly, in~\cite{Bleikertz&15} the authors have proposed a security system, \textit{Weatherman}, analysing changes in virtualised infrastructures with respect to security policies. The author achieve that by formalising the cloud management operations by use of graphs and graph transformations. Graphs representing the information flow are dispatched to GROOVE model checker~\cite{Ghamarian&10} for analysis against security policy properties. The authors note the graph based method provides an intuitive way of formalising cloud management operations, information flow and security policies.

% ~\cite{Cook&18b},~\cite{Alavizadeh&20},~\cite{Guanciale&16}

% \cite{Bera&10} \textbf{Formal verification of secure information flow in cloud computing:} Rather than developing their own infrastructure, companies often choose to host their IT systems using third-party cloud platforms such as those offered by Amazon, Microsoft and Google. However, there are situations where companies prefer to keep their data (or parts of it) in a more restricted environment. For example, to secure sensitive information by storing it in a privately owned cloud. This combination of secure private and less secure public clouds has led to the introduction of federated cloud systems, which is the management and deployment of cloud services located in different clouds. Although federated cloud system has the potential to reduce cost, it also complicates system security due to the movement of entities (data and services) between clouds. To address this, Zeng et al.\ propose an approach for analysing the dynamic behaviour of federated cloud systems~\cite{Zeng&16}. Specifically, by creating formal models that describe the security of information flow within federated clouds (by specifying security policies for how entities move between clouds) and where entities can change their security status. The formal model can be described using coloured Petri nets, which enables formal verification of the federated cloud system's dynamic behaviour using Petri nets tools.

%%% Local Variables:
%%% mode: latex
%%% TeX-master: "../../main.tex"
%%% End:

% \newpage
\subsubsection{Protocols}
% !TEX root = ../../main.tex

Enterprise computing has been moving towards the cloud. The move brought a need for secure infrastructure, identity and access management, improved connectivity (5G mobile networks), and new communication protocols. As shown in this section, FM have been widely applied in securing the protocols involved in the delivery of cloud services.

Amazon, the large cloud service provider, has used FM in security analysis of the Transport Layer Security (TLS) protocol. Amazon cloud services such as the AWS and Amazon S3~\cite{S3Amazon} use an open source version of TLS, the s2n~\cite{s2nAmazon}. In order to provide the enterprise customers with the high level of security assurance, Amazon has employed FM to prove the correctness of the s2n protocol. The s2n/TLS protocol uses the Hashed based Message Authentication Code (HMAC) to handle authentication of a message given a shared secret key to ensure that modifications to data in transit can be detected. In~\cite{Chudnov&18} the authors prove that HMAC is indistinguishable from a random generator given that the key is not known. The authors describe HMAC in the Cryptol specification language~\cite{Cryptol09} and use the Coq theorem prover to carry out the verification. The results of the verification are then connected to the implementation using the Software Analysis Workbench~\cite{Saw}. Similar proofs have been also made for other parts of the s2n protocol. It is important to note that the formal verification process at Amazon is applied continuously for components under constant development such as the s2n protocol.

Continuing in the domain of cloud, but with a focus on defining networks, the authors of~\cite{Jayaraman&2014} have created \textit{SecGuru}, a tool for automated analysis and debugging of connectivity protocols. The authors created the tool in order to help with the error-prone task of maintaining the network policy for large data centers. This is carried out by expressing the connectivity policies as logical formulas and using the Z3 SMT solver. \textit{SecGuru} is used by Microsoft to continuously check the integrity of hundreds of firewalls and routers within the Azure platform. The tool supports two modes of operation, 1) contract validation, where a network traffic pattern is accepted or rejected by a policy and 2) change impact, where the tool is used to compute a semantic difference between two network policies, determining an impact of a policy change. The authors note that \textit{SecGuru} is an important security component of the Azure platform.

Regarding new protocols, the work of \cite{Kumar&14} exemplifies how the application of FM tools, in the particular case the Alloy analyser, is able to identify vulnerabilities in SAML, one of the modern identity and access management protocols. SAML, which stands for Security Assertion Markup Language, enables different cloud providers to provide a single sign-on service. Similar to OAuth and OpenID, the protocol emerged as a web-based workflow and pose challenges to traditional analyses. The authors extend BAN logic to cover this, and were able to devise an attack on SAML ID linking.

Data centers and cloud systems are being increasingly used by enterprises to process data from small IoT devices with limited capabilities~\cite{Muhlbach&08}. In order to secure IoT device connectivity with the cloud, the authors of~\cite{Karla&15} have proposed a mutual authentication protocol based on a lightweight Elliptic Curve Cryptography (ECC). The secured protocol uses encrypted HTTP cookies within three stages. First, the device registers itself with the cloud. When computation is required, the device sends a login request. Finally, the device and the cloud mutually authenticate using ECC parameters. The authors have modelled the proposed protocol in the HLPSL language and used the AVISPA tools, OFMC and CL-AtSe to verify the security of their protocol against seven attacks ranging from eavesdropping to cookie theft. The verification showed that the protocol remains secure, % against the attacks and the authors note that
% this makes them confident
increasing confidence that the protocol is ready for practical application. Similar to IoT devices, many enterprises are deploying mobile devices to connect to the cloud for their enterprise services. In~\cite{Roy&17} the authors propose a lightweight mobile authentication scheme for mobile cloud computing. The scheme contains four properties, first a trusted third party is not involved during login phase, second a mobile user only has one set of unique credentials, third the authentication process avoids costly operations on a mobile device and fourth the mutual authentication uses lightweight cryptography. The authors provide a mutual authentication proof using BAN logic and verify their scheme against a Dolev-Yao attacker using ProVerif. The authors note that the scheme's high security and low cost is well suited for practical application.
A significant improvement to IoT infrastructure is the emergence of 5G, the fifth generation of mobile networks. We foresee several applications in this domain, given the novelty and the safety-critical aspect of such infrastructure. The work of \cite{Basin&18} shows how to use Tamarin, a protocol verification tool, to find and issues and suggest fixes to an Authentication Key Exchange protocol. Furthermore the author of~\cite{Aiash&15} analyses proposed authentication protocols for mobile devices. These protocols need to satisfy requirements of mutual authentication of a SIM card and a device and a user authentication against the device, preferably by biometric means. The author has considered two protocols, the authentication framework protocol~\cite{Zheng&05} and the mobile Ethernet protocol~\cite{Daisuke&06}. The author has specified six security properties ranging from mutual authentication to key freshness and modelled the protocols in CSP for analysis using the FDR model checker. In order to analyse more than the existing protocols, the author has proposed two protocols of their own. The analysis has discovered that the mobile Ethernet protocol is vulnerable to replay attacks, the authentication framework protocol did not show any vulnerabilities, however the author notes that due to large overhead of the protocol it might not be suitable for mobile devices. From the two protocols proposed by the author, one did not show any vulnerabilities and the author considers it a good candidate for mobile networks authentication, while the second proposed protocol did not satisfy the property of mutual authentication.

\subsubsection{Implementation}
% !TEX root = ../../main.tex

Enterprise computing of today consists of many applications that are implemented using different languages, technologies and frameworks. As the works below show, FM are a good fit for this heterogeneous landscape.

One of the examples of modern technologies in the enterprise domain is blockchain smart contracts as defined in~\cite{Hamida&17}. These smart contracts are digital contracts that provide verifiable and permanent agreement for satisfaction of common contractual conditions. The authors of~\cite{Daejun&18} have verified smart contracts within the blockchain network, Ethereum, where the smart contracts were compiled to the Ethereum Virtual Machine (EVM) bytecode. The authors have verified the functional correctness of high-profile smart contracts such as ERC20 token contracts~\cite{ERC20} to ensure security properties of these contracts. The verification has been carried out using the K-framework's reachability logic theorem prover~\cite{Stefanescu&16}, while the authors have introduced specific EVM lemmas to optimise the verification time. The authors discovered that the token implementation that diverges from the ERC20 specification contains several security vulnerabilities.

Web technology also finds a lot of utilisation in enterprise computing. Web applications and web services often rely on third-party frameworks and plugins to add new features. In~\cite{Nunes&15} the authors have developed a static code analysis tool for PHP plugins, called \textit{phpSAFE}. This tool is used to detect cross-site scripting and SQL injection vulnerabilities by creating a model using lexical and semantic analysis of the abstract syntax tree of the PHP code and following the flow of uncontrolled environment variables. The authors have applied their tool to 35 PHP plugins discovering over 580 vulnerabilities and note that about 40\% of the vulnerabilities discovered within the analysed PHP plugins are still present in their updated versions. Due to this, the authors encourage use of analysis tools on third party component integrations. Similarly, web applications can integrate third party APIs. The authors of~\cite{Xing&13} have created a tool, \textit{InteGuard} that uses invariant analysis~\cite{Hamlet&05} for in-the-loop malicious behaviour detection. This tool generates several types of invariants based on network traffic including the content of HTML files and Javascript code and analyses the traffic against these invariants. The authors have validated their tool against 11 real world exploits based on subtle logic flaws in hybrid web applications and note that the tool effectively detected these flaws without generation of false positives.

Security of hypervisors also stems from their implementation. The authors of~\cite{Vasudevan&16} created a framework, \textit{\"{u}berSpark}, for implementation of hypervisors written in C and Assembly with verified security properties. The framework uses logical components to express functionality and provide behavioural contracts. The implementation is carried out in CASM, a dialect of C with embedded assembly, and verified using FRAMA-C~\cite{Signoles&12} for static analysis of behavioural contracts, abstract variable assertions and control flow integrity. The authors reason formally about their framework by expressing its security properties as theorems and use it to construct a reference hypervisor demonstrating that a security verified hypervisor can have a performance close to unverified hypervisor. A few years prior, in~\cite{Vasudevan&13} the authors have created the eXtensible and Modular Hypervisor Framework (XMHF). The goal of this framework was to verify the security property of memory integrity within single-guest hypervisors. As a first step, the authors have expressed security properties and invariants. These properties were then dispatched to the CBMC model checker~\cite{Clarke&04} verifying the actual C implementation. As the CBMC model checker could not handle all C constructs, the authors automatically verified 5208 out of 6018 lines of code and audited the rest of the code manually. The authors have validated their framework by comparing its performance against general purpose hypervisors demonstrating similar performance values and note that the XMHF provides a good starting point for implementation of secure hypervisors.

%%% Local Variables:
%%% mode: latex
%%% TeX-master: "../../main.tex"
%%% End:

% \newpage
\subsubsection{Hardware}
% !TEX root = ../../main.tex

Enterprise computing requires significant hardware infrastructure and must provide assurances such as data confidentiality and computational security. This provides an opportunity to use FM to satisfy the required assurances.

Enterprise customers often consider a cloud provider as an untrusted entity, where the cloud administrators themselves could pose a security threat~\cite{Santos&09}. In this regard, the authors of~\cite{Seol&2015} created a cloud isolation system, isolating the user data from cloud administrators and limiting the operations that the administrators could take against a user's virtual machine. This is based on a hardware module, that the authors named \textit{Trusted Cloud Module} (TCM) that provides a limited set of interfaces to the cloud administrator, manages encryption keys and provides secure storage for the user. The module is built from off-the-shelf hardware components using the Scyther verification tool. They verify the security of communication between the TCM and the other components and the attestation that the TCM provides. Their analysis considered seven attack vectors, excluding the DoS attacks. The authors state that their system increases security for users while demonstrating reasonable I/O throughput. The basis of trusted computing is the Trusted Platform Module (TPM) co-processor providing secure storage and computing environment. Unfortunately the security of platforms using TPM is often not formally verified leading to vulnerabilities~\cite{Bruschi&05}. To mitigate this, the authors of~\cite{Bai&14} have proposed \textit{TRUSTFOUND} a formal modelling framework for model checking of trusted computing platforms. This framework provides a model of the TPM; a formalism for modelling hardware, communications, cryptography and trusted computing techniques. Furthermore the framework contains models of several attackers including a hardware attacker, network attacker and a system attacker. The formalisms used within the framework are a Trusted CSP\#, an extension of CSP\#, and LS$^2$~\cite{Datta&09}, where the PAT model checker is used for verification. The authors used their framework to analyse a cloud computing platform and an envelope protocol, focusing on attestation and confidentiality, detecting six implied assumptions and two severe logic flaws. The authors are also working on supporting the specification of TPM 2.0.

Sometimes, in order to provide strong authentication, small One Time Password (OTP) generation hardware is used by enterprises to authenticate users towards cloud services~\cite{Bonneau&12}. One such device is~\textit{Yubikey}, a USB OTP generator. In~\cite{Kunemann&13}, the authors have formally analysed the security of the Yubikey OTP and also a security of Hardware Security Module (HSM), developed by the same company in order to prevent attacks in case that the authentication server itself has been compromised. The authors first analyse the Yubikey OTP while the authentication server is not compromised proving the security of the OTP scheme. Secondly the verification included a case where the authentication server is compromised and secured using the HSM, uncovering two potential attacks that could lead to release of all secure keys. The verification has been carried out using the Tamarin theorem prover, where the authors provided several intermediate lemmas. Furthermore the authors proposed a scheme and changes to the HSM resolving the discovered security issues.

Another challenging issue of enterprise computing within the cloud is addressing CPU side-channel attacks. One of these attacks is a timing channel attack, where an attacker, possibly a virtual machine, could determine the algorithm executed by another virtual machine in a shared environment. To solve this, the authors of~\cite{Ferraiuolo&17a} have proposed \textit{Timing Compartments}, an isolation scheme implemented in hardware isolating timing information between parties sharing the resources. The authors have implemented their scheme to a quad-core CPU and performed information flow analysis using SecVerilog, demonstrating the effectiveness of their scheme. The authors note that while their scheme is effective, performance optimisations were needed in order to extract reasonable performance from the CPU.

% When investigating cybercrimes, it is important to deal with access control both at the virtual level (credentials to a remote printing service) and physical level (who has access to the computer at the janitor's office). The work in \cite{Probst&09} justifies such affirmation, and it develops a model and analysis approach to identify the credentials needed to reach locations in a system, and therefore deduce the identity of an attacker from log data and predict the system points vulnerable to insider attacks. \todo{This may be re-classified at the enterprise level.}

%%% Local Variables:
%%% mode: latex
%%% TeX-master: "../../main.tex"
%%% End:

% \newpage

% !TEX root = ../main.tex

\section{Future Outlook}

The survey has provided an overview of use of FM within security in several domains. Based on the research conducted within these domains it is expected that in some cases the use of FM will accelerate while, in other cases the use will increase with a slower pace. There is however a general trend of increase of adoption. In cases within the financial domain, it is clear that the use of FM comes with new financial technologies such as cryptocurrencies and smart contracts. This adoption could be seen in a survey aimed specifically at the smart contracts domain~\cite{Harz&18}. The use of mobile applications in the financial domain is also spurring a demand for high security assurance, that could be delivered by use of FM. Finally with the rise of cryptocurrencies, the hardware within the financial domain is being specialised to facilitate transactions. It is expected that the security of this hardware will continue to be scrutinised formally with increasing coverage and complexity.

The industrial domain faces its own set of unique challenges in the area of cyber security. It is expected that with increase in automation complexity and use of digital technologies in critical industrial installations, FM will play a crucial role. The trend was already presented in 2015 by~\cite{Kriaa&15} who have surveyed approaches for security and safety of industrial control systems, including informal approaches. As of now most works within this domain are of reactive nature, i.e. analysis of existing systems and protocols. However several works within the survey show a trend towards utilisation of FM early in the design process of new industrial installations and protocols. Another emerging trend within the industrial domain is integration of formal verification tools with the software development processes, this is clear primarily in terms of robotic applications and PLC code. It is expected that with increasing complexity of robotic applications and underlying hardware, FM will play a significant role in the future.

The domain of consumer computation is a rapidly evolving one. The consumer trends move fast, however a somewhat surprising amount of work is already put forward to use of FM in malware protection~\cite{Biondi&18}, going as far as creation of formally verified Internet browsers. Also, the shift of computation from computers to smartphones has brought new security challenges. In this area a lot of focus has already been put on analysis of the Android OS permission system. This area shows an increase in the amount of research and as long as the mobile OSs are in use by millions of users the formal verification will accelerate, possibly leading to full formal security verification of a popular mobile OS. Consumer hardware such as smartphones is putting into use TPMs, securing mobile computation. Also in this area the use of FM is increasing, specifically to ensure the properties of the secure TPM enclave.

Finally the domain of enterprise computation has uncovered several interesting trends. The first of these is a trend to use FM against virtualisation hypervisors in order to analyse security properties of existing virtual environments as well as use the knowledge to build fully formally verified hypervisors. Another important trend is the significant investment that major cloud computation providers are putting into formal verification of security of their products. To this end not only have existing tools been applied, but the cloud providers have turned towards development of their own FM tools. Both of these trends are expected not only to continue but also to accelerate due to the ever increasing popularity of cloud computing and virtualisation. Several of these trends were already mentioned in 2002 by~\cite{Jullig&02}.

It is important to note that several authors have expressed a wish for improvement of automated formal verification tools. This is in order to allow for simplified entry of non-practitioners to the world of FM. Both academia and industry is moving towards addressing this wish, with tools becoming similar to software development IDEs and in some cases integration of FM toolkit directly to an existing IDE. It is expected that knowledge of FM will become important for system and software engineering disciplines in the future and therefore collaborative projects between industry and academia shall provide experts in this domain. These issues have been discussed for several years now~\cite{Davis&13}.

When it comes to FM techniques, static analysis tools are becoming popular in software development, while model checking is moving strongly towards system design and protocol verification with model checking being designed for specific problems. Theorem proving is also showing a promise of playing a crucial role in the future, given that the perceived large learning curve could be minimised.

As the use of FM is accelerating within all of the different domains considered in this survey, it is imperative that a new survey is carried out as soon as in five to ten years. By then it is expected that the tools will reach the quality of commercial grade IDEs and integration with a wide variety of text editors~\cite{Jorgensen&19} and the techniques will become a known factor when developing and designing a new system, application or a hardware component.

%%% Local Variables:
%%% mode: latex
%%% TeX-master: "../../main.tex"
%%% End:

% !TEX root = ../main.tex
\section{Conclusions}

%\noindent%
More than 30 years ago, Burrows et al. published their pioneering work on the BAN logic for security protocol analysis~\cite{BurrowsAN1990}. Their work was not fully formal and was shown to permit approval of dangerous protocols. Nevertheless, they showed that their logic was good at revealing various subtle security flaws and drawbacks, specifically in authentication protocols. They set out to answer five questions:
\begin{enumerate}
\item Does this protocol work? 
\item Can it be made to work?
\item Exactly what does this protocol achieve?
\item Does this protocol need more assumptions than another protocol?
\item Does this protocol do anything unnecessary? 
\end{enumerate}
Their important paper inspired a generation of security researchers to use FM to devise and analyse security protocols and to answer similar questions. Furthermore in describing the benefits of formal verification for software engineering, Dijkstra famously quoted ``Testing shows the presence, not the absence of bugs''. More than 50 years later, in the setting of computer security, we might now have sufficient evidence to claim ``Formal methods show the presence, not the absence of security flaws''~\cite{Buxton&70}. That is, although FM provides rigorous tools and techniques for proving the absence of security flaws, this rigour comes with a proviso: proofs are only possible if the security flaw is documented and specified within a formal framework. Current limitations mean that FM cannot uncover any new flaws since we may not actually be looking for them.

Take for instance the recent Spectre and Meltdown attacks~\cite{Kocher&19}. Like many vulnerabilities, both attacks have been shown to exist for a range of processors that make use of speculative execution, and mitigation against these requires software-side interventions. However, despite the widespread nature of these vulnerabilities, the attacks themselves were not discovered through formal verification, but rather through a series of experiments over the training and timing of micro-architectural components. Fortunately, once a security flaw has been uncovered, even complex attacks such as Spectre and Meltdown can be formally characterised and isolated~\cite{Cheang&19}. Once this has been done, the next phase is a formal framework for reasoning about such issues, followed by more streamlined tools to scale verification to larger scale systems. Thus, as important as it is to continue research into the practical use of FM in security, it is equally important to expand our reasoning capabilities for FM in security through the study of theoretical aspects of the discipline. Without this, there is a possibility of a new type of security flaw that falls outside the realms of current day logics. Spectre and Meltdown, for instance, are instances of subset-closed hyper-properties~\cite{Cheang&19}. Hence without existing works on hyper-properties~\cite{Clarkson&10} and subsequent works on their verification, the specification and therefore use of FM to protect against Spectre and Meltdown would have been much more difficult. 

In this paper, we have shown how FM have had an impact on society so far and how this impact will increase in the future. In the past, security has been an optional extra that industry does not want to invest in during development. But times are changing. For example, security has become a core selling point for Amazon Web Services (see Section~\ref{sec:application}).  FM have been used successfully in the financial, industrial, consumer, and enterprise sectors (see Section~\ref{sec:survey}).

%  \begin{itemize}
%  \item Progress in the field of FM for security
%  \item Categorization for FM in security
%  \item Most used FM (model checking, theorem proving, validation) in security
%  \item Impact of FM in security so far
%  \item Most common area for usage of FM in security verification
%  \item Future trends in FM in security
%  \end{itemize}

\subsection*{Specification Languages and Associated Tools}

Our survey covers more than a decade of the use of FM in security. It reveals the rich variety of formal specification languages and their tools, theorem provers, model checkers, and verification frameworks. We have recorded more than 40 different specification languages and more than 40 different verification tools. These include the following.

\paragraph{Specification languages} %
AADL (Architecture Analysis \& Design Language) \cite{Cofer&18,Apvrille&16}, %
ASF (Anonymous Secure Framework) \cite{Kumar&17}, %
ASLan++ (AVANTSSAR Specification Language) \cite{Armando&12}, %
BAN logic \cite{BurrowsAN1990,Snekkenes1991}, %
Boogie \cite{BarnettCDJL05}, %
Boxed Ambients \cite{Jarraya&12}, %
CASM (ASM-based SL for compilers) \cite{Vasudevan&16}, %
CCS (Calculus of Communicating Systems) \cite{Kumar&19}, %
COVERT (compositional analysis of Android apps) \cite{Bagheri&15}, %
CSP (Communicating Sequential Systems) \cite{Heneghan&19}, %
CSP$\#$ (shared variables CSP) \cite{Sun&08}, %
CTL (Computation tree temporal logic) \cite{Song&12b}, %
%Circus (CSP + state + timing) \cite{Freitas2011}, %
Cloud Calculus \cite{Jarraya&12}, %
Cryptol \cite{Cryptol09}, %
Dynamic State Machine \cite{Nardone&16}, %
ERC20 token contracts \cite{Daejun&18}, %
Event-B \cite{Devyanin&14}, %
HLPSL (High Level Protocol Specification Language) \cite{Bojjagani&15}, %
Hoare logic \cite{Guanciale&16}, %
LS2 (Logic of Secure Systems) \cite{Bai&14}, %
LTL (linear-time temporal logic) \cite{Xiao&14}, %
Markov Decision Process \cite{Moshin&17}, %
%Mobile Ambients \cite{FranqueiraEWL09}, %
Petri nets \cite{Amoah&16}, %
$\pi$-calculus \cite{Blanchet16}, %
PlusCal \cite{AkhtarZ017}, %
Promela \cite{MartinelliMN18}, %
RTL (real-time logic) \cite{Guo&17}, %
SPDL (Security Protocol Description Language) \cite{Madhoun&16}, %
SysML-Sec \cite{Apvrille&16}, %
%TAME (Timed Automata Modeling Environment) \cite{Heitmeyer2008}, %
TLA+ (Temporal Logic of Actions) \cite{Cook18}, %
Trusted CSP$\#$ \cite{Bai&14}, %
\"{u}berSpark \cite{Vasudevan&16}, %
VDM \cite{Freitas&18}, %
Verilog \cite{Love&11}, %
VHDL \cite{Guo&16}, %
VML \cite{Skowyra&14}, %
vTRUST \cite{Hao&13}, %
XMHF (eXtensible and Modular Hypervisor Framework) \cite{Vasudevan&13}, %
Z \cite{WoodcockSCCJ2008}. %

\paragraph{Model checkers}
AVISPA (Automated Validation of Internet Security Protocols and Applications) \cite{Armando&05}, %
Alloy \cite{Devyanin&14}, %
CBMC (Bounded Model Checker for C and C++) \cite{ChattopadhyayR18}, %
CWB-NC (Concurrency Workbench of New Century) \cite{Kumar&19}, %
Cadence IFV (RTL block-level verifier) \cite{Guo&16}, %
FDR \cite{FDR42019}, %
GROOVE \cite{Ghamarian&10}, %
jKind \cite{Gacek&17}, %
NuSMV \cite{NuSMV}, %
OFMC (on-the-fly model checker) \cite{BasinMV05}, %
PAT (Process Analysis Toolkit for CSP$\#$) \cite{Sun&08}, %
PRISM (probabilistic model checker) \cite{Moshin&17}, %
SATMC (SAT-based model checker for security protocols) \cite{Bojjagani&15}, %
SPIN \cite{Tsukada&16}, %
TRUSTFOUND \cite{Bai&14}, %
UPPAAL \cite{Mercaldo&19}, %
UVHM (formal analysis scheme for hypervisors) \cite{Vasudevan&13}. %

\paragraph{Theorem provers}
Coq \cite{Cook18}, %
Isabelle/HOL \cite{KleinAEHCDEEKNSTW2010}, %
K-framework \cite{Daejun&18}, %
%PVS \cite{Miller&05}, %
TAMARIN \cite{Kunemann&13}, %
Why \cite{FilliatreM07}. %

\paragraph{Verification tools and frameworks}
AndroBugs (Framework For Android Vulnerability Scanning) \cite{Taylor&17}, %
Cl-Atse (protocol analyser) \cite{Karla&15}, %
FUDGE (Fuzz driver generator) \cite{BabicBCIKKLSW19}, %
Frama-C \cite{Vasudevan&16}, %
Kraka\-tau \cite{Kumar&19}, %
Maude (rewrite engine) \cite{Nigam&19}, %
MobSF (mobile security framework) \cite{Ibrar&17}, %
phpSAFE \cite{Nunes&15}, %
ProVerif \cite{Blanchet16}, %
Quark \cite{Dongseok&12}, %
SAW (Software Analysis Workbench) \cite{Cook18}, %
SMACK \cite{Cook18}, %
SecGuru \cite{Jayaraman&2014}, %
SecVeriLog \cite{Ferraiuolo&17a}, %
Sugar (SAT-based) \cite{Madi&18}, %
TTool (translator from SysML-Sec to $\pi$-calculus) \cite{Apvrille&16}, %
Z3 \cite{Athalye&19}. %

This shows how research and application in FM for security has developed since Burrows et al.'s seminal paper \cite{BurrowsAN1990}. Our survey concentrated in particular on the practical application of these techniques, especially on an industrial scale. Today, it seems inconceivable that a company would produce a commercial secure system without subjecting it to formal analysis. We also suspect that hackers use formal techniques to crack supposedly secure systems. 

As a final statement we need to acknowledge that a survey provides a snapshot in time within a developing field. It is therefore necessary to return to a survey work every decade, something we are planning on doing. Despite this shortcoming it is the opinion of the authors that a survey work is an important part of the research as it provides a starting point and a direction indicator for new and experienced practitioners looking for works under the large field of formal methods in security.

\section{Acknowledgements}
   This work is supported by the Manufacturing Academy of Denmark, for
   more information see \url{www.made.dk}. Brijesh Dongol is supported by
   grants ``FaCT: Faithful Composition of Trust'' and EPSRC grant
   EP/R032556/1. Steve Schneider is supported by EPSRC grants EP/P031811/1 and EP/R006938/1. Jim Woodcock is supported by the Poul Due Jensen
   Foundation and grants EP/M025756/1, EP/R025479/1, and IEC/NSFC/170319. We would also like to thank Nick Battle, Jaco van de Pol and Bas Spitters for reviews of earlier versions of this article. Finally, we would very much like to thank the anonymous reviewers of an earlier version of this article for their valuable input which definitely has improved it.

\bibliographystyle{alpha}
\bibliography{malware,refs}

\newcommand{\etalchar}[1]{$^{#1}$}
\begin{thebibliography}{BGWL{\etalchar{+}}18}

\bibitem[AAA{\etalchar{+}}12]{Armando&12}
Alessandro Armando, Wihem Arsac, Tigran Avanesov, Michele Barletta, Alberto
  Calvi, Alessandro Cappai, Roberto Carbone, Yannick Chevalier, Luca Compagna,
  Jorge Cu{\'e}llar, Gabriel Erzse, Simone Frau, Marius Minea, Sebastian
  M{\"o}dersheim, David von Oheimb, Giancarlo Pellegrino, Serena~Elisa Ponta,
  Marco Rocchetto, Michael Rusinowitch, Mohammad Torabi~Dashti, Mathieu
  Turuani, and Luca Vigan{\`o}.
\newblock The {AVANTSSAR} platform for the automated validation of trust and
  security of service-oriented architectures.
\newblock In Cormac Flanagan and Barbara K{\"o}nig, editors, {\em Tools and
  Algorithms for the Construction and Analysis of Systems}, pages 267--282,
  Berlin, Heidelberg, 2012. Springer Berlin Heidelberg.

\bibitem[ABB{\etalchar{+}}05]{Armando&05}
A.~Armando, D.~Basin, Y.~Boichut, Y.~Chevalier, L.~Compagna, J.~Cuellar,
  P.~Hankes Drielsma, P.~C. He{\'a}m, O.~Kouchnarenko, J.~Mantovani,
  S.~M{\"o}dersheim, D.~von Oheimb, M.~Rusinowitch, J.~Santiago, M.~Turuani,
  L.~Vigan{\`o}, and L.~Vigneron.
\newblock The {AVISPA} tool for the automated validation of internet security
  protocols and applications.
\newblock In Kousha Etessami and Sriram~K. Rajamani, editors, {\em Computer
  Aided Verification}, pages 281--285, Berlin, Heidelberg, 2005. Springer
  Berlin Heidelberg.

\bibitem[ABK{\etalchar{+}}19]{Athalye&19}
Anish Athalye, Adam Belay, M.~Frans Kaashoek, Robert Morris, and Nickolai
  Zeldovich.
\newblock {Notary}: {A} device for secure transaction approval.
\newblock In {\em Proceedings of the 27th ACM Symposium on Operating Systems
  Principles}, SOSP ’19, page 97–113, New York, NY, USA, 2019. Association
  for Computing Machinery.

\bibitem[Abr10]{Abrial&10}
Jean-Raymond Abrial.
\newblock {\em Modeling in {Event}-{B}: {System} and Software Engineering}.
\newblock Cambridge University Press, Cambridge, UK, 2010.

\bibitem[ABSW13]{Alsheri&13}
A.~{Alshehri}, J.~A. {Briffa}, S.~{Schneider}, and S.~{Wesemeyer}.
\newblock Formal security analysis of {NFC} {M}-coupon protocols using
  {Casper}/{FDR}.
\newblock In {\em 2013 5th International Workshop on Near Field Communication
  (NFC)}, pages 1--6, 2013.

\bibitem[ACD90]{Alur&90}
R.~{Alur}, C.~{Courcoubetis}, and D.~{Dill}.
\newblock Model-checking for real-time systems.
\newblock In {\em [1990] Proceedings. Fifth Annual IEEE Symposium on Logic in
  Computer Science}, pages 414--425, 445 Hoes Lane Piscataway, NJ 08854 USA,
  June 1990. IEEE.

\bibitem[ACF16]{Amoah&16}
Raphael Amoah, Seyit Camtepe, and Ernest Foo.
\newblock Formal modelling and analysis of {DNP3} secure authentication.
\newblock {\em Journal of Network and Computer Applications}, 59:345 -- 360,
  2016.

\bibitem[ADKT11]{Alglave&11}
Jade Alglave, Alastair~F. Donaldson, Daniel Kroening, and Michael Tautschnig.
\newblock Making software verification tools really work.
\newblock In Tevfik Bultan and Pao-Ann Hsiung, editors, {\em Automated
  Technology for Verification and Analysis}, pages 28--42, Berlin, Heidelberg,
  2011. Springer Berlin Heidelberg.

\bibitem[ADMM14]{Andrychowicz&14}
Marcin Andrychowicz, Stefan Dziembowski, Daniel Malinowski, and {\L}ukasz
  Mazurek.
\newblock Modeling bitcoin contracts by timed automata.
\newblock In Axel Legay and Marius Bozga, editors, {\em Formal Modeling and
  Analysis of Timed Systems}, pages 7--22, Cham, 2014. Springer International
  Publishing.

\bibitem[ADRP13]{Aarts&13}
Fides Aarts, Joeri De~Ruiter, and Erik Poll.
\newblock Formal models of bank cards for free.
\newblock In {\em 2013 IEEE Sixth International Conference on Software Testing,
  Verification and Validation Workshops}, pages 461--468, 445 Hoes Lane
  Piscataway, NJ 08854 USA, 2013. IEEE.

\bibitem[AE18]{Annenkov&18}
Danil Annenkov and Martin Elsman.
\newblock Certified compilation of financial contracts.
\newblock In {\em Proceedings of the 20th International Symposium on Principles
  and Practice of Declarative Programming}, PPDP ’18, New York, NY, USA,
  2018. Association for Computing Machinery.

\bibitem[AG96]{AdveG96}
Sarita~V. Adve and Kourosh Gharachorloo.
\newblock Shared memory consistency models: {A} tutorial.
\newblock {\em {IEEE} Computer}, 29(12):66--76, 1996.

\bibitem[AG99]{Abadi&99}
Mart\'{i}n Abadi and Andrew~D. Gordon.
\newblock A calculus for cryptographic protocols: {The} {Spi} calculus.
\newblock {\em Information and Computation}, 148(1):1 -- 70, 1999.

\bibitem[AGKK19]{Arapinis&19}
Myrto Arapinis, Andriana Gkaniatsou, Dimitris Karakostas, and Aggelos Kiayias.
\newblock A formal treatment of hardware wallets.
\newblock In Ian Goldberg and Tyler Moore, editors, {\em Financial Cryptography
  and Data Security}, pages 426--445, Cham, 2019. Springer International
  Publishing.

\bibitem[Aia15]{Aiash&15}
Mahdi Aiash.
\newblock A formal analysis of authentication protocols for mobile devices in
  next generation networks.
\newblock {\em Concurrency and Computation: Practice and Experience},
  27(12):2938--2953, 2015.

\bibitem[ALKH17]{Abbasi&17}
Imran~Hafeez Abbasi, Faiq~Khalid Lodhi, Awais~Mehmood Kamboh, and Osman Hasan.
\newblock Formal verification of gate-level multiple side channel parameters to
  detect hardware trojans.
\newblock In Cyrille Artho and Peter~Csaba {\"O}lveczky, editors, {\em Formal
  Techniques for Safety-Critical Systems}, pages 75--92, Cham, 2017. Springer
  International Publishing.

\bibitem[ALR16]{Apvrille&16}
L.~{Apvrille}, L.~{Li}, and Y.~{Roudier}.
\newblock Model-driven engineering for designing safe and secure embedded
  systems.
\newblock In {\em 2016 Architecture-Centric Virtual Integration (ACVI)}, pages
  4--7, 445 Hoes Lane Piscataway, NJ 08854 USA, April 2016. IEEE.

\bibitem[{Ama}19a]{S3Amazon}
{Amazon.com Inc.}
\newblock {{Amazon} {Simple} {Storage} {Service} ({S3})}.
\newblock \url{aws.amazon.com/s3/}, 2019.
\newblock Accessed February 7 2019.

\bibitem[{Ama}19b]{s2nAmazon}
{Amazon.com Inc.}
\newblock {s2n}.
\newblock \url{github.com/awslabs/s2n}, 2019.
\newblock Accessed February 7 2019.

\bibitem[AMM14]{Abughazalah&14}
S.~{Abughazalah}, K.~{Markantonakis}, and K.~{Mayes}.
\newblock Secure mobile payment on {NFC}-enabled mobile phones formally
  analysed using {CasperFDR}.
\newblock In {\em 2014 IEEE 13th International Conference on Trust, Security
  and Privacy in Computing and Communications}, pages 422--431, 445 Hoes Lane
  Piscataway, NJ 08854 USA, 2014. IEEE.

\bibitem[AUS12]{Ahamad2012}
Shakeel Ahamad, Siba Udgata, and V~Sastry.
\newblock A new mobile payment system with formal verification.
\newblock {\em International Journal of Internet Technology and Secured
  Transactions}, 4:71--103, 01 2012.

\bibitem[AZHL12]{Au&12}
Kathy Wain~Yee Au, Yi~Fan Zhou, Zhen Huang, and David Lie.
\newblock {PS}cout: {Analyzing} the {Android} permission specification.
\newblock In {\em Proceedings of the 2012 ACM Conference on Computer and
  Communications Security}, CCS '12, pages 217--228, New York, NY, USA, 2012.
  ACM.

\bibitem[AZP17]{AkhtarZ017}
Sabina Akhtar, Ehtesham Zahoor, and Olivier Perrin.
\newblock Formal verification of authorization policies for enterprise social
  networks using pluscal-2.
\newblock In Imed Romdhani, Lei Shu, Takahiro Hara, Zhangbing Zhou, Timothy~J.
  Gordon, and Deze Zeng, editors, {\em Collaborative Computing: Networking,
  Applications and Worksharing - 13th International Conference, CollaborateCom
  2017, Edinburgh, UK, December 11-13, 2017, Proceedings}, volume 252 of {\em
  Lecture Notes of the Institute for Computer Sciences, Social Informatics and
  Telecommunications Engineering}, pages 530--540. Springer, 2017.

\bibitem[BAN90]{BurrowsAN1990}
Michael Burrows, Mart{\'{\i}}n Abadi, and Roger~M. Needham.
\newblock A logic of authentication.
\newblock {\em {ACM} Trans. Comput. Syst.}, 8(1):18--36, 1990.

\bibitem[Bar12]{Barnes2012}
John Barnes.
\newblock {\em {Spark}: {The} Proven Approach to High Integrity Software}.
\newblock Altran Praxis, UK, 2012.

\bibitem[BBB{\etalchar{+}}19]{Barbosa&19}
Manuel Barbosa, Gilles Barthe, Karthik Bhargavan, Bruno Blanchet, Cas Cremers,
  Kevin Liao, and Bryan Parno.
\newblock {SoK}: {Computer}-aided cryptography.
\newblock Cryptology ePrint Archive, Report 2019/1393, 2019.
\newblock \url{eprint.iacr.org/2019/1393}.

\bibitem[BBC{\etalchar{+}}19a]{BabicBCIKKLSW19}
Domagoj Babic, Stefan Bucur, Yaohui Chen, Franjo Ivancic, Tim King, Markus
  Kusano, Caroline Lemieux, L{\'{a}}szl{\'{o}} Szekeres, and Wei Wang.
\newblock {FUDGE:} fuzz driver generation at scale.
\newblock In Marlon Dumas, Dietmar Pfahl, Sven Apel, and Alessandra Russo,
  editors, {\em Proceedings of the {ACM} Joint Meeting on European Software
  Engineering Conference and Symposium on the Foundations of Software
  Engineering, {ESEC/SIGSOFT} {FSE} 2019, Tallinn, Estonia, August 26-30,
  2019}, pages 975--985. {ACM}, 2019.

\bibitem[BBC{\etalchar{+}}19b]{Babic&19}
Domagoj Babi\'{c}, Stefan Bucur, Yaohui Chen, Franjo Ivan\v{c}i\'{c}, Tim King,
  Markus Kusano, Caroline Lemieux, L\'{a}szl\'{o} Szekeres, and Wei Wang.
\newblock {FUDGE}: {Fuzz} driver generation at scale.
\newblock In {\em Proceedings of the 2019 27th ACM Joint Meeting on European
  Software Engineering Conference and Symposium on the Foundations of Software
  Engineering}, ESEC/FSE 2019, page 975–985, New York, NY, USA, 2019.
  Association for Computing Machinery.

\bibitem[BCB01]{Blackburn&01}
M.R. Blackburn, Ramaswamy Chandramouli, and Robert Busser.
\newblock Model-based approach to security test automation.
\newblock {\em Quality Week}, 01 2001.

\bibitem[BCD{\etalchar{+}}05]{BarnettCDJL05}
Michael Barnett, Bor{-}Yuh~Evan Chang, Robert DeLine, Bart Jacobs, and
  K.~Rustan~M. Leino.
\newblock {Boogie}: {A} modular reusable verifier for object-oriented programs.
\newblock In Frank~S. de~Boer, Marcello~M. Bonsangue, Susanne Graf, and
  Willem~P. de~Roever, editors, {\em Formal Methods for Components and Objects,
  4th International Symposium, {FMCO} 2005, Amsterdam, The Netherlands,
  November 1-4, 2005, Revised Lectures}, volume 4111 of {\em Lecture Notes in
  Computer Science}, pages 364--387, Berlin, Heidelberg, 2005. Springer.

\bibitem[BCJ{\etalchar{+}}06]{BarnesCJWCE2006}
Janet Barnes, Rod Chapman, Randy Johnson, James Widmaier, David Cooper, and
  Bill Everett.
\newblock Engineering the {Tokeneer} enclave protection system.
\newblock In {\em Proceedings of the 1st {IEEE} International Symposium on
  Secure Software Engineering}, page~10, 445 Hoes Lane Piscataway, NJ 08854
  USA, 2006. IEEE Computer Society Press.

\bibitem[BCLM05]{Bruschi&05}
D.~{Bruschi}, L.~{Cavallaro}, A.~{Lanzi}, and M.~{Monga}.
\newblock Replay attack in {TCG} specification and solution.
\newblock In {\em 21st Annual Computer Security Applications Conference
  (ACSAC'05)}, pages 11 pp.--137, 445 Hoes Lane Piscataway, NJ 08854 USA, Dec
  2005. IEEE.

\bibitem[BCMS05]{Bugliesi&05}
Michele Bugliesi, Silvia Crafa, Massimo Merro, and V.~Sassone.
\newblock Communication and mobility control in boxed ambients.
\newblock {\em Information and Computation}, 202:39--86, 10 2005.

\bibitem[BDH{\etalchar{+}}18]{Basin&18}
David Basin, Jannik Dreier, Lucca Hirschi, Sa{\v{s}}a Radomirovic, Ralf Sasse,
  and Vincent Stettler.
\newblock A formal analysis of {5G} authentication.
\newblock In {\em Proceedings of the 2018 ACM SIGSAC Conference on Computer and
  Communications Security}, pages 1383--1396, 2018.

\bibitem[BDSS08]{BratusDSS08}
Sergey Bratus, Nihal D'Cunha, Evan~R. Sparks, and Sean~W. Smith.
\newblock {TOCTOU}, traps, and trusted computing.
\newblock In Peter Lipp, Ahmad{-}Reza Sadeghi, and Klaus{-}Michael Koch,
  editors, {\em Trusted Computing - Challenges and Applications, First
  International Conference on Trusted Computing and Trust in Information
  Technologies, Trust 2008, Villach, Austria, March 11-12, 2008, Proceedings},
  volume 4968 of {\em Lecture Notes in Computer Science}, pages 14--32, Berlin,
  Heidelberg, 2008. Springer.

\bibitem[BGV11]{Benabbas&11}
Siavosh Benabbas, Rosario Gennaro, and Yevgeniy Vahlis.
\newblock Verifiable delegation of computation over large datasets.
\newblock In Phillip Rogaway, editor, {\em Advances in Cryptology -- CRYPTO
  2011}, pages 111--131, Berlin, Heidelberg, 2011. Springer Berlin Heidelberg.

\bibitem[BGWL{\etalchar{+}}18]{Biondi&18}
Fabrizio Biondi, Thomas Given-Wilson, Axel Legay, Cassius Puodzius, and Jean
  Quilbeuf.
\newblock Tutorial: {An} overview of malware detection and evasion techniques.
\newblock In Tiziana Margaria and Bernhard Steffen, editors, {\em Leveraging
  Applications of Formal Methods, Verification and Validation. Modeling}, pages
  565--586, Cham, 2018. Springer International Publishing.

\bibitem[BHBLT17]{Hamida&17}
Elyes Ben~Hamida, Kei~Leo Brousmiche, Hugo Levard, and Eric Thea.
\newblock {Blockchain for Enterprise: {Overview}, Opportunities and
  Challenges}.
\newblock In {\em {The Thirteenth International Conference on Wireless and
  Mobile Communications (ICWMC 2017)}}, page~17, Nice, France, July 2017. IARIA
  XPS Press.

\bibitem[BHvOS12]{Bonneau&12}
J.~{Bonneau}, C.~{Herley}, P.~C. v.~{Oorschot}, and F.~{Stajano}.
\newblock The quest to replace passwords: {A} framework for comparative
  evaluation of web authentication schemes.
\newblock In {\em 2012 IEEE Symposium on Security and Privacy}, pages 553--567,
  445 Hoes Lane Piscataway, NJ 08854 USA, 2012. IEEE.

\bibitem[BHW{\etalchar{+}}14]{Bai&14}
Guangdong Bai, Jianan Hao, Jianliang Wu, Yang Liu, Zhenkai Liang, and Andrew
  Martin.
\newblock {TrustFound}: {Towards} a formal foundation for model checking
  trusted computing platforms.
\newblock In Cliff Jones, Pekka Pihlajasaari, and Jun Sun, editors, {\em FM
  2014: Formal Methods}, pages 110--126, Cham, 2014. Springer International
  Publishing.

\bibitem[BKMJ18]{Bagheri&18}
Hamid Bagheri, Eunsuk Kang, Sam Malek, and Daniel Jackson.
\newblock A formal approach for detection of security flaws in the {Android}
  permission system.
\newblock {\em Formal Aspects of Computing}, 30(5):525--544, Sep 2018.

\bibitem[Bla16]{Blanchet16}
Bruno Blanchet.
\newblock Modeling and verifying security protocols with the applied pi
  calculus and proverif.
\newblock {\em Found. Trends Priv. Secur.}, 1(1-2):1--135, 2016.

\bibitem[BML{\etalchar{+}}11]{Bohn&11}
R.~Bohn, John Messina, Fang Liu, Jin Tong, and Jian Mao.
\newblock {NIST} cloud computing reference architecture.
\newblock pages 594--596, 07 2011.

\bibitem[BMV05]{BasinMV05}
David~A. Basin, Sebastian M{\"{o}}dersheim, and Luca Vigan{\`{o}}.
\newblock {OFMC:} {A} symbolic model checker for security protocols.
\newblock {\em Int. J. Inf. Sec.}, 4(3):181--208, 2005.

\bibitem[BMV14]{Boureanu&14}
Ioana Boureanu, Aikaterini Mitrokotsa, and Serge Vaudenay.
\newblock Towards secure distance bounding.
\newblock In Shiho Moriai, editor, {\em Fast Software Encryption}, pages
  55--67, Berlin, Heidelberg, 2014. Springer Berlin Heidelberg.

\bibitem[BR70]{Buxton&70}
J.~N. Buxton and B.~Randell.
\newblock {\em Software Engineering Techniques: Report of a Conference
  Sponsored by the NATO Science Committee, Rome, Italy, 27-31 Oct. 1969,
  Brussels, Scientific Affairs Division, NATO}.
\newblock 1970.

\bibitem[BS15]{Bojjagani&15}
Sriramulu Bojjagani and V.~N. Sastry.
\newblock {SSMBP}: {A} secure {SMS}-based mobile banking protocol with formal
  verification.
\newblock In {\em WiMob}, pages 252--259, 445 Hoes Lane Piscataway, NJ 08854
  USA, 2015. IEEE Computer Society.

\bibitem[BSCS18]{Blanchet&18}
Bruno Blanchet, Ben Smyth, Vincent Cheval, and Marc Sylvestre.
\newblock {\em {ProVerif} 2.00: {Automatic} Cryptographic Protocol Verifier,
  User Manual and Tutorial}.
\newblock INRIA, 2018.
\newblock {Originally appeared as Bruno Blanchet and Ben Smyth (2011) ProVerif
  1.85: Automatic Cryptographic Protocol Verifier, User Manual and Tutorial.}

\bibitem[BSGM15]{Bagheri&15}
H.~{Bagheri}, A.~{Sadeghi}, J.~{Garcia}, and S.~{Malek}.
\newblock {COVERT}: Compositional analysis of {Android} inter-app permission
  leakage.
\newblock {\em IEEE Transactions on Software Engineering}, 41(9):866--886,
  September 2015.

\bibitem[BSNRN14]{Bruni&14}
Alessandro Bruni, Michal Sojka, Flemming Nielson, and Hanne Riis~Nielson.
\newblock Formal security analysis of the {MaCAN} protocol.
\newblock In Elvira Albert and Emil Sekerinski, editors, {\em Integrated Formal
  Methods}, pages 241--255, Cham, 2014. Springer International Publishing.

\bibitem[BSW{\etalchar{+}}18]{Bracho&18}
Alejandro Bracho, Can Saygin, HungDa Wan, Yooneun Lee, and Alireza Zarreh.
\newblock A simulation-based platform for assessing the impact of cyber-threats
  on smart manufacturing systems.
\newblock {\em Procedia Manufacturing}, 26:1116--1127, 2018.
\newblock 46th SME North American Manufacturing Research Conference, NAMRC 46,
  Texas, USA.

\bibitem[BTW{\etalchar{+}}13]{Busold&13}
Christoph Busold, Ahmed Taha, Christian Wachsmann, Alexandra Dmitrienko,
  Herv\'{e} Seudi\'{e}, Majid Sobhani, and Ahmad-Reza Sadeghi.
\newblock Smart keys for cyber-cars: {Secure} smartphone-based {NFC}-enabled
  car immobilizer.
\newblock In {\em Proceedings of the Third ACM Conference on Data and
  Application Security and Privacy}, CODASPY ’13, page 233–242, New York,
  NY, USA, 2013. Association for Computing Machinery.

\bibitem[BVGM15]{Bleikertz&15}
S\"{o}ren Bleikertz, Carsten Vogel, Thomas Gro\ss{}, and Sebastian
  M\"{o}dersheim.
\newblock Proactive security analysis of changes in virtualized
  infrastructures.
\newblock In {\em Proceedings of the 31st Annual Computer Security Applications
  Conference}, ACSAC 2015, page 51–60, New York, NY, USA, 2015. Association
  for Computing Machinery.

\bibitem[Cad20]{Cadence}
{Cadence} {IFV} model checker.
\newblock
  \url{www.cadence.com/en_US/home/tools/system-design-and-verification/formal-and-static-verification/jasper-gold-verification-platform.html},
  2020.
\newblock Accessed: 2020-02-21.

\bibitem[{Can}01]{Canetti&01}
R.~{Canetti}.
\newblock Universally composable security: {A} new paradigm for cryptographic
  protocols.
\newblock In {\em Proceedings 42nd IEEE Symposium on Foundations of Computer
  Science}, pages 136--145, 445 Hoes Lane Piscataway, NJ 08854 USA, 2001. IEEE.

\bibitem[CCC{\etalchar{+}}18]{Chudnov&18}
Andrey Chudnov, Nathan Collins, Byron Cook, Joey Dodds, Brian Huffman, Colm
  MacC{\'a}rthaigh, Stephen Magill, Eric Mertens, Eric Mullen, Serdar Tasiran,
  Aaron Tomb, and Eddy Westbrook.
\newblock Continuous formal verification of {Amazon} s2n.
\newblock In Hana Chockler and Georg Weissenbacher, editors, {\em Computer
  Aided Verification}, pages 430--446, Cham, 2018. Springer International
  Publishing.

\bibitem[CCD{\etalchar{+}}17]{Cohn-Gordon&17}
K.~{Cohn-Gordon}, C.~{Cremers}, B.~{Dowling}, L.~{Garratt}, and D.~{Stebila}.
\newblock A formal security analysis of the {Signal} {Messaging} {Protocol}.
\newblock In {\em 2017 IEEE European Symposium on Security and Privacy (EuroS
  P)}, pages 451--466, 445 Hoes Lane Piscataway, NJ 08854 USA, April 2017.
  IEEE.

\bibitem[CCGR99]{NuSMV}
Alessandro Cimatti, Edmund~M. Clarke, Fausto Giunchiglia, and Marco Roveri.
\newblock {NUSMV}: {A} new symbolic model verifier.
\newblock In {\em Proceedings of the 11th International Conference on Computer
  Aided Verification}, CAV ’99, page 495–499, Berlin, Heidelberg, 1999.
  Springer-Verlag.

\bibitem[CCR06]{CCRA2006}
Common Criteria Recognition~Agreement CCRA.
\newblock Common criteria for information technology security evaluation.
  {Part} 1: {Introduction} and general model.
\newblock Tech. Rep. CCMB-2006-09-001, Version 3.1, Revision 1, Sept 2006.

\bibitem[CFM16]{Chen&16}
S.~{Chen}, H.~{Fu}, and H.~{Miao}.
\newblock Formal verification of security protocols using {Spin}.
\newblock In {\em 2016 IEEE/ACIS 15th International Conference on Computer and
  Information Science (ICIS)}, pages 1--6, 445 Hoes Lane Piscataway, NJ 08854
  USA, June 2016. IEEE.

\bibitem[CG00]{Cardelli&00}
Luca Cardelli and Andrew~D. Gordon.
\newblock Mobile ambients.
\newblock {\em Theoretical Computer Science}, 240(1):177 -- 213, 2000.

\bibitem[CGB{\etalchar{+}}18]{Cofer&18}
D.~{Cofer}, A.~{Gacek}, J.~{Backes}, M.~W. {Whalen}, L.~{Pike}, A.~{Foltzer},
  M.~{Podhradsky}, G.~{Klein}, I.~{Kuz}, J.~{ Andronick}, G.~{Heiser}, and
  D.~{Stuart}.
\newblock A formal approach to constructing secure air vehicle software.
\newblock {\em Computer}, 51(11):14--23, Nov 2018.

\bibitem[CGDR{\etalchar{+}}15]{Chothia&15}
Tom Chothia, Flavio~D Garcia, Joeri De~Ruiter, Jordi Van Den~Breekel, and
  Matthew Thompson.
\newblock Relay cost bounding for contactless {EMV} payments.
\newblock In {\em International Conference on Financial Cryptography and Data
  Security}, pages 189--206, Berlin, Heidelberg, 2015. Springer.

\bibitem[CGHS17]{Chothia&17}
Tom Chothia, Flavio~D Garcia, Chris Heppel, and Chris~McMahon Stone.
\newblock Why banker bob (still) can’t get {TLS} right: {A} security analysis
  of {TLS} in leading {UK} banking apps.
\newblock In {\em International Conference on Financial Cryptography and Data
  Security}, pages 579--597, Berlin, Heidelberg, 2017. Springer.

\bibitem[CKK{\etalchar{+}}12]{CuoqKKPSY12}
Pascal Cuoq, Florent Kirchner, Nikolai Kosmatov, Virgile Prevosto, Julien
  Signoles, and Boris Yakobowski.
\newblock {Frama}-{C} --- {A} software analysis perspective.
\newblock In George Eleftherakis, Mike Hinchey, and Mike Holcombe, editors,
  {\em Software Engineering and Formal Methods - 10th International Conference,
  {SEFM} 2012, Thessaloniki, Greece, October 1-5, 2012. Proceedings}, volume
  7504 of {\em Lecture Notes in Computer Science}, pages 233--247, Berlin,
  Heidelberg, 2012. Springer.

\bibitem[CKL04]{Clarke&04}
Edmund Clarke, Daniel Kroening, and Flavio Lerda.
\newblock A tool for checking {ANSI-C} programs.
\newblock In Kurt Jensen and Andreas Podelski, editors, {\em Tools and
  Algorithms for the Construction and Analysis of Systems}, pages 168--176,
  Berlin, Heidelberg, 2004. Springer Berlin Heidelberg.

\bibitem[CLW{\etalchar{+}}14]{Chen&14}
Xiaofeng Chen, Jin Li, Jian Weng, Jianfeng Ma, and Wenjing Lou.
\newblock Verifiable computation over large database with incremental updates.
\newblock In Miros{\l}aw Kuty{\l}owski and Jaideep Vaidya, editors, {\em
  Computer Security - ESORICS 2014}, pages 148--162, Cham, 2014. Springer
  International Publishing.

\bibitem[CM12]{Cremers&12}
Cas Cremers and Sjouke Mauw.
\newblock {\em Operational Semantics and Verification of Security Protocols}.
\newblock Springer, Berlin, Heidelberg, 2012.

\bibitem[CMN{\etalchar{+}}18]{Cimitile&18}
Aniello Cimitile, Francesco Mercaldo, Vittoria Nardone, Antonella Santone, and
  Corrado~Aaron Visaggio.
\newblock {Talos}: {No} more ransomware victims with formal methods.
\newblock {\em International Journal of Information Security}, 17(6):719--738,
  November 2018.

\bibitem[Coo18]{Cook18}
Byron Cook.
\newblock {Formal Reasoning About the Security of {Amazon} {Web} {Services}}.
\newblock In Hana Chockler and Georg Weissenbacher, editors, {\em Computer
  Aided Verification}, pages 38--47, Cham, 2018. Springer International
  Publishing.

\bibitem[CR18]{ChattopadhyayR18}
Sudipta Chattopadhyay and Abhik Roychoudhury.
\newblock Symbolic verification of cache side-channel freedom.
\newblock {\em {IEEE} Trans. Comput. Aided Des. Integr. Circuits Syst.},
  37(11):2812--2823, 2018.

\bibitem[Cre08]{Cremers&08}
Cas J.~F. Cremers.
\newblock The {Scyther} tool: {Verification}, falsification, and analysis of
  security protocols.
\newblock In Aarti Gupta and Sharad Malik, editors, {\em Computer Aided
  Verification}, pages 414--418, Berlin, Heidelberg, 2008. Springer Berlin
  Heidelberg.

\bibitem[CRSS19]{Cheang&19}
K.~{Cheang}, C.~{Rasmussen}, S.~{Seshia}, and P.~{Subramanyan}.
\newblock A formal approach to secure speculation.
\newblock In {\em 2019 IEEE 32nd Computer Security Foundations Symposium
  (CSF)}, pages 288--28815, 2019.

\bibitem[CS96]{Cleaveland&96}
Rance Cleaveland and Steve Sims.
\newblock The {NCSU} concurrency workbench.
\newblock In Rajeev Alur and Thomas~A. Henzinger, editors, {\em Computer Aided
  Verification}, pages 394--397, Berlin, Heidelberg, 1996. Springer Berlin
  Heidelberg.

\bibitem[CS10]{Clarkson&10}
Michael~R. Clarkson and Fred~B. Schneider.
\newblock Hyperproperties.
\newblock {\em J. Comput. Secur.}, 18(6):1157–1210, September 2010.

\bibitem[DA07]{Dominikus&07}
S.~{Dominikus} and M.~{Aigner}.
\newblock {mCoupons}: {An} application for near field communication ({NFC}).
\newblock In {\em 21st International Conference on Advanced Information
  Networking and Applications Workshops (AINAW'07)}, volume~2, pages 421--428,
  2007.

\bibitem[DCC{\etalchar{+}}13]{Davis&13}
Jennifer~A. Davis, Matthew Clark, Darren Cofer, Aaron Fifarek, Jacob Hinchman,
  Jonathan Hoffman, Brian Hulbert, Steven~P. Miller, and Lucas Wagner.
\newblock Study on the barriers to the industrial adoption of formal methods.
\newblock In Charles Pecheur and Michael Dierkes, editors, {\em Formal Methods
  for Industrial Critical Systems}, pages 63--77, Berlin, Heidelberg, 2013.
  Springer Berlin Heidelberg.

\bibitem[DDSW11]{Davi&11}
Lucas Davi, Alexandra Dmitrienko, Ahmad-Reza Sadeghi, and Marcel Winandy.
\newblock Privilege escalation attacks on {Android}.
\newblock In Mike Burmester, Gene Tsudik, Spyros Magliveras, and Ivana
  Ili{\'{c}}, editors, {\em Information Security}, pages 346--360, Berlin,
  Heidelberg, 2011. Springer Berlin Heidelberg.

\bibitem[DF89]{Desmedt&89}
Yvo Desmedt and Yair Frankel.
\newblock Threshold cryptosystems.
\newblock In {\em Conference on the Theory and Application of Cryptology},
  pages 307--315, Berlin, Heidelberg, 1989. Springer.

\bibitem[DFGK09]{Datta&09}
Anupam Datta, Jason Franklin, Deepak Garg, and Dilsun Kaynar.
\newblock A logic of secure systems and its application to trusted computing.
\newblock In {\em Proceedings - IEEE Symposium on Security and Privacy}, pages
  221--236, 445 Hoes Lane Piscataway, NJ 08854 USA, 10 2009. IEEE.

\bibitem[DFLO19]{Distefano&19}
Dino Distefano, Manuel F{\"a}hndrich, Francesco Logozzo, and Peter~W O'Hearn.
\newblock Scaling static analyses at {Facebook}.
\newblock {\em Communications of the ACM}, 62(8):62--70, 2019.

\bibitem[DKK{\etalchar{+}}14]{Devyanin&14}
Petr~N. Devyanin, Alexey~V. Khoroshilov, Victor~V. Kuliamin, Alexander~K.
  Petrenko, and Ilya~V. Shchepetkov.
\newblock Formal verification of {OS} security model with {Alloy} and
  {Event}-{B}.
\newblock In Yamine Ait~Ameur and Klaus-Dieter Schewe, editors, {\em Abstract
  State Machines, Alloy, B, TLA, VDM, and Z}, pages 309--313, Berlin,
  Heidelberg, 2014. Springer Berlin Heidelberg.

\bibitem[dMB08]{deMoura&08}
Leonardo de~Moura and Nikolaj Bj{\o}rner.
\newblock {Z3}: {An} efficient {SMT} solver.
\newblock In C.~R. Ramakrishnan and Jakob Rehof, editors, {\em Tools and
  Algorithms for the Construction and Analysis of Systems}, pages 337--340,
  Berlin, Heidelberg, 2008. Springer Berlin Heidelberg.

\bibitem[DMC{\etalchar{+}}18]{Duan&18}
Zhangbo Duan, Hongliang Mao, Zhidong Chen, Xiaomin Bai, Kai Hu, and Jean-Pierre
  Talpin.
\newblock Formal modeling and verification of blockchain system.
\newblock In {\em Proceedings of the 10th International Conference on Computer
  Modeling and Simulation}, ICCMS 2018, page 231–235, New York, NY, USA,
  2018. Association for Computing Machinery.

\bibitem[DPP{\etalchar{+}}17]{Dreier&17}
Jannik Dreier, Maxime Puys, Marie-Laure Potet, Pascal Lafourcade, and
  Jean-Louis Roch.
\newblock Formally verifying flow properties in industrial systems.
\newblock In {\em {SECRYPT 2017 - 14th International Conference on Security and
  Cryptography}}, Proceedings of the 14th International Joint Conference on
  e-Business and Telecommunications (ICETE 2017) - Volume 4: SECRYPT, Madrid,
  Spain, July 24-26, 2017., pages 55--66, Portugal, July 2017. SCITEPRESS
  Science And Technology Publications.

\bibitem[DRR17]{Denzel&17}
Michael Denzel, Mark Ryan, and Eike Ritter.
\newblock A malware-tolerant, self-healing industrial control system framework.
\newblock In Sabrina De~Capitani~di Vimercati and Fabio Martinelli, editors,
  {\em ICT Systems Security and Privacy Protection}, pages 46--60, Cham, 2017.
  Springer International Publishing.

\bibitem[DS81]{Denning&81}
Dorothy~E. Denning and Giovanni~Maria Sacco.
\newblock Timestamps in key distribution protocols.
\newblock {\em Commun. ACM}, 24(8):533–536, August 1981.

\bibitem[DT17]{Dam&17}
Khanh-Huu-The Dam and Tayssir Touili.
\newblock Learning {Android} malware.
\newblock In {\em Proceedings of the 12th International Conference on
  Availability, Reliability and Security}, ARES '17, pages 59:1--59:9, New
  York, NY, USA, 2017. ACM.

\bibitem[DT18]{Dam&18}
Khanh Huu~The Dam and Tayssir Touili.
\newblock Learning malware using generalized graph kernels.
\newblock In {\em Proceedings of the 13th International Conference on
  Availability, Reliability and Security}, ARES 2018, pages 28:1--28:6, New
  York, NY, USA, 2018. ACM.

\bibitem[DY81]{DolevY81}
Danny Dolev and Andrew~Chi{-}Chih Yao.
\newblock On the security of public key protocols (extended abstract).
\newblock In {\em 22nd Annual Symposium on Foundations of Computer Science,
  Nashville, Tennessee, USA, 28-30 October 1981}, pages 350--357, 445 Hoes Lane
  Piscataway, NJ 08854 USA, 1981. {IEEE} Computer Society.

\bibitem[EAP14]{Apvrille&14}
Andrea Enrici, Ludovic Apvrille, and Renaud Pacalet.
\newblock {TTool}/{DiplodocusDF}: {A} {UML} environment for hardware/software
  co-design of data-dominated systems-on-chip, 2014.

\bibitem[EM09]{Cryptol09}
Levent Erk{\"o}k and John Matthews.
\newblock {Pragmatic equivalence and safety checking in {Cryptol}}.
\newblock In {\em Proceedings of the 3rd workshop on Programming Languages
  meets Program Verification}, pages 73--82, New York, NY, USA, 2009. ACM, ACM.

\bibitem[{Fab}20]{ERC20}
{Fabian Vogelsteller and Vitalik Buterin}.
\newblock {{ERC20} Token Standard}.
\newblock \url{github.com/ethereum/EIPs/blob/master/EIPS/eip-20.md}, 2020.
\newblock Accessed March 24 2020.

\bibitem[FCH{\etalchar{+}}11]{Felt&11}
Adrienne~Porter Felt, Erika Chin, Steve Hanna, Dawn Song, and David Wagner.
\newblock {Android} permissions demystified.
\newblock In {\em Proceedings of the 18th ACM Conference on Computer and
  Communications Security}, CCS '11, pages 627--638, New York, NY, USA, 2011.
  ACM.

\bibitem[FHB89]{FlynnHB89}
Mike Flynn, Tim Hoverd, and David Brazier.
\newblock {Formaliser} --- {An} interactive support tool for {Z}.
\newblock In John~E. Nicholls, editor, {\em Proceedings of the Fourth Annual
  {Z} User Meeting, Oxford, UK, December 15, 1989}, Workshops in Computing,
  pages 128--141, Berlin, Heidelberg, 1989. Springer.

\bibitem[FHK19]{Fett&19}
D.~{Fett}, P.~{Hosseyni}, and R.~{Küsters}.
\newblock An extensive formal security analysis of the {OpenID} financial-grade
  {API}.
\newblock In {\em 2019 IEEE Symposium on Security and Privacy (SP)}, pages
  453--471, 445 Hoes Lane Piscataway, NJ 08854 USA, May 2019. IEEE.

\bibitem[FKS14]{Fett&14}
D.~{Fett}, R.~{Küsters}, and G.~{Schmitz}.
\newblock An expressive model for the web infrastructure: {Definition} and
  application to the {Browser} {ID} {SSO} {System}.
\newblock In {\em 2014 IEEE Symposium on Security and Privacy}, pages 673--688,
  445 Hoes Lane Piscataway, NJ 08854 USA, 2014. IEEE.

\bibitem[FL09]{Fitzgerald&09}
John Fitzgerald and Peter~Gorm Larsen.
\newblock {\em {Modelling Systems -- {Practical} Tools and Techniques in
  Software Development}}.
\newblock Cambridge University Press, The Edinburgh Building, Cambridge CB2
  2RU, UK, {Second} edition, 2009.
\newblock {ISBN 0-521-62348-0}.

\bibitem[FLR17]{Fisher&17}
Kathleen Fisher, John Launchbury, and Raymond Richards.
\newblock The {HACMS} program: {Using} formal methods to eliminate exploitable
  bugs.
\newblock {\em Philosophical Transactions of the Royal Society A: Mathematical,
  Physical and Engineering Sciences}, 375(2104):20150401, 2017.

\bibitem[FM07]{FilliatreM07}
Jean{-}Christophe Filli{\^{a}}tre and Claude March{\'{e}}.
\newblock The {Why}/{Krakatoa}/{Caduceus} platform for deductive program
  verification.
\newblock In Werner Damm and Holger Hermanns, editors, {\em Computer Aided
  Verification, 19th International Conference, {CAV} 2007, Berlin, Germany,
  July 3-7, 2007, Proceedings}, volume 4590 of {\em Lecture Notes in Computer
  Science}, pages 173--177, Berlin, Heidelberg, 2007. Springer.

\bibitem[Fre18]{Freitas&18}
Leo Freitas.
\newblock {VDM} at large: {Modelling} the {EMV}{\textregistered} $2^{nd}$
  generation kernel.
\newblock In {\em Brazilian Symposium on Formal Methods}, pages 109--125,
  Berlin, Heidelberg, 2018. Springer.

\bibitem[FWX{\etalchar{+}}17]{Ferraiuolo&17a}
Andrew Ferraiuolo, Yao Wang, Rui Xu, Danfeng Zhang, Andrew~C. Myers, and
  G.~Edward Suh.
\newblock Full-processor timing channel protection with applications to secure
  hardware compartments.
\newblock Technical report, Cornell University Library, 2017.

\bibitem[FXZ{\etalchar{+}}17]{Ferraiuolo&17}
Andrew Ferraiuolo, Rui Xu, Danfeng Zhang, Andrew~C. Myers, and G.~Edward Suh.
\newblock Verification of a practical hardware security architecture through
  static information flow analysis.
\newblock {\em SIGARCH Comput. Archit. News}, 45(1):555–568, April 2017.

\bibitem[{Gal}19]{Saw}
{Galois Inc.}
\newblock {The Software Analysis Workbench}.
\newblock \url{saw.galois.com/index.html}, 2019.
\newblock Accessed February 7 2019.

\bibitem[GBW{\etalchar{+}}17]{Gacek&17}
Andrew Gacek, John Backes, Mike Whalen, Lucas~G. Wagner, and Elaheh Ghassabani.
\newblock The {JKind} model checker, 2017.

\bibitem[GDMJ16a]{Guo&17}
X.~{Guo}, R.~G. {Dutta}, P.~{Mishra}, and Y.~{Jin}.
\newblock Automatic {RTL}-to-formal code converter for {IP} security formal
  verification.
\newblock In {\em 2016 17th International Workshop on Microprocessor and SOC
  Test and Verification (MTV)}, pages 35--38, 445 Hoes Lane Piscataway, NJ
  08854 USA, Dec 2016. IEEE.

\bibitem[GDMJ16b]{Guo&16}
X.~{Guo}, R.~G. {Dutta}, P.~{Mishra}, and Y.~{Jin}.
\newblock Scalable {SoC} trust verification using integrated theorem proving
  and model checking.
\newblock In {\em 2016 IEEE International Symposium on Hardware Oriented
  Security and Trust (HOST)}, pages 124--129, 445 Hoes Lane Piscataway, NJ
  08854 USA, May 2016. IEEE.

\bibitem[GDMJ17]{Guo&2017hw}
X.~{Guo}, R.~G. {Dutta}, P.~{Mishra}, and Y.~{Jin}.
\newblock Automatic code converter enhanced {PCH} framework for {SoC} trust
  verification.
\newblock {\em IEEE Transactions on Very Large Scale Integration (VLSI)
  Systems}, 25(12):3390--3400, Dec 2017.

\bibitem[GdR{\etalchar{+}}10]{Ghamarian&10}
A.H. Ghamarian, M.J. {de Mol}, Arend Rensink, Eduardo Zambon, and M.V.
  Zimakova.
\newblock {\em Modelling and Analysis Using {GROOVE}}.
\newblock Number TR-CTIT-10-18 in CTIT Technical Report Series. Centre for
  Telematics and Information Technology (CTIT), Netherlands, 4 2010.

\bibitem[GFLS11]{Guha&11}
Arjun Guha, Matthew Fredrikson, Benjamin Livshits, and Nikhil Swamy.
\newblock Verified security for browser extensions.
\newblock In {\em 2011 IEEE symposium on security and privacy}, pages 115--130,
  445 Hoes Lane Piscataway, NJ 08854 USA, 2011. IEEE.

\bibitem[GGS10]{Gutierrez-Garcia&10}
J.~Octavio Gutierrez-Garcia and Kwang Sim.
\newblock Agent-based service composition in cloud computing.
\newblock volume 121, pages 1--10, 01 2010.

\bibitem[GM84]{gm84}
Shafi Goldwasser and Silvio Micali.
\newblock Probabilistic encryption.
\newblock {\em Journal of computer and system sciences}, 28(2):270--299, 1984.

\bibitem[GNDB16]{Guanciale&16}
Roberto Guanciale, Hamed Nemati, Mads Dam, and Christoph Baumann.
\newblock Provably secure memory isolation for {Linux} on {ARM}: {Submission}
  to special issue on verified information flow security.
\newblock {\em Journal of Computer Security}, 24:793--837, 12 2016.

\bibitem[GR19]{FDR42019}
Thomas Gibson-Robinson.
\newblock {\em {FDR4}: {The} {CSP} Refinement Checker}.
\newblock Oxford University Department of Computer Science,
  \url{www.cs.ox.ac.uk/projects/fdr/}, 2019.

\bibitem[GSM{\etalchar{+}}11]{Gandhi&11}
R.~{Gandhi}, A.~{Sharma}, W.~{Mahoney}, W.~{Sousan}, Q.~{Zhu}, and
  P.~{Laplante}.
\newblock Dimensions of cyber-attacks: {Cultural}, social, economic, and
  political.
\newblock {\em IEEE Technology and Society Magazine}, 30(1):28--38, Spring
  2011.

\bibitem[Hal05]{Hall&05}
Anthony Hall.
\newblock Realising the benefits of formal methods.
\newblock In Kung-Kiu Lau and Richard Banach, editors, {\em Formal Methods and
  Software Engineering}, pages 1--4, Berlin, Heidelberg, 2005. Springer Berlin
  Heidelberg.

\bibitem[Ham05]{Hamlet&05}
Dick Hamlet.
\newblock Invariants and state in testing and formal methods.
\newblock {\em SIGSOFT Softw. Eng. Notes}, 31(1):48–51, September 2005.

\bibitem[HB12a]{Hartung&12}
Daniel Hartung and Christoph Busch.
\newblock Biometric transaction authentication protocol: {Formal} model
  verification and ``four-eyes'' principle extension.
\newblock In George Danezis, Sven Dietrich, and Kazue Sako, editors, {\em
  Financial Cryptography and Data Security}, pages 88--103, Berlin, Heidelberg,
  2012. Springer Berlin Heidelberg.

\bibitem[HB12b]{Hartung&10}
Daniel Hartung and Christoph Busch.
\newblock Biometric transaction authentication protocol: {Formal} model
  verification and ``four-eyes'' principle extension.
\newblock In George Danezis, Sven Dietrich, and Kazue Sako, editors, {\em
  Financial Cryptography and Data Security}, pages 88--103, Berlin, Heidelberg,
  2012. Springer Berlin Heidelberg.

\bibitem[HH18]{Hailesellasie&18}
Muluken Hailesellasie and Syed~Rafay Hasan.
\newblock Intrusion detection in {PLC}-based industrial control systems using
  formal verification approach in conjunction with graphs.
\newblock {\em Journal of Hardware and Systems Security}, 2(1):1--14, Mar 2018.

\bibitem[HK18]{Harz&18}
Dominik Harz and William Knottenbelt.
\newblock Towards safer smart contracts: {A} survey of languages and
  verification methods, 2018.

\bibitem[HLC{\etalchar{+}}13]{Hao&13}
Jianan Hao, Yang Liu, Wentong Cai, Guangdong Bai, and Jun Sun.
\newblock {vTRUST}: {A} formal modeling and verification framework for
  virtualization systems.
\newblock In Lindsay Groves and Jing Sun, editors, {\em Formal Methods and
  Software Engineering}, pages 329--346, Berlin, Heidelberg, 2013. Springer
  Berlin Heidelberg.

\bibitem[Hoa78]{Hoare&78}
C.~A.~R. Hoare.
\newblock {Communicating} {Sequential} {Processes}.
\newblock {\em Commun. ACM}, 21(8):666–677, August 1978.

\bibitem[Hoa85]{Hoare85}
C.~A.~R. Hoare.
\newblock {\em {Communicating} {Sequential} {Processes}}.
\newblock Prentice-Hall, USA, 1985.

\bibitem[HRG{\etalchar{+}}18]{Huang&18}
Bo-Yuan Huang, Sayak Ray, Aarti Gupta, Jason~M. Fung, and Sharad Malik.
\newblock Formal security verification of concurrent firmware in {SoC}s using
  instruction-level abstraction for hardware.
\newblock In {\em Proceedings of the 55th Annual Design Automation Conference},
  DAC '18, pages 91:1--91:6, New York, NY, USA, 2018. ACM.

\bibitem[HSB{\etalchar{+}}19]{Heneghan&19}
J.~{Heneghan}, S.~A. {Shaikh}, J.~{Bryans}, M.~{Cheah}, and P.~{Wooderson}.
\newblock Enabling security checking of automotive {ECU}s with formal {CSP}
  models.
\newblock In {\em 2019 49th Annual IEEE/IFIP International Conference on
  Dependable Systems and Networks Workshops (DSN-W)}, pages 90--97, 445 Hoes
  Lane Piscataway, NJ 08854 USA, June 2019. IEEE.

\bibitem[HSR{\etalchar{+}}18]{Hildenbrandt&18}
E.~{Hildenbrandt}, M.~{Saxena}, N.~{Rodrigues}, X.~{Zhu}, P.~{Daian},
  D.~{Guth}, B.~{Moore}, D.~{Park}, Y.~{Zhang}, A.~{Stefanescu}, and G.~{Rosu}.
\newblock {KEVM}: {A} complete formal semantics of the {Ethereum} virtual
  machine.
\newblock In {\em 2018 IEEE 31st Computer Security Foundations Symposium
  (CSF)}, pages 204--217, 445 Hoes Lane Piscataway, NJ 08854 USA, 2018. IEEE.

\bibitem[IK06]{Daisuke&06}
Daisuke Inoue and Masahiro Kuroda.
\newblock Secure service framework on mobile ethernet.
\newblock {\em Journal of the National Institute of Information and
  Communications Technology}, 53:61--71, 12 2006.

\bibitem[IMMS19]{Iadarola&19}
G.~{Iadarola}, F.~{Martinelli}, F.~{Mercaldo}, and A.~{Santone}.
\newblock Formal methods for {Android} banking malware analysis and detection.
\newblock In {\em 2019 Sixth International Conference on Internet of Things:
  Systems, Management and Security (IOTSMS)}, pages 331--336, 445 Hoes Lane
  Piscataway, NJ 08854 USA, 2019. IEEE.

\bibitem[ISCM17]{Ibrar&17}
Fahad Ibrar, Hamza Saleem, Sam Castle, and Muhammad~Zubair Malik.
\newblock A study of static analysis tools to detect vulnerabilities of
  branchless banking applications in developing countries.
\newblock In {\em Proceedings of the Ninth International Conference on
  Information and Communication Technologies and Development}, ICTD ’17, New
  York, NY, USA, 2017. Association for Computing Machinery.

\bibitem[ITS91]{ITSEC2006}
ITSEC.
\newblock {Information} {Technology} {Security} {Evaluation} {Criteria}
  ({ITSEC}): {Preliminary} harmonised criteria.
\newblock Document COM(90) 314, Version 1.2, Commission of the European
  Communities, June 1991.

\bibitem[Jac00]{Jackson&00}
Daniel Jackson.
\newblock Automating first-order relational logic.
\newblock {\em Proceedings of the ACM SIGSOFT Symposium on the Foundations of
  Software Engineering}, 25, 09 2000.

\bibitem[Jac12]{Jackson12}
Daniel Jackson.
\newblock {\em Software Abstractions: Logic, Language, and Analysis}.
\newblock The MIT Press, Cambridge, Massachusetts, United States, 2012.

\bibitem[JBOK14]{Jayaraman&2014}
Karthick Jayaraman, Nikolaj Bj{\o}rner, Geoff Outhred, and Charlie Kaufman.
\newblock {Automated Analysis and Debugging of Network Connectivity Policies}.
\newblock Technical report, MSR, Seattle, WA, USA, Tech. Rep. MSR-TR-2014-102,
  2014.

\bibitem[JED{\etalchar{+}}12]{Jarraya&12}
Y.~{Jarraya}, A.~{Eghtesadi}, M.~{Debbabi}, Y.~{Zhang}, and M.~{Pourzandi}.
\newblock Cloud calculus: {Security} verification in elastic cloud computing
  platform.
\newblock In {\em 2012 International Conference on Collaboration Technologies
  and Systems (CTS)}, pages 447--454, 445 Hoes Lane Piscataway, NJ 08854 USA,
  May 2012. IEEE.

\bibitem[JK09]{Jensen&09}
Kurt Jensen and Lars~M. Kristensen.
\newblock {\em Coloured {Petri} Nets: {Modelling} and Validation of Concurrent
  Systems}.
\newblock Springer Publishing Company, Incorporated, Berlin, Heidelberg, 1st
  edition, 2009.

\bibitem[JKW07]{Jensen&07}
Kurt Jensen, {Lars Michael} Kristensen, and {Lisa Marie} Wells.
\newblock Coloured {Petri} nets and {CPN} tools for modelling and validation of
  concurrent systems.
\newblock {\em International Journal on Software Tools for Technology
  Transfer}, 9(3/4):213--254, 2007.

\bibitem[JRVP08]{Rancz&08}
Katalin~Tünde Jánosi-Rancz, Viorica Varga, and Janos Puskas.
\newblock A software tool for data analysis based on formal concept analysis.
\newblock {\em Studia Univ. Babes-Bolyai, Informatica}, 53(2):67--78, 01 2008.

\bibitem[JTL12a]{Jang&12}
Dongseok Jang, Zachary Tatlock, and Sorin Lerner.
\newblock Establishing browser security guarantees through formal shim
  verification.
\newblock In {\em Proceedings of the 21st USENIX Conference on Security
  Symposium}, Security’12, page~8, USA, 2012. USENIX Association.

\bibitem[JTL12b]{Dongseok&12}
Dongseok Jang, Zachary Tatlock, and Sorin Lerner.
\newblock Establishing browser security guarantees through formal shim
  verification.
\newblock In {\em Presented as part of the 21st {USENIX} Security Symposium
  ({USENIX} Security 12)}, pages 113--128, Bellevue, WA, 2012. {USENIX}.

\bibitem[J{\"u}l02]{Jullig&02}
Richard J{\"u}llig.
\newblock Formal methods in enterprise computing.
\newblock In Chris George and Huaikou Miao, editors, {\em Formal Methods and
  Software Engineering}, pages 22--23, Berlin, Heidelberg, 2002. Springer
  Berlin Heidelberg.

\bibitem[JW96]{Jackson&96c}
Daniel Jackson and Jeanette Wing.
\newblock Lightweight formal methods.
\newblock {\em IEEE Computer}, 29(4):22--23, April 1996.

\bibitem[KAE{\etalchar{+}}10]{KleinAEHCDEEKNSTW2010}
Gerwin Klein, June Andronick, Kevin Elphinstone, Gernot Heiser, David Cock,
  Philip Derrin, Dhammika Elkaduwe, Kai Engelhardt, Rafal Kolanski, Michael
  Norrish, Thomas Sewell, Harvey Tuch, and Simon Winwood.
\newblock {seL4}: {Formal} verification of an operating-system kernel.
\newblock {\em Commun. {ACM}}, 53(6):107--115, 2010.

\bibitem[KB10]{Kramer&10}
Simon Kramer and Julian~C Bradfield.
\newblock A general definition of malware.
\newblock {\em Journal in computer virology}, 6(2):105--114, 2010.

\bibitem[KBG{\etalchar{+}}17]{Kumar&17}
P.~{Kumar}, A.~{Braeken}, A.~{Gurtov}, J.~{Iinatti}, and P.~H. {Ha}.
\newblock Anonymous secure framework in connected smart home environments.
\newblock {\em IEEE Transactions on Information Forensics and Security},
  12(4):968--979, April 2017.

\bibitem[KCKB14]{Kallenberg&14}
Corey Kallenberg, Sam Cornwell, Xeno Kovah, and John Butterworth.
\newblock Setup for failure: {Defeating} secure boot.
\newblock {\em The MITRE Corporation}, 2014.

\bibitem[Ken15]{Kenney&15}
Michael Kenney.
\newblock Cyber-terrorism in a post-{Stuxnet} world.
\newblock {\em Orbis}, 59, 12 2015.

\bibitem[KHF{\etalchar{+}}19]{Kocher&19}
P.~{Kocher}, J.~{Horn}, A.~{Fogh}, D.~{Genkin}, D.~{Gruss}, W.~{Haas},
  M.~{Hamburg}, M.~{Lipp}, S.~{Mangard}, T.~{Prescher}, M.~{Schwarz}, and
  Y.~{Yarom}.
\newblock Spectre attacks: {Exploiting} speculative execution.
\newblock In {\em 2019 IEEE Symposium on Security and Privacy (SP)}, pages
  1--19, 445 Hoes Lane Piscataway, NJ 08854 USA, 2019. IEEE.

\bibitem[KHLF10]{Khurana&10}
H.~{Khurana}, M.~{Hadley}, N.~{Lu}, and D.~A. {Frincke}.
\newblock Smart-grid security issues.
\newblock {\em IEEE Security Privacy}, 8(1):81--85, Jan 2010.

\bibitem[KI16]{Kotipalli&16}
Srinivasa~Rao Kotipalli and Mohammed~A Imran.
\newblock {\em Hacking {Android}}.
\newblock Packt Publishing Ltd, UK, 2016.

\bibitem[KKG19]{Kumar&19}
N.~{Kumar}, V.~{Kumar}, and M.~{Gaur}.
\newblock Banking trojans {APK} detection using formal methods.
\newblock In {\em 2019 4th International Conference on Information Systems and
  Computer Networks (ISCON)}, pages 606--609, 445 Hoes Lane Piscataway, NJ
  08854 USA, 2019. IEEE.

\bibitem[KKHE17]{Kottler&17}
S.~{Kottler}, M.~{Khayamy}, S.~R. {Hasan}, and O.~{Elkeelany}.
\newblock Formal verification of ladder logic programs using {NuSMV}.
\newblock In {\em SoutheastCon 2017}, pages 1--5, 445 Hoes Lane Piscataway, NJ
  08854 USA, March 2017. IEEE.

\bibitem[KLSB11]{Kupferschmid&11}
Stefan Kupferschmid, Matthew Lewis, Tobias Schubert, and Bernd Becker.
\newblock Incremental preprocessing methods for use in {BMC}.
\newblock {\em Formal Methods in System Design}, 39(2):185--204, 2011.

\bibitem[KNP11]{Prism}
M.~Kwiatkowska, G.~Norman, and D.~Parker.
\newblock {PRISM} 4.0: {Verification} of probabilistic real-time systems.
\newblock In G.~Gopalakrishnan and S.~Qadeer, editors, {\em Proc. 23rd
  International Conference on Computer Aided Verification (CAV'11)}, volume
  6806 of {\em LNCS}, pages 585--591. Springer, 2011.

\bibitem[Koz83]{Kozen&83}
Dexter Kozen.
\newblock Results on the propositional $\mu$-calculus.
\newblock {\em Theoretical Computer Science}, 27(3):333 -- 354, 1983.
\newblock Special Issue Ninth International Colloquium on Automata, Languages
  and Programming (ICALP) Aarhus, Summer 1982.

\bibitem[KPCBH15]{Kriaa&15}
Siwar Kriaa, Ludovic Pietre-Cambacedes, Marc Bouissou, and Yoran Halgand.
\newblock A survey of approaches combining safety and security for industrial
  control systems.
\newblock {\em Reliability Engineering \& System Safety}, 139:156 -- 178, 2015.

\bibitem[KS13]{Kunemann&13}
Robert K{\"u}nnemann and Graham Steel.
\newblock {YubiSecure}? {Formal} security analysis results for the {Yubikey}
  and {YubiHSM}.
\newblock In Audun J{\o}sang, Pierangela Samarati, and Marinella Petrocchi,
  editors, {\em Security and Trust Management}, pages 257--272, Berlin,
  Heidelberg, 2013. Springer Berlin Heidelberg.

\bibitem[KS15]{Karla&15}
Sheetal Kalra and Sandeep~K. Sood.
\newblock Secure authentication scheme for {IoT} and cloud servers.
\newblock {\em Pervasive and Mobile Computing}, 24:210 -- 223, 2015.
\newblock Special Issue on Secure Ubiquitous Computing.

\bibitem[KsM02]{Khu-smith&02}
Vorapranee Khu-smith and Chris~J. Mitchell.
\newblock Using {EMV} cards to protect e-commerce transactions.
\newblock In {\em Proceedings of the Third International Conference on
  E-Commerce and Web Technologies}, EC-WEB ’02, page 388–399, Berlin,
  Heidelberg, 2002. Springer-Verlag.

\bibitem[KTJB19]{Kulik&19}
Tomas Kulik, Peter W.~V. Tran-J{\o}rgensen, and Jalil Boudjadar.
\newblock Formal security analysis of cloud-connected industrial control
  systems.
\newblock In Jean-Louis Lanet and Cristian Toma, editors, {\em Innovative
  Security Solutions for Information Technology and Communications}, pages
  71--84, Cham, 2019. Springer International Publishing.

\bibitem[Kum14]{Kumar&14}
Apurva Kumar.
\newblock A lightweight formal approach for analyzing security of web
  protocols.
\newblock In {\em International Workshop on Recent Advances in Intrusion
  Detection}, pages 192--211. Springer, 2014.

\bibitem[Lam02]{TLA}
Leslie Lamport.
\newblock {\em {Specifying Systems: {The} {TLA+} Language and Tools for
  Hardware and Software Engineers}}.
\newblock Addison-Wesley Longman Publishing Co., Inc., Boston, MA, USA, 2002.

\bibitem[Lam09]{PlusCal}
Leslie Lamport.
\newblock {The {PlusCal} Algorithm Language}.
\newblock In Martin Leucker and Carroll Morgan, editors, {\em Theoretical
  Aspects of Computing - ICTAC 2009}, pages 36--60, Berlin, Heidelberg, 2009.
  Springer Berlin Heidelberg.

\bibitem[LBF{\etalchar{+}}10]{Larsen&10}
Peter~Gorm Larsen, Nick Battle, Miguel Ferreira, John Fitzgerald, Kenneth
  Lausdahl, and Marcel Verhoef.
\newblock The {Overture} initiative integrating tools for {VDM}.
\newblock {\em ACM SIGSOFT Software Engineering Notes}, 35(1):1--6, 2010.

\bibitem[LCH{\etalchar{+}}16]{Letan&16}
Thomas Letan, Pierre Chifflier, Guillaume Hiet, Pierre Neron, and Benjamin
  Morin.
\newblock {SpecCert}: {Specifying} and verifying hardware-based security
  enforcement.
\newblock volume 9995, pages 496--512, 11 2016.

\bibitem[LFBP14]{Lerner&14}
Lee~W. Lerner, Zane~R. Franklin, William~T. Baumann, and Cameron~D. Patterson.
\newblock Using high-level synthesis and formal analysis to predict and preempt
  attacks on industrial control systems.
\newblock In {\em Proceedings of the 2014 ACM/SIGDA International Symposium on
  Field-Programmable Gate Arrays}, FPGA ’14, page 209–212, New York, NY,
  USA, 2014. Association for Computing Machinery.

\bibitem[LHB{\etalchar{+}}96]{ISOVDM96}
P.~G. Larsen, B.~S. Hansen, H.~Brunn, N.~Plat, H.~Toetenel, D.~J. Andrews,
  J.~Dawes, G.~Parkin, et~al.
\newblock {{Information} technology -- {Programming} languages, their
  environments and system software interfaces -- {Vienna} {Development}
  {Method} -- {Specification} Language -- {Part} 1: {Base} language}, December
  1996.

\bibitem[Lin15]{Lin&15}
Yu-Cheng Lin.
\newblock {Androbugs} framework: {An} android application security
  vulnerability scanner, 2015.

\bibitem[LJM11]{Love&11}
E.~{Love}, Y.~{Jin}, and Y.~{Makris}.
\newblock Enhancing security via provably trustworthy hardware intellectual
  property.
\newblock In {\em 2011 IEEE International Symposium on Hardware-Oriented
  Security and Trust}, pages 12--17, 445 Hoes Lane Piscataway, NJ 08854 USA,
  June 2011. IEEE.

\bibitem[LLB10]{Larsen&10c}
Peter~Gorm Larsen, Kenneth Lausdahl, and Nick Battle.
\newblock {Combinatorial Testing for {VDM}}.
\newblock In {\em Proceedings of the 2010 8th IEEE International Conference on
  Software Engineering and Formal Methods}, SEFM '10, pages 278--285,
  Washington, DC, USA, September 2010. IEEE Computer Society.
\newblock {ISBN 978-0-7695-4153-2}.

\bibitem[Low95]{Lowe1995}
Gavin Lowe.
\newblock An attack on the {Needham}-{Schroeder} public-key authentication
  protocol.
\newblock {\em Inf. Process. Lett.}, 56(3):131--133, 1995.

\bibitem[LPY97]{Larsen&97}
Kim~G. Larsen, Paul Pettersson, and Wang Yi.
\newblock {Uppaal} in a nutshell.
\newblock {\em International Journal on Software Tools for Technology
  Transfer}, 1(1):134--152, Dec 1997.

\bibitem[LQL12]{LalQL12}
Akash Lal, Shaz Qadeer, and Shuvendu~K. Lahiri.
\newblock A solver for reachability modulo theories.
\newblock In P.~Madhusudan and Sanjit~A. Seshia, editors, {\em Computer Aided
  Verification - 24th International Conference, {CAV} 2012, Berkeley, CA, USA,
  July 7-13, 2012 Proceedings}, volume 7358 of {\em Lecture Notes in Computer
  Science}, pages 427--443, Berlin, Heidelberg, 2012. Springer.

\bibitem[LS99]{Lu&99}
S.~{Lu} and S.~A. {Smolka}.
\newblock Model checking the secure electronic transaction ({SET}) protocol.
\newblock In {\em MASCOTS '99. Proceedings of the Seventh International
  Symposium on Modeling, Analysis and Simulation of Computer and
  Telecommunication Systems}, pages 358--364, 445 Hoes Lane Piscataway, NJ
  08854 USA, 1999. IEEE.

\bibitem[MAH15]{Mayo&15}
Jackson~R Mayo, Robert~C Armstrong, and Geoffrey~C Hulette.
\newblock Digital system robustness via design constraints: {The} lesson of
  formal methods.
\newblock In {\em 2015 Annual IEEE Systems Conference (SysCon) Proceedings},
  pages 109--114, 445 Hoes Lane Piscataway, NJ 08854 USA, 2015. IEEE.

\bibitem[MAHvM16]{Mehrnezhad&16}
Maryam Mehrnezhad, Mohammed~Aamir Ali, Feng Hao, and Aad van Moorsel.
\newblock {NFC} payment spy: {A} privacy attack on contactless payments.
\newblock In Lidong Chen, David McGrew, and Chris Mitchell, editors, {\em
  Security Standardisation Research}, pages 92--111, Cham, 2016. Springer
  International Publishing.

\bibitem[MCJ18]{Miller&18}
Andrew Miller, Zhicheng Cai, and Somesh Jha.
\newblock Smart contracts and opportunities for formal methods.
\newblock In Tiziana Margaria and Bernhard Steffen, editors, {\em Leveraging
  Applications of Formal Methods, Verification and Validation. Industrial
  Practice}, pages 280--299, Cham, 2018. Springer International Publishing.

\bibitem[Mes00]{Meseguer00}
Jos{\'e} Meseguer.
\newblock Rewriting logic and {Maude}: {A} wide-spectrum semantic framework for
  object-based distributed systems.
\newblock In Scott~F. Smith and Carolyn~L. Talcott, editors, {\em Formal
  Methods for Open Object-Based Distributed Systems IV}, pages 89--117, Boston,
  MA, 2000. Springer US.

\bibitem[MGP16]{Madhoun&16}
N.~E. {Madhoun}, F.~{Guenane}, and G.~{Pujolle}.
\newblock An online security protocol for {NFC} payment: {Formally} analyzed by
  the {Scyther} tool.
\newblock In {\em 2016 Second International Conference on Mobile and Secure
  Services (MobiSecServ)}, pages 1--7, 445 Hoes Lane Piscataway, NJ 08854 USA,
  2016. IEEE.

\bibitem[Mil89]{Milner&89}
R.~Milner.
\newblock {\em Communication and Concurrency}.
\newblock Prentice-Hall, Inc., USA, 1989.

\bibitem[MJA{\etalchar{+}}18]{Madi&18}
Taous Madi, Yosr Jarraya, Amir Alimohammadifar, Suryadipta Majumdar, Yushun
  Wang, Makan Pourzandi, Lingyu Wang, and Mourad Debbabi.
\newblock {ISOTOP}: {Auditing} virtual networks isolation across cloud layers
  in {OpenStack}.
\newblock {\em ACM Trans. Priv. Secur.}, 22(1), October 2018.

\bibitem[MLHK14]{Masmoudi&14}
Fatma Masmoudi, Monia Loulou, and Ahmed Hadj~Kacem.
\newblock Formal security framework for agent based cloud systems.
\newblock 11 2014.

\bibitem[MML12]{Moreno&12}
R.~{Moreno-Vozmediano}, R.~S. {Montero}, and I.~M. {Llorente}.
\newblock {IaaS} cloud architecture: {From} virtualized datacenters to
  federated cloud infrastructures.
\newblock {\em Computer}, 45(12):65--72, 2012.

\bibitem[MMN18]{MartinelliMN18}
Fabio Martinelli, Francesco Mercaldo, and Vittoria Nardone.
\newblock Identifying insecure features in android applications using model
  checking.
\newblock In Paolo Mori, Steven Furnell, and Olivier Camp, editors, {\em
  Proceedings of the 4th International Conference on Information Systems
  Security and Privacy, {ICISSP} 2018, Funchal, Madeira - Portugal, January
  22-24, 2018}, pages 589--596. SciTePress, 2018.

\bibitem[MMS19]{Mercaldo&19}
F.~{Mercaldo}, F.~{Martinelli}, and A.~{Santone}.
\newblock Real-time {SCADA} attack detection by means of formal methods.
\newblock In {\em 2019 IEEE 28th International Conference on Enabling
  Technologies: Infrastructure for Collaborative Enterprises (WETICE)}, pages
  231--236, 445 Hoes Lane Piscataway, NJ 08854 USA, June 2019. IEEE.

\bibitem[MNSV16]{Mercaldo&16}
Francesco Mercaldo, Vittoria Nardone, Antonella Santone, and Corrado~Aaron
  Visaggio.
\newblock Download malware? {No}, thanks: {How} formal methods can block update
  attacks.
\newblock In {\em Proceedings of the 4th FME Workshop on Formal Methods in
  Software Engineering}, FormaliSE '16, pages 22--28, New York, NY, USA, 2016.
  ACM.

\bibitem[MO15]{Macedo&15}
Hugo~Daniel Macedo and Jos{\'e}~Nuno Oliveira.
\newblock A linear algebra approach to {OLAP}.
\newblock {\em Formal Aspects of Computing}, 27(2):283--307, 2015.

\bibitem[MPS19]{Marcedone&19}
Antonio Marcedone, Rafael Pass, and Abhi Shelat.
\newblock Minimizing trust in hardware wallets with two factor signatures.
\newblock In Ian Goldberg and Tyler Moore, editors, {\em Financial Cryptography
  and Data Security}, pages 407--425, Cham, 2019. Springer International
  Publishing.

\bibitem[MPX{\etalchar{+}}13]{Mai&13}
Haohui Mai, Edgar Pek, Hui Xue, Samuel~Talmadge King, and Parthasarathy
  Madhusudan.
\newblock Verifying security invariants in {ExpressOS}.
\newblock In {\em Proceedings of the Eighteenth International Conference on
  Architectural Support for Programming Languages and Operating Systems},
  ASPLOS ’13, page 293–304, New York, NY, USA, 2013. Association for
  Computing Machinery.

\bibitem[MSCB13]{Sharygina&13}
Simon Meier, Benedikt Schmidt, Cas Cremers, and David Basin.
\newblock The {TAMARIN} prover for the symbolic analysis of security protocols.
\newblock In Natasha Sharygina and Helmut Veith, editors, {\em Computer Aided
  Verification: 25th International Conference, CAV 2013, Saint Petersburg,
  Russia, July 13-19, 2013, Proceedings}, volume 8044 of {\em Lecture Notes in
  Computer Science}, pages 696 -- 701, Berlin, 2013. Springer.
\newblock Computer Aided Verification: 25th International Conference (CAV
  2013); Conference Location: Saint Petersburg, Russia; Conference Date: July
  13-19, 2013; .

\bibitem[MSHA17]{Moshin&17}
M.~{Mohsin}, M.~U. {Sardar}, O.~{Hasan}, and Z.~{Anwar}.
\newblock {IoTRiskAnalyzer}: {A} probabilistic model checking based framework
  for formal risk analytics of the internet of things.
\newblock {\em IEEE Access}, 5:5494--5505, 2017.

\bibitem[MT13]{Macedo&13}
Hugo~Daniel Macedo and Tayssir Touili.
\newblock Mining malware specifications through static reachability analysis.
\newblock In {\em European Symposium on Research in Computer Security}, pages
  517--535, Berlin, Heidelberg, 2013. Springer, Springer Berlin Heidelberg.

\bibitem[MTT{\etalchar{+}}12]{Morrisett&12}
Greg Morrisett, Gang Tan, Joseph Tassarotti, Jean-Baptiste Tristan, and Edward
  Gan.
\newblock {RockSalt}: {Better}, faster, stronger {SFI} for the {X86}.
\newblock In {\em Proceedings of the 33rd ACM SIGPLAN Conference on Programming
  Language Design and Implementation}, PLDI ’12, page 395–404, New York,
  NY, USA, 2012. Association for Computing Machinery.

\bibitem[MW08]{Muhlbach&08}
Sascha M{\"u}hlbach and Sebastian Wallner.
\newblock Secure communication in microcomputer bus systems for embedded
  devices.
\newblock {\em Journal of Systems Architecture}, 54(11):1065 -- 1076, 2008.
\newblock Embedded Systems: Architectures, Modeling and Simulation.

\bibitem[Nec11]{Necula11}
George~C. Necula.
\newblock Proof-carrying code.
\newblock In Henk C.~A. van Tilborg and Sushil Jajodia, editors, {\em
  Encyclopedia of Cryptography and Security, 2nd Ed}, pages 984--986. Springer,
  Berlin, Heidelberg, 2011.

\bibitem[NFV15]{Nunes&15}
P.~J.~C. {Nunes}, J.~{Fonseca}, and M.~{Vieira}.
\newblock {phpSAFE}: {A} security analysis tool for {OOP} web application
  plugins.
\newblock In {\em 2015 45th Annual IEEE/IFIP International Conference on
  Dependable Systems and Networks}, pages 299--306, 445 Hoes Lane Piscataway,
  NJ 08854 USA, June 2015. IEEE.

\bibitem[NGB{\etalchar{+}}16]{Nardone&16b}
Roberto Nardone, Ugo Gentile, Massimo Benerecetti, Adriano Peron, Valeria
  Vittorini, Stefano Marrone, and Nicola Mazzocca.
\newblock Modeling railway control systems in {Promela}.
\newblock In Cyrille Artho and Peter~Csaba {\"O}lveczky, editors, {\em Formal
  Techniques for Safety-Critical Systems}, pages 121--136, Cham, 2016. Springer
  International Publishing.

\bibitem[NGP{\etalchar{+}}15]{Nardone&15}
Roberto Nardone, Ugo Gentile, Adriano Peron, Massimo Benerecetti, Valeria
  Vittorini, Stefano Marrone, Renato De~Guglielmo, Nicola Mazzocca, and Luigi
  Velardi.
\newblock Dynamic state machines for formalizing railway control system
  specifications.
\newblock In Cyrille Artho and Peter~Csaba {\"O}lveczky, editors, {\em Formal
  Techniques for Safety-Critical Systems}, pages 93--109, Cham, 2015. Springer
  International Publishing.

\bibitem[NK14]{NipkowK2014}
Tobias Nipkow and Gerwin Klein.
\newblock {\em Concrete Semantics --- With {Isabelle}/{HOL}}.
\newblock Springer, Berlin, 2014.

\bibitem[NMB{\etalchar{+}}16]{Ngabonziza&16}
B.~{Ngabonziza}, D.~{Martin}, A.~{Bailey}, H.~{Cho}, and S.~{Martin}.
\newblock Trustzone explained: {Architectural} features and use cases.
\newblock In {\em 2016 IEEE 2nd International Conference on Collaboration and
  Internet Computing (CIC)}, pages 445--451, 445 Hoes Lane Piscataway, NJ 08854
  USA, Nov 2016. IEEE.

\bibitem[NPYG09]{Pincetic&09}
Matias Negrete-Pincetic, Felipe Yoshida, and George Gross.
\newblock Towards quantifying the impacts of cyber attacks in the competitive
  electricity market environment.
\newblock In {\em 2009 IEEE Bucharest PowerTech}, pages 1--8, 445 Hoes Lane
  Piscataway, NJ 08854 USA, 08 2009. IEEE.

\bibitem[NRM16]{Nardone&16}
R.~{Nardone}, R.~J. {Rodríguez}, and S.~{Marrone}.
\newblock Formal security assessment of {Modbus} protocol.
\newblock In {\em 2016 11th International Conference for Internet Technology
  and Secured Transactions (ICITST)}, pages 142--147, 445 Hoes Lane Piscataway,
  NJ 08854 USA, Dec 2016. IEEE.

\bibitem[NS78]{NeedhamS1978}
Roger~M. Needham and Michael~D. Schroeder.
\newblock Using encryption for authentication in large networks of computers.
\newblock {\em Commun. {ACM}}, 21(12):993--999, 1978.

\bibitem[NT19]{Nigam&19}
V.~{Nigam} and C.~{Talcott}.
\newblock Formal security verification of {Industry} 4.0 applications.
\newblock In {\em 2019 24th IEEE International Conference on Emerging
  Technologies and Factory Automation (ETFA)}, pages 1043--1050, Zaragoza,
  Spain, Sep. 2019. IEEE.

\bibitem[OH18]{OHearn&18}
Peter~W. O~Hearn.
\newblock Continuous reasoning: {Scaling} the impact of formal methods.
\newblock In {\em Proceedings of the 33rd Annual ACM/IEEE Symposium on Logic in
  Computer Science}, LICS ’18, page 13–25, New York, NY, USA, 2018.
  Association for Computing Machinery.

\bibitem[OKQ17]{Obaid2017}
Iqra Obaid, Syed Kazmi, and Awais Qasim.
\newblock Modeling and verification of payment system in {E}-banking.
\newblock {\em International Journal of Advanced Computer Science and
  Applications}, 8, 01 2017.

\bibitem[PPL16]{Puys&16}
Maxime Puys, Marie-Laure Potet, and Pascal Lafourcade.
\newblock Formal analysis of security properties on the {OPC}-{UA} {SCADA}
  protocol.
\newblock In Amund Skavhaug, J{\'e}r{\'e}mie Guiochet, and Friedemann Bitsch,
  editors, {\em Computer Safety, Reliability, and Security}, pages 67--75,
  Cham, 2016. Springer International Publishing.

\bibitem[Pri06]{Priss&06}
Uta Priss.
\newblock Formal concept analysis in information science.
\newblock {\em Annual Review of Information Science and Technology}, 40, 01
  2006.

\bibitem[Put94]{Puterman&94}
Martin~L. Puterman.
\newblock {\em {Markov} Decision Processes: {Discrete} Stochastic Dynamic
  Programming}.
\newblock John Wiley \& Sons, Inc., USA, 1st edition, 1994.

\bibitem[PZS{\etalchar{+}}18a]{Park&18}
Daejun Park, Yi~Zhang, Manasvi Saxena, Philip Daian, and Grigore Ro\c{s}u.
\newblock A formal verification tool for {Ethereum} {VM} bytecode.
\newblock In {\em Proceedings of the 2018 26th ACM Joint Meeting on European
  Software Engineering Conference and Symposium on the Foundations of Software
  Engineering}, ESEC/FSE 2018, page 912–915, New York, NY, USA, 2018.
  Association for Computing Machinery.

\bibitem[PZS{\etalchar{+}}18b]{Daejun&18}
Daejun Park, Yi~Zhang, Manasvi Saxena, Philip Daian, and Grigore Roundefinedu.
\newblock A formal verification tool for {Ethereum} {VM} bytecode.
\newblock In {\em Proceedings of the 2018 26th ACM Joint Meeting on European
  Software Engineering Conference and Symposium on the Foundations of Software
  Engineering}, ESEC/FSE 2018, page 912–915, New York, NY, USA, 2018.
  Association for Computing Machinery.

\bibitem[QCG{\etalchar{+}}09]{ROS}
Morgan Quigley, Ken Conley, Brian~P. Gerkey, Josh Faust, Tully Foote, Jeremy
  Leibs, Rob Wheeler, and Andrew~Y. Ng.
\newblock {ROS}: {An} open-source robot operating system, 2009.

\bibitem[QPP{\etalchar{+}}17]{Quarta&17}
Davide Quarta, Marcello Pogliani, Mario Polino, Federico Maggi, Andrea~Maria
  Zanchettin, and Stefano Zanero.
\newblock An experimental security analysis of an industrial robot controller.
\newblock In {\em 2017 IEEE Symposium on Security and Privacy (SP)}, pages
  268--286, 445 Hoes Lane Piscataway, NJ 08854 USA, 05 2017. IEEE.

\bibitem[RCD{\etalchar{+}}17]{Roy&17}
S.~{Roy}, S.~{Chatterjee}, A.~K. {Das}, S.~{Chattopadhyay}, N.~{Kumar}, and
  A.~V. {Vasilakos}.
\newblock On the design of provably secure lightweight remote user
  authentication scheme for mobile cloud computing services.
\newblock {\em IEEE Access}, 5:25808--25825, 2017.

\bibitem[RCD{\etalchar{+}}19]{cybok}
Awais Rashid, Howard Chivers, George Danezis, Emil Lupu, and Andrew Martin,
  editors.
\newblock {\em The Cyber Security Body of Knowledge}.
\newblock \url{www.cybok.org}, 2019.

\bibitem[RE14]{RakamaricE14}
Zvonimir Rakamaric and Michael Emmi.
\newblock {SMACK:} {Decoupling} source language details from verifier
  implementations.
\newblock In Armin Biere and Roderick Bloem, editors, {\em Computer Aided
  Verification - 26th International Conference, {CAV} 2014, Held as Part of the
  Vienna Summer of Logic, {VSL} 2014, Vienna, Austria, July 18-22, 2014.
  Proceedings}, volume 8559 of {\em Lecture Notes in Computer Science}, pages
  106--113, Berlin, Heidelberg, 2014. Springer.

\bibitem[RLS13]{Roland&13}
M.~{Roland}, J.~{Langer}, and J.~{Scharinger}.
\newblock Applying relay attacks to {Google} {Wallet}.
\newblock In {\em 2013 5th International Workshop on Near Field Communication
  (NFC)}, pages 1--6, 445 Hoes Lane Piscataway, NJ 08854 USA, 2013. IEEE.

\bibitem[RRS13]{Rysavy&13}
O.~{Rysavy}, J.~{Rab}, and M.~{Sveda}.
\newblock Improving security in {SCADA} systems through firewall policy
  analysis.
\newblock In {\em 2013 Federated Conference on Computer Science and Information
  Systems}, pages 1435--1440, 445 Hoes Lane Piscataway, NJ 08854 USA, Sep.
  2013. IEEE.

\bibitem[RSB{\etalchar{+}}15]{Rana&15}
R.~{Rana}, M.~{Staron}, C.~{Berger}, A.~{Nilsson}, R.~{Scandariato},
  A.~{Weilenmann}, and M.~{Rydmark}.
\newblock On the role of cross-disciplinary research and {SSE} in addressing
  the challenges of the digitalization of society.
\newblock In {\em 2015 6th IEEE International Conference on Software
  Engineering and Service Science (ICSESS)}, pages 1106--1109, Beijing, China,
  September 2015. IEEE.

\bibitem[RSG{\etalchar{+}}01]{MASP}
Peter Y.~A. Ryan, Steve Schneider, Michael Goldsmith, Gavin Lowe, and Bill
  Roscoe.
\newblock {\em Modelling and analysis of security protocols}.
\newblock Addison-Wesley-Longman, USA, 2001.

\bibitem[RT16]{Rochetto&16}
Marco Rocchetto and Nils~Ole Tippenhauer.
\newblock {CPDY:} extending the dolev-yao attacker with physical-layer
  interactions, 2016.

\bibitem[RT17]{Rocchetto&17}
Marco Rocchetto and Nils~Ole Tippenhauer.
\newblock Towards formal security analysis of industrial control systems.
\newblock In {\em Proceedings of the 2017 ACM on Asia Conference on Computer
  and Communications Security}, ASIA CCS '17, pages 114--126, New York, NY,
  USA, 2017. ACM.

\bibitem[RZR{\etalchar{+}}13]{Rieke&13}
R.~{Rieke}, M.~{Zhdanova}, J.~{Repp}, R.~{Giot}, and C.~{Gaber}.
\newblock Fraud detection in mobile payments utilizing process behavior
  analysis.
\newblock In {\em 2013 International Conference on Availability, Reliability
  and Security}, pages 662--669, 445 Hoes Lane Piscataway, NJ 08854 USA, 2013.
  IEEE.

\bibitem[SAE{\etalchar{+}}18]{Sadowski&18}
Caitlin Sadowski, Edward Aftandilian, Alex Eagle, Liam Miller-Cushon, and Ciera
  Jaspan.
\newblock Lessons from building static analysis tools at {Google}.
\newblock {\em Commun. ACM}, 61(4):58–66, March 2018.

\bibitem[SAS{\etalchar{+}}18]{Sepulveda&18}
J.~{Sepulveda}, D.~{Aboul-Hassan}, G.~{Sigl}, B.~{Becker}, and M.~{Sauer}.
\newblock Towards the formal verification of security properties of a
  network-on-chip router.
\newblock In {\em 2018 IEEE 23rd European Test Symposium (ETS)}, pages 1--6,
  445 Hoes Lane Piscataway, NJ 08854 USA, May 2018. IEEE.

\bibitem[SBL18]{Souaf&18}
S.~{Souaf}, P.~{Berthome}, and F.~{Loulergue}.
\newblock A cloud brokerage solution: {Formal} methods meet security in cloud
  federations.
\newblock In {\em 2018 International Conference on High Performance Computing
  Simulation (HPCS)}, pages 691--699, 2018.

\bibitem[SC16]{Smith&16}
Eric Smith and Alessandro Coglio.
\newblock {Android} platform modeling and {Android} app verification in the
  {ACL2} theorem prover.
\newblock In Arie Gurfinkel and Sanjit~A. Seshia, editors, {\em Verified
  Software: Theories, Tools, and Experiments}, pages 183--201, Cham, 2016.
  Springer International Publishing.

\bibitem[SCK{\etalchar{+}}12]{Signoles&12}
Julien Signoles, Pascal Cuoq, Florent Kirchner, Nikolai Kosmatov, Virgile
  Prevosto, and Boris Yakobowski.
\newblock {Frama-C}: {A} software analysis perspective.
\newblock {\em Formal Aspects of Computing}, 27, 10 2012.

\bibitem[SCW00]{StepneyCW2000}
Susan Stepney, David Cooper, and Jim Woodcock.
\newblock An electronic purse: {Specification}, refinement, and proof.
\newblock Technical Monograph PRG-126, Oxford University Computing Laboratory,
  July 2000.

\bibitem[SGR09]{Santos&09}
N.~Santos, Krishna~P. Gummadi, and Rodrigo Rodrigues.
\newblock Towards trusted cloud computing.
\newblock In {\em HotCloud’09: Proceedings of the 2009 Conference on Hot
  Topics in Cloud Computing}, number~9, page~3, USA, 01 2009. USENIX
  Association.

\bibitem[SIR13]{Santone&13}
Antonella Santone, Valentina Intilangelo, and Domenico Raucci.
\newblock Efficient formal verification in banking processes.
\newblock In {\em 2013 IEEE Ninth World Congress on Services}, pages 325--332,
  445 Hoes Lane Piscataway, NJ 08854 USA, 2013. IEEE.

\bibitem[SJL{\etalchar{+}}15]{Seol&2015}
Jinho Seol, Seongwook Jin, Daewoo Lee, Jaehyuk Huh, and Seungryoul Maeng.
\newblock A trusted iaas environment with hardware security module.
\newblock {\em IEEE Transactions on Services Computing}, 9:1--1, 01 2015.

\bibitem[SKMT12]{Sasse&12}
Ralf Sasse, Samuel~T King, Jos{\'e} Meseguer, and Shuo Tang.
\newblock {IBOS}: {A} correct-by-construction modular browser.
\newblock In {\em International Workshop on Formal Aspects of Component
  Software}, pages 224--241, Berlin, Heidelberg, 2012. Springer.

\bibitem[SLBK14]{Skowyra&14}
R.~{Skowyra}, A.~{Lapets}, A.~{Bestavros}, and A.~{Kfoury}.
\newblock A verification platform for {SDN}-enabled applications.
\newblock In {\em 2014 IEEE International Conference on Cloud Engineering},
  pages 337--342, 445 Hoes Lane Piscataway, NJ 08854 USA, March 2014. IEEE.

\bibitem[SLD08]{Sun&08}
Jun Sun, Yang Liu, and Jin~Song Dong.
\newblock Model checking {CSP} revisited: {Introducing} a process analysis
  toolkit.
\newblock In Tiziana Margaria and Bernhard Steffen, editors, {\em Leveraging
  Applications of Formal Methods, Verification and Validation}, pages 307--322,
  Berlin, Heidelberg, 2008. Springer Berlin Heidelberg.

\bibitem[SLDC09]{Sun&09}
J.~{Sun}, Y.~{Liu}, J.~S. {Dong}, and C.~{Chen}.
\newblock Integrating specification and programs for system modeling and
  verification.
\newblock In {\em 2009 Third IEEE International Symposium on Theoretical
  Aspects of Software Engineering}, pages 127--135, 445 Hoes Lane Piscataway,
  NJ 08854 USA, 2009. IEEE.

\bibitem[SLZ13]{She&2013}
Yuchao She, Hui Li, and Hui Zhu.
\newblock {UVHM}: {Model} checking based formal analysis scheme for
  hypervisors.
\newblock In Khabib Mustofa, Erich~J. Neuhold, A.~Min Tjoa, Edgar Weippl, and
  Ilsun You, editors, {\em Information and Communication Technology}, pages
  300--305, Berlin, Heidelberg, 2013. Springer Berlin Heidelberg.

\bibitem[SMX18]{Shrestha&18}
Roshan Shrestha, Hoda Mehrpouyan, and Dianxiang Xu.
\newblock Model checking of security properties in industrial control systems
  ({ICS}).
\newblock In {\em Proceedings of the Eighth ACM Conference on Data and
  Application Security and Privacy}, CODASPY ’18, page 164–166, New York,
  NY, USA, 2018. Association for Computing Machinery.

\bibitem[{Sne}91]{Snekkenes1991}
E.~{Snekkenes}.
\newblock Exploring the {BAN} approach to protocol analysis.
\newblock In {\em Proceedings. 1991 IEEE Computer Society Symposium on Research
  in Security and Privacy}, pages 171--181, Oakland, CA, USA, may 1991. IEEE.

\bibitem[Spi89]{Spivey1989}
J.M. Spivey.
\newblock {\em The {Z} Notation: {A} Reference Manual}.
\newblock Prentice-Hall, USA, 1989.

\bibitem[SPSK17]{Siddavatam&17}
Irfan Siddavatam, Sachin Parekh, Tanay Shah, and Faruk Kazi.
\newblock Testing and validation of {Modbus}/{TCP} protocol for secure {SCADA}
  communication in {CPS} using formal methods.
\newblock {\em Scalable Computing: Practice and Experience}, 18, 11 2017.

\bibitem[SPY{\etalchar{+}}16]{Stefanescu&16}
Andrei Stefanescu, Daejun Park, Shijiao Yuwen, Yilong Li, and Grigore
  Roundefinedu.
\newblock Semantics-based program verifiers for all languages.
\newblock In {\em Proceedings of the 2016 ACM SIGPLAN International Conference
  on Object-Oriented Programming, Systems, Languages, and Applications}, OOPSLA
  2016, page 74–91, New York, NY, USA, 2016. Association for Computing
  Machinery.

\bibitem[ST12]{Song&12b}
Fu~Song and Tayssir Touili.
\newblock Efficient malware detection using model-checking.
\newblock In {\em International Symposium on Formal Methods}, pages 418--433,
  Berlin, Heidelberg, 2012. Springer, Springer Berlin Heidelberg.

\bibitem[ST14a]{Song&14b}
Fu~Song and Tayssir Touili.
\newblock Model-checking for {Android} malware detection.
\newblock In Jacques Garrigue, editor, {\em Programming Languages and Systems},
  pages 216--235, Cham, 2014. Springer International Publishing.

\bibitem[ST14b]{Song&14}
Fu~Song and Tayssir Touili.
\newblock Pushdown model checking for malware detection.
\newblock {\em International Journal on Software Tools for Technology
  Transfer}, 16(2):147--173, 2014.

\bibitem[SWA{\etalchar{+}}12]{Steward&12}
C.~{Steward Jr.}, L.~A. {Wahsheh}, A.~{Ahmad}, J.~M. {Graham}, C.~V. {Hinds},
  A.~T. {Williams}, and S.~J. {DeLoatch}.
\newblock Software security: {The} dangerous afterthought.
\newblock In {\em 2012 Ninth International Conference on Information Technology
  - New Generations}, pages 815--818, 445 Hoes Lane Piscataway, NJ 08854 USA,
  April 2012. IEEE.

\bibitem[Tea19]{Coq19}
The Coq~Development Team.
\newblock {\em The Coq Reference Manual}.
\newblock LogiCal Project, January 2019.
\newblock Version 8.9.1.

\bibitem[TJ07]{Torlak&07}
Emina Torlak and Daniel Jackson.
\newblock {Kodkod}: {A} relational model finder.
\newblock In Orna Grumberg and Michael Huth, editors, {\em Tools and Algorithms
  for the Construction and Analysis of Systems}, pages 632--647, Berlin,
  Heidelberg, 2007. Springer Berlin Heidelberg.

\bibitem[TJK19]{Jorgensen&19}
{Peter W{\"u}rtz Vinther} Tran-J{\o}rgensen and Tomas Kulik.
\newblock Migrating {Overture} to a different {IDE}.
\newblock In {Carl } Gamble and Luis {Diogo Couto}, editors, {\em Proceedings
  of the 17th Overture Workshop}, number CS-TR- 1530 - 2019 in Technical Report
  Series, pages 32--47, UK, 2019. Newcastle University.

\bibitem[TJKBL19]{Tran-Joergensen&19}
Peter W.~V. Tran-J\o{}rgensen, Tomas Kulik, Jalil Boudjadar, and Peter~Gorm
  Larsen.
\newblock Security analysis of cloud-connected industrial control systems using
  combinatorial testing.
\newblock In {\em Proceedings of the 17th ACM-IEEE International Conference on
  Formal Methods and Models for System Design}, MEMOCODE ’19, New York, NY,
  USA, 2019. Association for Computing Machinery.

\bibitem[TM17]{Taylor&17}
Vincent~F. Taylor and Ivan Martinovic.
\newblock Short paper: {A} longitudinal study of financial apps in the {Google}
  {Play} {Store}.
\newblock In Aggelos Kiayias, editor, {\em Financial Cryptography and Data
  Security}, pages 302--309, Cham, 2017. Springer International Publishing.

\bibitem[TP16]{Tabrizi&16}
Farid~Molazem Tabrizi and Karthik Pattabiraman.
\newblock Formal security analysis of smart embedded systems.
\newblock In {\em Proceedings of the 32nd Annual Conference on Computer
  Security Applications}, ACSAC ’16, page 1–15, New York, NY, USA, 2016.
  Association for Computing Machinery.

\bibitem[TSS16]{Tsukada&16}
K.~{Tsukada}, K.~{Sawada}, and S.~{Shin}.
\newblock A toolchain on model checking {SPIN} via {Kalman} decomposition for
  control system software.
\newblock In {\em 2016 IEEE International Conference on Automation Science and
  Engineering (CASE)}, pages 300--305, 445 Hoes Lane Piscataway, NJ 08854 USA,
  Aug 2016. IEEE.

\bibitem[TTB08]{Tamura&08}
Naoyuki Tamura, Tomoya Tanjo, and Mutsunori Banbara.
\newblock System description of a {SAT}-based {CSP} solver {Sugar}, 01 2008.

\bibitem[Tur06]{Turuani&06}
Mathieu Turuani.
\newblock The {CL}-{Atse} protocol analyser.
\newblock In Frank Pfenning, editor, {\em Term Rewriting and Applications},
  pages 277--286, Berlin, Heidelberg, 2006. Springer Berlin Heidelberg.

\bibitem[TVR16]{Turuan&16}
Mathieu Turuani, Thomas Voegtlin, and Michael Rusinowitch.
\newblock Automated verification of {Electrum} wallet.
\newblock In {\em International Conference on Financial Cryptography and Data
  Security}, pages 27--42, Berlin, Heidelberg, 2016. Springer.

\bibitem[UA19]{Urbach&19}
Nils Urbach and Frederik Ahlemann.
\newblock {\em Digitalization as a Risk: {Security} and Business Continuity
  Management Are Central Cross-Divisional Functions of the Company}, pages
  85--92.
\newblock Springer International Publishing, Cham, 2019.

\bibitem[VCJ{\etalchar{+}}13]{Vasudevan&13}
A.~{Vasudevan}, S.~{Chaki}, L.~{Jia}, J.~{McCune}, J.~{Newsome}, and
  A.~{Datta}.
\newblock Design, implementation and verification of an {eXtensible} and
  modular hypervisor framework.
\newblock In {\em 2013 IEEE Symposium on Security and Privacy}, pages 430--444,
  445 Hoes Lane Piscataway, NJ 08854 USA, May 2013. IEEE.

\bibitem[VCM{\etalchar{+}}16]{Vasudevan&16}
Amit Vasudevan, Sagar Chaki, Petros Maniatis, Limin Jia, and Anupam Datta.
\newblock {{\"u}berSpark}: {Enforcing} verifiable object abstractions for
  automated compositional security analysis of a hypervisor.
\newblock In {\em 25th {USENIX} Security Symposium ({USENIX} Security 16)},
  pages 87--104, Austin, TX, August 2016. {USENIX} Association.

\bibitem[vOM12]{vonOheimb&12}
David von Oheimb and Sebastian M{\"o}dersheim.
\newblock {ASLan++} --- {A} formal security specification language for
  distributed systems.
\newblock In Bernhard~K. Aichernig, Frank~S. de~Boer, and Marcello~M.
  Bonsangue, editors, {\em Formal Methods for Components and Objects}, pages
  1--22, Berlin, Heidelberg, 2012. Springer Berlin Heidelberg.

\bibitem[WANC16]{Williams&16}
M.~{Williams}, L.~{Axon}, J.~R.~C. {Nurse}, and S.~{Creese}.
\newblock Future scenarios and challenges for security and privacy.
\newblock In {\em 2016 IEEE 2nd International Forum on Research and
  Technologies for Society and Industry Leveraging a better tomorrow (RTSI)},
  pages 1--6, 445 Hoes Lane Piscataway, NJ 08854 USA, September 2016. IEEE.

\bibitem[WD96]{WoodcockD1996}
Jim Woodcock and Jim Davies.
\newblock {\em Using {Z}: {Specification}, Refinement, and Proof}.
\newblock Prentice-Hall, USA, 1996.

\bibitem[Wei11]{Weinberger&11}
Sharon Weinberger.
\newblock Computer security: {Is} this the start of cyberwarfare?
\newblock {\em Nature}, 474:142--5, 06 2011.

\bibitem[WGS{\etalchar{+}}19]{Wang&19}
R.~{Wang}, Y.~{Guan}, H.~{Song}, X.~{Li}, X.~{Li}, Z.~{Shi}, and X.~{Song}.
\newblock A formal model-based design method for robotic systems.
\newblock {\em IEEE Systems Journal}, 13(1):1096--1107, March 2019.

\bibitem[{Win}90]{Wing&90}
J.~M. {Wing}.
\newblock A specifier's introduction to formal methods.
\newblock {\em Computer}, 23(9):8--22, September 1990.

\bibitem[Win98]{Wing&98}
Jeannette~M Wing.
\newblock A symbiotic relationship between formal methods and security.
\newblock In {\em Proceedings Computer Security, Dependability, and Assurance:
  From Needs to Solutions (Cat. No. 98EX358)}, pages 26--38, 445 Hoes Lane
  Piscataway, NJ 08854 USA, 1998. IEEE.

\bibitem[WMPO16]{Wardell&16}
Dean~C. Wardell, Robert~F. Mills, Gilbert~L. Peterson, and Mark~E. Oxley.
\newblock {A Method for Revealing and Addressing Security Vulnerabilities in
  Cyber-physical Systems by Modeling Malicious Agent Interactions with Formal
  Verification}.
\newblock {\em Procedia Computer Science}, 95:24--31, 2016.
\newblock Complex Adaptive Systems Los Angeles, CA November 2-4, 2016.

\bibitem[WNH17]{Wich&17}
Tobias Wich, Daniel Nemmert, and Detlef H{\"{u}}hnlein.
\newblock Towards secure and standard-compliant implementations of the {PSD2}
  directive.
\newblock In Lothar Fritsch, Heiko Ro{\ss}nagel, and Detlef H{\"{u}}hnlein,
  editors, {\em Open Identity Summit 2017, October 5-6, 2017, Karlstad
  University, Sweden}, volume {P-277} of {\em {LNI}}, pages 63--80, Bonn, DE,
  2017. Gesellschaft f{\"{u}}r Informatik.

\bibitem[WSC{\etalchar{+}}08]{WoodcockSCCJ2008}
Jim Woodcock, Susan Stepney, David Cooper, John Clark, and Jeremy Jacob.
\newblock The certification of the {Mondex} electronic purse to {ITSEC} {Level}
  {E6}.
\newblock {\em Formal Asp. Comput.}, 20(1):5--19, 2008.

\bibitem[WSC17]{Wang&17}
T.~{Wang}, Q.~{Su}, and T.~{Chen}.
\newblock Formal analysis of security properties of cyber-physical system based
  on timed automata.
\newblock In {\em 2017 IEEE Second International Conference on Data Science in
  Cyberspace (DSC)}, pages 534--540, 2017.

\bibitem[WZM12]{Wang&12}
W.~{Wang}, Q.~{Zeng}, and A.~P. {Mathur}.
\newblock A security assurance framework combining formal verification and
  security functional testing.
\newblock In {\em 2012 12th International Conference on Quality Software},
  pages 136--139, 2012.

\bibitem[XCWC13]{Xing&13}
Luyi Xing, Yangyi Chen, XiaoFeng Wang, and Shuo Chen.
\newblock {InteGuard}: {Toward} automatic protection of third-party web service
  integrations.
\newblock In {\em 20th Annual Network and Distributed System Security
  Symposium, {NDSS} 2013, San Diego, California, USA, February 24-27, 2013},
  page~17, USA, 2013. The Internet Society.

\bibitem[XWL14]{Xiao&14}
Meihua Xiao, Zilong Wan, and Hongling Liu.
\newblock The formal verification and improvement of simplified {SET} protocol.
\newblock {\em Journal of Software}, 9, 09 2014.

\bibitem[YHXH05]{Zheng&05}
{Yu Zheng}, D.~{He}, {Xiaohu Tang}, and {Hongxia Wang}.
\newblock {AKA} and authorization scheme for {4G} mobile networks based on
  trusted mobile platform.
\newblock In {\em 2005 5th International Conference on Information
  Communications Signal Processing}, pages 976--980, 2005.

\bibitem[YJS{\etalchar{+}}19]{Yoo&19}
J.~{Yoo}, Y.~{Jung}, D.~{Shin}, M.~{Bae}, and E.~{Jee}.
\newblock Formal modeling and verification of a federated byzantine agreement
  algorithm for blockchain platforms.
\newblock In {\em 2019 IEEE International Workshop on Blockchain Oriented
  Software Engineering (IWBOSE)}, pages 11--21, 445 Hoes Lane Piscataway, NJ
  08854 USA, 2019. IEEE.

\bibitem[ZKWG16]{Zeng&16}
Wen Zeng, Maciej Koutny, Paul Watson, and Vasileios Germanos.
\newblock Formal verification of secure information flow in cloud computing.
\newblock {\em Journal of Information Security and Applications}, 27:103--116,
  2016.

\bibitem[ZMSZ12]{Zhang&12}
Wei Zhang, Wenke Ma, Huiling Shi, and Fu-qiang Zhu.
\newblock Model checking and verification of the internet payment system with
  {SPIN}.
\newblock {\em JSW}, 7(9):1941--1949, 2012.

\bibitem[ZRM14]{Zonouz&14}
S.~{Zonouz}, J.~{Rrushi}, and S.~{McLaughlin}.
\newblock Detecting industrial control malware using automated {PLC} code
  analytics.
\newblock {\em IEEE Security Privacy}, 12(6):40--47, Nov 2014.

\bibitem[ZWSM15]{Zhang&15}
Danfeng Zhang, Yao Wang, G.~Edward Suh, and Andrew~C. Myers.
\newblock A hardware design language for timing-sensitive information-flow
  security.
\newblock In {\em Proceedings of the Twentieth International Conference on
  Architectural Support for Programming Languages and Operating Systems},
  ASPLOS ’15, page 503–516, New York, NY, USA, 2015. Association for
  Computing Machinery.

\end{thebibliography}

\newpage
\section*{APPENDIX}
\label{sec:table}
\setcounter{section}{1}
% !TEX root = ../../main.tex
This appendix provides an overview of the several surveyed research works in the form of a cross-tabulation according to \cite{Macedo&15}. We group the 115 covered references by Domain, Level Of Abstraction (LOA) and Approach (App.). 
\scriptsize
\begin{longtable}{lllp{8.5cm}}
\toprule
Domain & LOA & App. &  Works                                                                                                                                   \\
\midrule
\endhead
\midrule
\multicolumn{4}{r}{{Continued on next page}} \\
\midrule
\endfoot

\bottomrule
\endlastfoot
\multirow{12}{*}{Financial} & \multirow{2}{*}{Application} & MC &                                                       {\cite{Chothia&17}} {\cite{Kumar&19}} {\cite{Iadarola&19}} {\cite{Turuan&16}} \\
           &          & TP &                                                                                               {\cite{Annenkov&18}} {\cite{Park&18}} \\
\cline{2-4}
           & \multirow{2}{*}{System} & LW &  {\cite{Aarts&13}} \\
           &          & MC &                                                                    {\cite{Ahamad2012}} {\cite{Obaid2017}} {\cite{Zhang&12}} {\cite{Santone&13}} {\cite{Yoo&19}} {\cite{Duan&18}} \\
\cline{2-4}
           & \multirow{2}{*}{Protocol} & MC &                                                  {\cite{Abughazalah&14}} {\cite{Xiao&14}} {\cite{Hartung&12}} {\cite{Alsheri&13}} {\cite{Bojjagani&15}} \\
           &          & TP &                                                                                                                 {\cite{Madhoun&16}} \\
\cline{2-4}
           & \multirow{3}{*}{Implementation} & LW &                                                                            {\cite{Ibrar&17}} {\cite{Taylor&17}} {\cite{Freitas&18}} \\
           &          & MC &                                                                                                            {\cite{Andrychowicz&14}} \\
           &          & TP &                                                                                                                    {\cite{Fett&19}} \\
\cline{2-4}
           & \multirow{2}{*}{Hardware} & MC &                                                                                             {\cite{Athalye&19}} {\cite{Chothia&15}} \\
           &          & TP &                                                                                          {\cite{Marcedone&19}} {\cite{Arapinis&19}} \\
\cline{1-4}
\cline{2-4}
\multirow{10}{*}{Industrial} & Application & MC &                 {\cite{Heneghan&19}} {\cite{Nigam&19}} {\cite{Wang&19}} {\cite{Mercaldo&19}} {\cite{Zonouz&14}} {\cite{Kottler&17}} \\ \cline{2-4}
           & \multirow{2}{*}{System} & LW &                                                                                                         {\cite{Tran-Joergensen&19}} \\
           &          & MC &                             {\cite{Rocchetto&17}} {\cite{Denzel&17}} {\cite{Hailesellasie&18}} {\cite{Rysavy&13}} {\cite{Kulik&19}} {\cite{Wang&17}} \\
\cline{2-4}
           & \multirow{3}{*}{Protocol} & LW &                                                                                            {\cite{Siddavatam&17}} {\cite{Amoah&16}} \\
           &          & MC &                                                                              {\cite{Nardone&16}} {\cite{Puys&16}} {\cite{Bruni&14}} \\
           &          & TP &                                                                                                                  {\cite{Dreier&17}} \\
\cline{2-4}
           & \multirow{2}{*}{Implementation} & LW &     {\cite{OHearn&18}} \\
           &          & MC &                                                     {\cite{Apvrille&16}} {\cite{Cofer&18}} {\cite{Shrestha&18}} {\cite{Tsukada&16}} \\
           &          & TP &                                                                                           {\cite{Cofer&18}} \\
\cline{2-4}
           & \multirow{2}{*}{Hardware} & MC &                                                                                                                  {\cite{Abbasi&17}} \\
           &          & TP &                                                               {\cite{Huang&18}} {\cite{Lerner&14}} {\cite{Love&11}} {\cite{Guo&17}} \\
\cline{1-4}
\cline{2-4}
\multirow{13}{*}{Consumer} & \multirow{3}{*}{Application} & LW &               {\cite{Macedo&13}} {\cite{Dam&17}} {\cite{Dam&18}} {\cite{Mayo&15}} \\
           &          & MC &                                         {\cite{Song&12b}} {\cite{Song&14}} {\cite{Song&14b}} {\cite{Mercaldo&16}} \\
           &          & TP &                                                                                                                   {\cite{Smith&16}} \\
\cline{2-4}
           & \multirow{2}{*}{System} & MC &                {\cite{Bagheri&15}} {\cite{Bagheri&18}} {\cite{Devyanin&14}} {\cite{Kumar&17}} {\cite{Moshin&17}} {\cite{Busold&13}} \\
           &          & TP &                                                                                                                     {\cite{Mai&13}} \\
\cline{2-4}
           & \multirow{2}{*}{Protocol} & MC &                                     {\cite{Chen&16}} \\
           &          & TP &                                                                                                             {\cite{Cohn-Gordon&17}} \\
\cline{2-4}
           & \multirow{3}{*}{Implementation} & LW &                                                                                             {\cite{Babic&19}} {\cite{Distefano&19}} \\
           &          & MC &                                                                                                                   {\cite{Sasse&12}} \\
           &          & TP &                                                                             {\cite{Jang&12}} {\cite{Guha&11}} {\cite{Morrisett&12}} \\
\cline{2-4}
           & \multirow{3}{*}{Hardware} & LW &                                                                                                              {\cite{Ferraiuolo&17}} \\
           &          & MC &                                                                           {\cite{Tabrizi&16}} {\cite{Guo&16}} {\cite{Sepulveda&18}} \\
           &          & TP &                                                                                                                {\cite{Guo&16}} {\cite{Letan&16}} \\
\cline{1-4}
\cline{2-4}
\multirow{13}{*}{Enterprise} & \multirow{2}{*}{Application} & MC &                                                                           {\cite{Skowyra&14}} {\cite{Armando&12}} {\cite{She&2013}} {\cite{Cook18}} \\
           &          & TP &                                                                                               {\cite{Chen&14}} {\cite{Cook18}} \\
\cline{2-4}
           & \multirow{3}{*}{System} & LW &                                                                                                                    {\cite{Zeng&16}} \\
           &          & MC &                                                           {\cite{Madi&18}} {\cite{Hao&13}} {\cite{Bleikertz&15}} {\cite{Souaf&18}} \\
           &          & TP &                                                           {\cite{Wang&12}} {\cite{Jarraya&12}} {\cite{Masmoudi&14}} \\
\cline{2-4}
           & \multirow{2}{*}{Protocol} & MC &                                          {\cite{Aiash&15}} {\cite{Kumar&14}} {\cite{Karla&15}} {\cite{Roy&17}} \\
           &          & TP &                                                           {\cite{Basin&18}} {\cite{Chudnov&18}} {\cite{Jayaraman&2014}} \\
\cline{2-4}
           & \multirow{3}{*}{Implementation} & LW &                                                                            {\cite{Nunes&15}} {\cite{Xing&13}} {\cite{Vasudevan&16}} \\
           &          & MC &                                                                                                               {\cite{Vasudevan&13}} \\
           &          & TP &                                                                                                                  {\cite{Daejun&18}} \\
\cline{2-4}
           & \multirow{3}{*}{Hardware} & LW &                                                                                                             {\cite{Ferraiuolo&17a}} \\
           &          & MC &                                                                                                                     {\cite{Bai&14}} \\
           &          & TP &                                                                                             {\cite{Seol&2015}} {\cite{Kunemann&13}} \\
           \midrule
           \caption{The sorting of the 120 works leading to the taxonomy displayed in Figure \ref{fig:taxonomy}.}
\end{longtable}

\end{document}